\documentclass[twocolumn]{aastex631}
\usepackage[T1]{fontenc}
\usepackage{xspace}
\newcommand{\fluxcgs}{erg~s$^{-1}$~cm$^{-2}$}
\newcommand{\source}{4U~1728$-$34}
\newcommand{\nicer}{\emph{NICER}\xspace}
\newcommand{\fa}{$f_a$~}
\newcommand{\flencecgs}{erg~cm$^{-2}$}
\extrafloats{100} 
\DeclareUnicodeCharacter{2212}{-}
\submitjournal{ApJ}

\graphicspath{{./}{figures/}}
\begin{document}

\title{NICER observations of thermonuclear bursts from 4U 1728−34: Detection of oscillations prior to the onset of two bursts}

\author[0000-0002-5665-3452]{Z. Funda Bostanc\i~}
\affiliation{Istanbul University, Science Faculty, Department of Astronomy and Space Sciences, Beyaz\i t, 34119, \.Istanbul, T\"urkiye}
\affiliation{Istanbul University Observatory Research and Application Center, Istanbul University 34119, \.Istanbul T\"urkiye}

\author[0000-0002-4729-1592]{Tu\u{g}ba Boztepe}
\affiliation{Istanbul University, Graduate School of Sciences, Department of Astronomy and Space Sciences, Beyaz\i t, 34119, \.Istanbul, T\"urkiye}

\author[0000-0002-3531-9842]{Tolga G\"uver}
\affiliation{Istanbul University, Science Faculty, Department of Astronomy and Space Sciences, Beyaz\i t, 34119, \.Istanbul, T\"urkiye}
\affiliation{Istanbul University Observatory Research and Application Center, Istanbul University 34119, \.Istanbul T\"urkiye}

\author[0000-0001-7681-5845]{Tod E. Strohmayer}
\affiliation{Astrophysics Science Division and Joint Space-Science Institute, NASA's Goddard Space Flight Center, Greenbelt, MD 20771, USA}

\author[0000-0002-6447-3603]{Yuri Cavecchi}
\affiliation{Departament de Fs\'{i}ca, EEBE, Universitat Polite\'{c}nica de Catalunya, Av. Eduard Maristany 16, 08019 Barcelona, Spain}

\author[0000-0002-5274-6790]{Ersin G\"o\u{g}\"u\c{s}}
\affiliation{Faculty of Engineering and Natural Sciences, Sabanc\i~University, Orhanl\i-Tuzla 34956, \.Istanbul, T\"urkiye }

\author[0000-0002-3422-0074]{Diego Altamirano} 
\affiliation{School of Physics and Astronomy, University of Southampton, Southampton, SO17 1BJ, UK}

\author[0000-0002-7252-0991]{Peter Bult}
\affiliation{Department of Astronomy, University of Maryland, College Park, MD 20742, USA}
\affiliation{Astrophysics Science Division, NASA Goddard Space Flight Center, Greenbelt, MD 20771, USA}

\author[0000-0001-8804-8946]{Deepto~Chakrabarty}
\affil{MIT Kavli Institute for Astrophysics and Space Research, Massachusetts Institute of Technology, Cambridge, MA 02139, USA}

\author[0000-0002-6449-106X]{Sebastien~Guillot}
\affil{Institut de Recherche en Astrophysique et Plan\'{e}tologie, UPS-OMP, CNRS, CNES, 9 avenue du Colonel Roche, BP 44346, F-31028 Toulouse Cedex 4, France}

\author[0000-0002-6789-2723]{Gaurava K. Jaisawal}
\affil{DTU Space, Technical University of Denmark, Elektrovej 327-328, DK-2800 Lyngby, Denmark}

\author[0000-0002-0380-0041]{Christian~Malacaria} 
\affil{International Space Science Institute, Hallerstrasse 6, 3012 Bern, Switzerland}

\author[0000-0001-9822-6937]{Giulio C. Mancuso} 
\affiliation{Instituto Argentino de Radioastronom\'{\i}a (CCT-La Plata, CONICET; CICPBA), C.C. No. 5, 1894 Villa Elisa, Argentina}
\affiliation{Facultad de Ciencias Astron\'omicas y Geof\'{\i}sicas, Universidad Nacional de La Plata, Paseo del Bosque s/n, 1900 La Plata, Argentina}

\author[0000-0002-0118-2649]{Andrea Sanna}
\affil{Dipartimento di Fisica, Universit'a degli Studi di Cagliari, SP Monserrato-Sestu km 0.7, Monserrato 09042, Italy}

\author[0000-0001-7079-9338]{Jean H. Swank}
\affil{Astrophysics Science Division, NASA Goddard Space Flight Center, Greenbelt, MD 20771, USA}

%% Note that the \and command from previous versions of AASTeX is now
%% depreciated in this version as it is no longer necessary. AASTeX 
%% automatically takes care of all commas and "and"s between authors names.

%% AASTeX 6.31 has the new \collaboration and \nocollaboration commands to
%% provide the collaboration status of a group of authors. These commands 
%% can be used either before or after the list of corresponding authors. The
%% argument for \collaboration is the collaboration identifier. Authors are
%% encouraged to surround collaboration identifiers with ()s. The 
%% \nocollaboration command takes no argument and exists to indicate that
%% the nearby authors are not part of surrounding collaborations.

%% Mark off the abstract in the ``abstract'' environment. 
\begin{abstract}

We present temporal and time-resolved spectral analyses of all the thermonuclear X-ray bursts observed from the neutron star low-mass X-ray binary~(LMXB) \source~with \nicer from June 2017 to September 2019. In total, we detected 11 X-ray bursts from the source and performed time-resolved spectroscopy. Unlike some of the earlier results for other bursting sources from \nicer, our spectral results indicate that the use of a scaling factor for the persistent emission is not statistically necessary. This is primarily a result of the strong interstellar absorption in the line of sight towards \source, which causes the count rates to be significantly lower at low energies. We also searched for burst oscillations and detected modulations in six different bursts at around the previously known burst oscillation frequency of 363~Hz. Finally, we report the detection of oscillations prior to two bursts at 356 and 359~Hz, respectively. This is the first time in the literature where burst oscillations are detected before the rapid rise in X-ray flux, from any known burster. These oscillations disappear as soon as the burst rise starts and occur at a somewhat lower frequency than the oscillations we detect during the bursts.

\end{abstract}

%% Keywords should appear after the \end{abstract} command. 
%% The AAS Journals now uses Unified Astronomy Thesaurus concepts:
%% https://astrothesaurus.org
%% You will be asked to selected these concepts during the submission process
%% but this old "keyword" functionality is maintained in case authors want
%% to include these concepts in their preprints.

%% From the front matter, we move on to the body of the paper.
%% Sections are demarcated by \section and \subsection, respectively.
%% Observe the use of the LaTeX \label
%% command after the \subsection to give a symbolic KEY to the
%% subsection for cross-referencing in a \ref command.
%% You can use LaTeX's \ref and \label commands to keep track of
%% cross-references to sections, equations, tables, and figures.
%% That way, if you change the order of any elements, LaTeX will
%% automatically renumber them.
%%
%% We recommend that authors also use the natbib \citep
%% and \citet commands to identify citations.  The citations are
%% tied to the reference list via symbolic KEYs. The KEY corresponds
%% to the KEY in the \bibitem in the reference list below. 

\section{Introduction}

Thermonuclear X-ray bursts (hereafter X-ray bursts) are flashes in X-rays, observed from numerous neutron star low-mass X-ray binary systems \citep{2020ApJS..249...32G}. These flashes result from the unstable nuclear burning of the accreted material accumulated on the surface of the neutron star \citep{1975ApJ...195..735H,1978ApJ...220..291L}. During such an event, the observed X-ray intensity increases by a factor of $\sim10$ within $\sim0.5-5$~s, and then decreases exponentially ($\sim10-100$ s) as the surface of the star cools down. The energy released during a burst is typically 10$^{39}$ $-$ 10$^{40}$ ergs. The peak flux, duration, evolution and other properties of the bursts depend on the chemical composition of matter and the proportion of material deposited per unit star surface area, hence on the accretion rate \citep{tugba_boztepe_2022_7600607}. Since the amount of material deposited on the neutron star may evolve through different accretion rates for different bursts, in principle different burning regimes may be observed from the same source \citep[see, e.g.,][]{2006csxs.book..113S}. 
The spectral and timing properties of X-ray bursts can be a useful tool for understanding neutron star parameters 
\citep[such as radius, mass, and the equation of state,][]{2010AdSpR..45..949B,2016ApJ...820...28O,2016ARA&A..54..401O,2019ApJ...887L..26B}. However, a comprehensive understanding of the interaction between burst emission and the surrounding environment is equally crucial for such studies. For example, recent findings from \nicer, as well as some earlier results from RXTE \citep[see, e.g.,][]{2013ApJ...772...94W,2015ApJ...801...60W,2018ApJ...855L...4K,2018ApJ...856L..37K,2019ApJ...885L...1B,2020MNRAS.499..793B,2022MNRAS.510.1577G,2022ApJ...935..154G,2022ApJ...940...81B}, indicate that the persistent emission of a source may increase by up to an order of magnitude, especially around the peaks of the bursts. This excess emission is observed in the soft X-ray band (mostly below 3.0~keV), affecting the results obtained with instruments sensitive in the low energy bandpass. These findings are further supported by simulations showing an increase in the mass accretion rate onto the neutron star due to the combined effects of Poynting-Robertson drag, and reflection~\citep{2018ApJ...867L..28F,2020NatAs...4..541F,2022MNRAS.509.1736S}.

Nearly two decades after the first discovery of an X-ray burst from \source\, \citep{1976ApJ...210L..13H}, temporary oscillations during some of the bursts from this source were first discovered at 363 Hz by \cite{1996ApJ...469L...9S}. 
Since then these burst oscillations have been firmly confirmed in approximately 20\% of all known Type I X-ray bursters \footnote{\url{https://personal.sron.nl/~jeanz/bursterlist.html}} \citep{2012ARA&A..50..609W,2019ApJS..245...19B}. The observed frequencies typically range from $\sim250$ Hz to $\sim600$ Hz, and they are attributed to the spin frequency of the neutron star.
Burst oscillations are likely a result of rotational modulations caused by an asymmetric temperature distribution on the neutron star surface \citep{1996ApJ...469L...9S,1997ApJ...487L..77S,2003Natur.424...42C}. They are generally observed to occur at the rise/decay of some of the X-ray bursts \citep{Watts_2005, 2012ARA&A..50..609W}. Although the oscillation frequencies remain relatively consistent, there might be slight shifts of a few Hz during the typical duration of the burst, which lasts only a few seconds. Additionally, the oscillations occasionally vanish and then reappear throughout the burst~\citep{2002ApJ...580.1048M,2002ApJ...581..550M}. 

\source~(a.k.a. the Slow Burster or MXB 1728--34) stands among the earliest discovered and most extensively studied bursting LMXBs. Its bursts were first explored by SAS-3 and Uhuru \citep{1971ApJ...169L..99K, 1976IAUC.2922....1L}. It is known for its regular X-ray bursts \citep[see, e.g.,][]{2016MNRAS.455.2004Z,2017A&A...599A..89K, 2018ApJ...860...88B} and burst oscillations \citep[see, e.g.,][]{1996ApJ...469L...9S,2001ApJ...551..907V,2001ApJ...554..340F,2017ApJ...841...41V,2019ApJ...878..145M}, with a total of 96 bursts reported by \cite{1984ApJ...281..337B}. According to the Multi-INstrument Burst ARchive~\citep[MINBAR\footnote{\url{https://burst.sci.monash.edu/minbar/}},][]{2020ApJS..249...32G} a total of 1173 bursts have been detected with multiple instruments (RXTE/PCA, BeppoSAX/WFCs, INTEGRAL/JEMX), and no event with short recurrence has been reported. The source is thought to be an ultracompact X-ray binary~\citep{2003ApJ...593L..35S,2008ApJS..179..360G} inferred from the burst behavior with a H-poor donor. However, \cite{2020MNRAS.495L..37V} and \cite{Vincentelli2023} found that the orbital period of \source~ must be greater than $\sim1.1$~h or even $\sim3$~h, based on infrared observations of the delay between X-ray bursts and its reflection off disc and companion. Such an estimate contradicts the proposed ultra-compact nature and suggests that the companion should be a helium star. The distance to the source is estimated to range from 4.4 to 5.1 kpc using Eddington limit estimations of the photospheric radius expansion bursts~\citep{2000ApJ...542.1034D,2001ApJ...551..907V,2003ApJ...590..999G,2019MNRAS.487.1626Q}. 

In this paper, we report the detection and the X-ray time-resolved spectral and temporal analysis of the bursts from \source,~as observed with \nicer. In total, we have identified 11 X-ray bursts in the archive, seven of which have previously been reported by~\cite{2019ApJ...878..145M}. We detect burst oscillations in six of these events, three of which were also found by~\cite{2019ApJ...878..145M}. Finally, we devote particular attention to two bursts that exhibited oscillations immediately before the observed increase in the count rate.

%%%%%%%%%%%%%%%%%%%%%%%%%%%%%%%%%%%%%%%%%%%%%
\section{Observations and Data Analysis}

\source \, was observed with the \nicer X-ray Telescope Instrument \citep[XTI,][]{2016SPIE.9905E..4XO,2016SPIE.9905E..1HG} 
on board the International Space Station. The source was monitored from June 2017 to September 2019 for a total unfiltered and cleaned exposure of 230~ks and 163~ks, respectively. We used all public observations available through HEASARC \footnote{\url{https://heasarc.gsfc.nasa.gov}}. These observations are gathered under ObsIDs starting with 0050150106, 1050150102 $-$ 1050150158, and 2587010101 $-$ 2587010104. We processed the data using \texttt{NICERDAS} v8c with \texttt{HEASoft} version v6.29c and used \texttt{ftool} \texttt{XSELECT} to extract light curves and spectra following the standard criteria of the \emph{nicerl2}\footnote{\url{https://heasarc.gsfc.nasa.gov/lheasoft/ftools/headas/nimaketime.html}} tool. We used the task \texttt{barycorr} to apply barycentric corrections for the analysis assuming the source coordinates (J2000) as 17$^{h}$ 31$^{m}$ 57.73$^{s}$ and $-$33$\degr$ 50$\arcmin$ 02.5$\arcsec$. 

In order to identify X-ray bursts, we generated 0.25~s binned light curves in the 0.5$-$10~keV energy range and searched for the characteristic fast-rise exponential-decay features \citep{2020ApJS..249...32G}. In total, we have identified 11 X-ray bursts across all the observations. The light curves of these bursts are shown in \autoref{fig:burst_lc2}, where BID denotes the burst number. Following \cite{2022ApJ...935..154G} and using 0.5~s light curves including all the events in the 0.5$-$10~keV range, we defined the start time of a burst when the count rate is 4$\sigma$ above the persistent rate (see Section~\ref{sec:analysis}). On the other hand, the rise time is defined as the interval between the burst start and the first moment when the count rate reached 98\% of the burst's peak value, which we label as the peak time. Regarding the decay phase of the bursts, we offer two definitions: the e-folding time is defined as the time when the count rate decreases by a factor of $e$ after the peak moment, and the decay length is the time when the count rate decreases to 10\% of the peak. The peak-rate, rise-time, pre-burst rate, e-folding time, and the decay time lengths of all bursts are listed in \autoref{tab:bursts}. 

\begin{figure}
 	\includegraphics[scale=0.5]{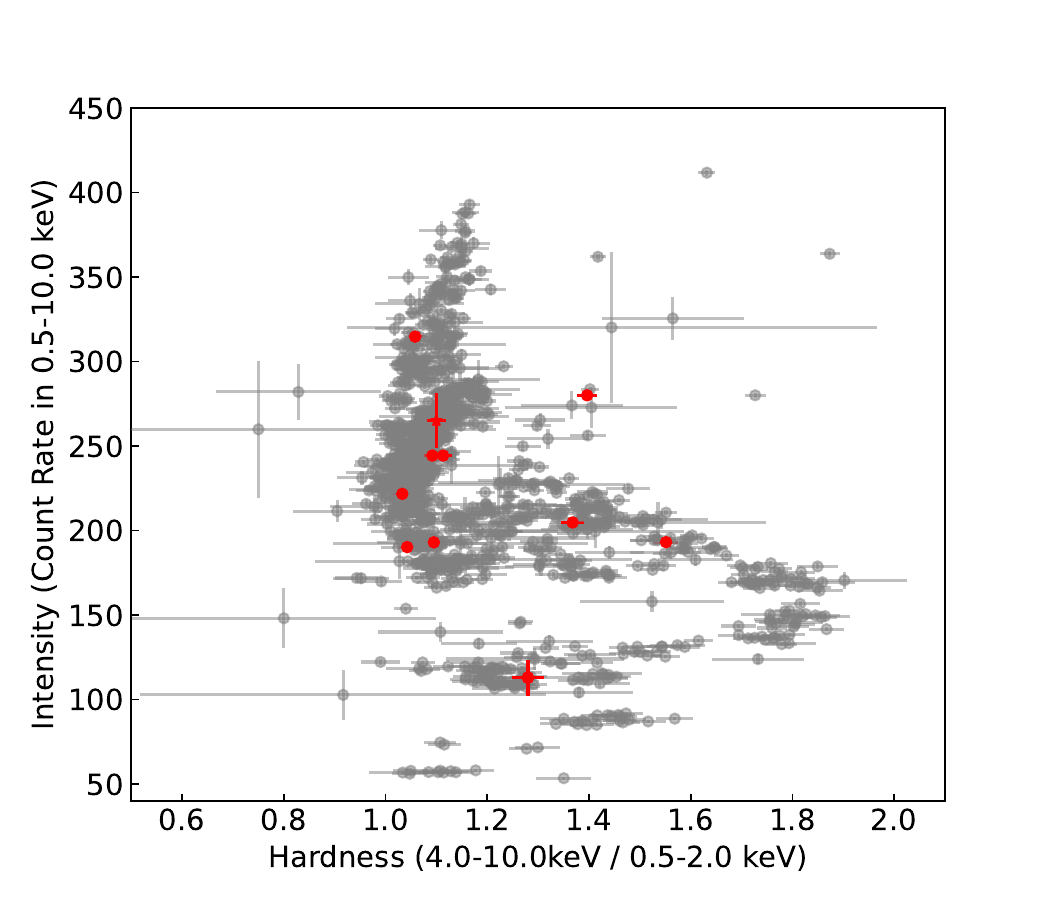}
    \caption{Hardness$-$Intensity diagram showing all \nicer\,observations of \source~from 2017 to 2019. 
    Observations in which an X-ray burst was detected are indicated by red filled circles.}
    \label{fig:burst_hd}
\end{figure}

\source~is classified as an atoll source based on the shape of the tracks in the color–color and hardness-intensity diagrams \citep{1989A&A...225...79H}. In order to determine the spectral state of the system when a burst was observed we constructed a hardness-intensity diagram. For this purpose, we generated light curves in the 0.5$-$2 and 4$-$10~keV bands with a time resolution of 128~s from clean event files \citep[see, e.g.,][]{2021ApJ...910...37G,2022MNRAS.510.1577G,2022ApJ...935..154G}. The resulting hardness-intensity diagram is shown in \autoref{fig:burst_hd}. Instances of X-ray bursts are indicated with red filled circles. From the hardness-intensity diagram, we see that the intensity primarily varies when the hardness is relatively low ($\sim$ 1.05) and bursts seem to happen when the hardness ratio is between 1.04 to 1.6. Notably, our observation dataset predominantly encompasses count rates $\gtrsim$ 220 counts s$^{-1}$ in the 0.5$-$10~keV range. On the other hand, during bursts 2 and 4, the intensity of \source~ exhibits slight deviations with a count rate in the 0.5$-$10~keV of 331 counts s$^{-1}$ and 112 counts s$^{-1}$, respectively. 

Although the low orbit of \nicer often prevents a conclusive analysis, we also examined the burst recurrence times, which is defined as the interval since the previous burst. Only for bursts 9 and 10 we can establish a limit on the recurrence as they both happen during the same observation. In that case, the recurrence time we measure is $\sim$ 4.56~hours. \cite{2020ApJS..249...32G} present the distribution of recurrence times for this source in the MINBAR catalog. The minimum recurrence observed from this source with RXTE/PCA is reported as 1.8 hours, while the maximum is 7.9 hours, with an average of 3.7~hours. The value we infer is compatible with this range although it remains slightly on the longer side of the distribution.

%%%%%%%%%%%%%%%%%%%%%%%%%%%%%%%%%%%%
\begin{table*}
\caption{Some characteristic properties of all thermonuclear X-ray bursts from \source~detected with \nicer. Parameters are derived from 0.5$-$10~keV light curves with a time resolution of 0.5~s, therefore the uncertainties in the rise and decay times are 0.5~s. BID shows the observed burst number.}
\begin{tabular}{cccccccc}
 \hline
  \hline
BID & MJD (TDB)  & OBSID & Peak-Rate$^a$ & Pre-burst Rate$^b$ & Rise-Time & e-folding time & Decay Time$^c$\\
 & & & counts/s & counts/s & s & s & s \\
 \hline
1 & 57940.82411458&0050150106&2265$\pm$71&223.3$\pm$1.6&1.75 & 5.5 & 15.0\\
2 & 57953.11721065&1050150102&2805$\pm$79&331.3$\pm$1.7&1.50& 6.5 & 16.5 \\
3 & 57979.45513310&1050150111&1857$\pm$65&220.8$\pm$1.6&1.50& 5.5 & 14.5\\
4 & 57998.09995718&1050150127&2289$\pm$69&111.4$\pm$3.2&2.25 & 6.5 & 23.5\\
5 & 58006.66877199&1050150134&1999$\pm$67&243.1$\pm$1.7&1.75 & 5.5 & 17.0\\
6 & 58010.09684144&1050150137&4204$\pm$94&232.1$\pm$1.5&0.75 & 5.0 & --$^{d}$\\
7 & 58156.69664468&1050150149&2265$\pm$70&207.2$\pm$1.5&1.00& 8.0 & 23.0\\
8 & 58308.89758681&1050150158&1921$\pm$65&203.5$\pm$1.4&1.00 & 10.5 & 21.5\\
9 & 58724.34034375&2587010101&2840$\pm$79&262.4$\pm$1.8&1.00 & 7.5 & 19.5\\
10 & 58724.53015278&2587010101&2022$\pm$67&240.1$\pm$6.7&1.00 & 9.0 & 36.0\\
11 & 58727.84581481&2587010104&2920$\pm$80&246.3$\pm$1.7&0.75& 5.5 & 14.0\\
\hline
\end{tabular}
\label{tab:bursts}\\
\footnotesize{$^a$ Pre-burst count rates are subtracted. }\\
\footnotesize{$^b$ Calculated as the average count rate 100~s prior to the burst start time. Uncertainties reflect the standard error of the average of all the count rates used.}\\
\footnotesize{$^c$ The time for the count rate to reach 10\% of the peak value.}\\
\footnotesize{$^d$ Good time interval (GTI) ended before reaching the criteria.}

\end{table*}
%%%%%%%%%%%%%%%%%%%%%%%%%%%%%%%%%%%%
%%%%%%%%%%%%%%%%%%%%%%%%%%%%%%%%%%%%

\subsection{Time Resolved Spectral Analysis}\label{sec:analysis}

The first step for the time-resolved spectral analysis is modeling the persistent emission of the source, since the persistent emission acts as a background during the burst. For that purpose we extracted a spectrum from the data obtained 100~s prior to each burst. In the case of burst 4 unfortunately there is only a 24~s interval available before or after the burst. We therefore used this data for estimating the persistent spectrum of the source before the burst. In the case of burst 10, the pre-burst data had only an exposure time of 30~s so we used postburst  X-ray spectra as our background. In burst 5, overshoot\footnote{\url{https://heasarc.gsfc.nasa.gov/docs/nicer/analysis_threads/overshoot-intro/}} rates are very high and show significant variations throughout the burst, which mostly affected our estimation of the energy distribution of the persistent emission. Therefore we did not include this burst in our spectral analysis \citep[see, e.g.,][] {2019ApJ...885L...1B,2022ApJ...935..154G}.

For each observation, we also generated background spectra using the \texttt{nibackgen3C50} tool \citep{2022AJ....163..130R} as well as the response matrix files (RMFs) and ancillary response files (ARFs) using \texttt{nicerrmf} and \texttt{nicerarf}, respectively.~We analyzed spectral data in the 1$-$10~keV range utilizing \emph{Sherpa} \citep{2001SPIE.4477...76F} with custom python scripts aided by Astropy  \citep{2018AJ....156..123A}, NumPy \citep{van2011numpy}, Matplotlib \citep{Hunter:2007}, and Pandas \citep{mckinney-proc-scipy-2010}. Following \cite{2021ApJ...910...37G,2022MNRAS.510.1577G,2022ApJ...935..154G}, we tried several modeling options including thermal (blackbody and disk blackbody models) and non-thermal (power-law, Compton scattering models) components. The resulting fits indicate that the pre-burst spectra of \source~can often be modeled assuming a simple absorbed power-law component. For the interstellar absorption, we used the \emph{tbabs} model \citep{2000ApJ...542..914W} assuming ISM abundances \citep{2000ApJ...542..914W} and cross-sections as presented by \cite{1995A&AS..109..125V}. Initially, we allowed the hydrogen column density values to be free before each burst. We then calculated the error weighted average of all the best-fit values and used the resulting value, N$\rm{_H}$ = 4.18 $\times10^{22}~\rm{cm}^{-2}$, as a fixed parameter for further analysis. To determine the average Hydrogen column density we excluded the pre-burst data from burst 4, where the exposure time for that spectrum was much shorter. Our best-fit hydrogen column density is in agreement with \cite{2016ApJ...827..134S,2017MNRAS.466.4991M,2019MNRAS.484.3004W} who reported values in the range N$\rm{_H}$ $\sim$ 3.9 $-$ 4.6 $\times10^{22}~\rm{cm}^{-2}$. However, it is important to note that in most instances, the hydrogen column density towards \source~is found to be much lower i.e, N$\rm{_H}$ $\sim$  2.6 $-$ 2.9 $\times10^{22}~\rm{cm}^{-2}$ \citep[see, e.g.,][] {2006A&A...448..817D,2008arXiv0810.0007W,2011A&A...530A..99E,2013ApJ...772...94W}. We present the best-fit results of this analysis in \autoref{tab:pers} where we also provide the 1$-$10~keV unabsorbed fluxes using the \emph{sample\_flux} command within \emph{sherpa}. The uncertainties in the fluxes are calculated by drawing 10000 samples from a normal distribution whose mean and the standard deviation equals the best fit parameter value and its 1$\sigma$ uncertainty.
 %%%%%%%%%%%%%%%%%%%%%%%%%%%%%%%%%%%%%%%%%%%%%%%%%%%%%%%%%%%%%%%%%%%%%%%%
\begin{table}
\centering
\caption{Best fit model results for pre-burst X-ray spectra of \source~using a single power-law model. $\gamma$ values are also provided assuming an Eddington limit of 4.04$\times10^{-8}$\fluxcgs~ as described in \autoref{sec:time_res_res}.}
\begin{tabular}{ccccccccc}
\hline 
\hline 
BID&$\Gamma$ & Flux* & $\chi^2$ / dof & $\gamma$\\
& && & \\
\hline    
1&1.52$\pm$0.02&3.61$\pm$0.11&478.44/332&0.09\\
2&1.86$\pm$0.01&4.85$\pm$0.11&381.17/389&0.12\\
3&1.49$\pm$0.02&3.48$\pm$0.10&312.01/326&0.09\\
4&1.80$\pm$0.08&0.67$\pm$0.01&11.54/18&0.02\\
6&1.86$\pm$0.02&3.39$\pm$0.09&341.23/323&0.08\\
7&1.71$\pm$0.02&3.12$\pm$0.09&328.59/302&0.08\\
8&1.83$\pm$0.02&2.98$\pm$0.09&313.32/290&0.07\\
9&1.81$\pm$0.01&3.88$\pm$0.10&378.87/357&0.10\\
10&1.85$\pm$0.01&4.13$\pm$0.10&450.21/362&0.10\\
11&1.81$\pm$0.02&3.84$\pm$0.10&315.86/342&0.95\\
\hline
\end{tabular}
\label{tab:pers}\\
\footnotesize{$^*$ Unabsorbed 1$-$10~keV flux in units of $\times10^{-9}$~\fluxcgs.}\\
\footnotesize{Note : We fixed the N$\rm{_H}$ = 4.18$\times10^{22}~\rm{cm}^{-2}$.}\\
\end{table}
%%%%%%%%%%%%%%%%%%%%%%%%%%%%%%%%%%%%%%%%%%%%%%%%%%%%%%%%%%%%%%%%%%%%%%%%

To track the spectral evolution throughout the bursts, we generated X-ray spectra following the methods outlined by \cite{2008ApJS..179..360G, 2012ApJ...747...76G, 2021ApJ...910...37G,2022MNRAS.510.1577G,2022ApJ...935..154G} by adaptively determining the exposure time. We started our exposure times for each spectrum from 0.125~s and increased following the change in the observed count rates, to be able to keep the uncertainty in the inferred spectral parameters as comparable as possible. A typical average count rate is $\sim$ 560 counts/s. For each X-ray spectrum,  we initially used the best-fit model with fixed parameters for the persistent emission and subtracted only the background generated by the \emph{nibackgen3C50} tool. If statistically required, we then added a blackbody component to account for the additional emission from the X-ray burst and followed its spectral evolution. We also calculated the bolometric X-ray flux of the blackbody component using the \emph{sample\_flux} command within \emph{sherpa} in the 0.01 $-$ 200~keV range for each modeled burst spectrum. In addition to this approach, we also tried to add a scaling factor, \fa \citep[following][]{2013ApJ...772...94W,2015ApJ...801...60W,2022MNRAS.510.1577G,2022ApJ...935..154G}, to the persistent emission model. However, as shown in Section~\ref{sec:time_res_res}, contrary to previous findings from \nicer, in the case of \source~this approach did not yield statistically significant improvements for most spectra.

\subsection{Search for Burst Oscillations}
We performed a timing analysis based on Z$^{2}_{n}$ statistics to search for burst oscillations across 11 bursts. The Z$^{2}$ statistic is defined as follows:
\begin{equation}
Z_n^2=\frac{2}{N_\gamma} \sum_{k=1}^n\left[\left(\sum_{j=1}^{N_\gamma} \cos k \nu t_j\right)^2+\left(\sum_{j=1}^{N_\gamma} \sin k \nu t_j\right)^2\right],
\end{equation}
where Z$^{2}$ represents the measured power of the signal, $n$ is the number of harmonics ($k = 1...n$ is the index), $N_\gamma$ is the number of photons used in the time bin, $\nu$ denotes the frequency under consideration, and $t_j$ is the arrival time of the $j$th count relative to some reference time. In the absence of a coherent signal Z$^{2}_{n}$ powers follow a $\chi^{2}$ distribution with 2n degrees of freedom \citep{buccheri}. 

We selected $n$  = 1 for our search. We constructed dynamical power spectra using search intervals of 2~s and 4~s. These time windows are then shifted with a step size of 1/32~s. Since the reported signals for burst oscillations in \source~range between 358 and 367 Hz according to \cite{2020ApJS..249...32G}, we considered frequencies between 355 and 370~Hz with a frequency step of 0.1~Hz. We searched for burst oscillations in three different energy bands, 0.5$-$12, 0.5$-$6, and 6$-$12~keV in order to compare our results with those of \cite{2019ApJ...878..145M}. We identify signals with the highest powers and then computed the probability of the signals assuming a Poisson noise distributed as $\chi^{2}$ with two degrees of freedom.

We also computed the fractional rms amplitude of candidate oscillations in each burst from phase-folded light curves obtained in the time interval of the light curve in which the signal is significant. Then we fitted the phase-folded light curves with a sinusoidal model defined as $A + B \sin(2 \pi \nu t - \phi_0)$, and from the best fitting parameters we calculated the fractional rms defined as $B/(\sqrt{2}A)$ \citep[see, e.g.,][]{2019ApJS..245...19B}. 
%%%%%%%%%%%%%%%%%%%%%%%%%%%%%%%%%%%%

\section{Results \& Discussion}

Below we present the main findings of our analysis on the spectral and temporal properties of the detected thermonuclear bursts from \source.

%%%%%%%%%%%%%%%%%%%%%%%%%%%%%%%
\subsection{Spectral Results}
\label{sec:time_res_res}
\begin{figure*}
    \centering
    \includegraphics[scale=0.5]{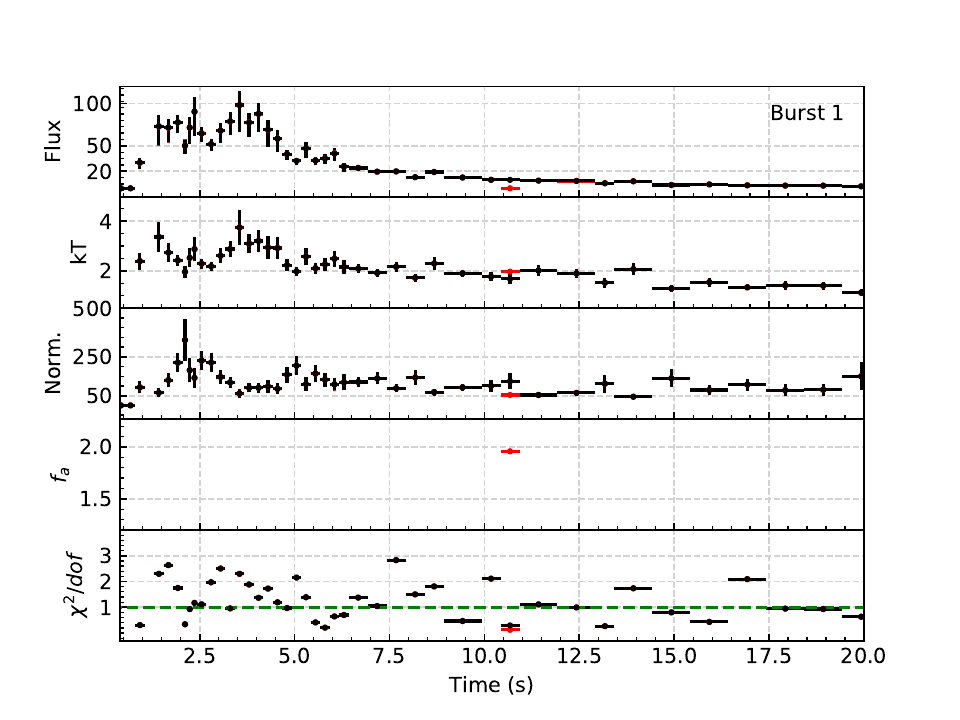}
    \includegraphics[scale=0.5]{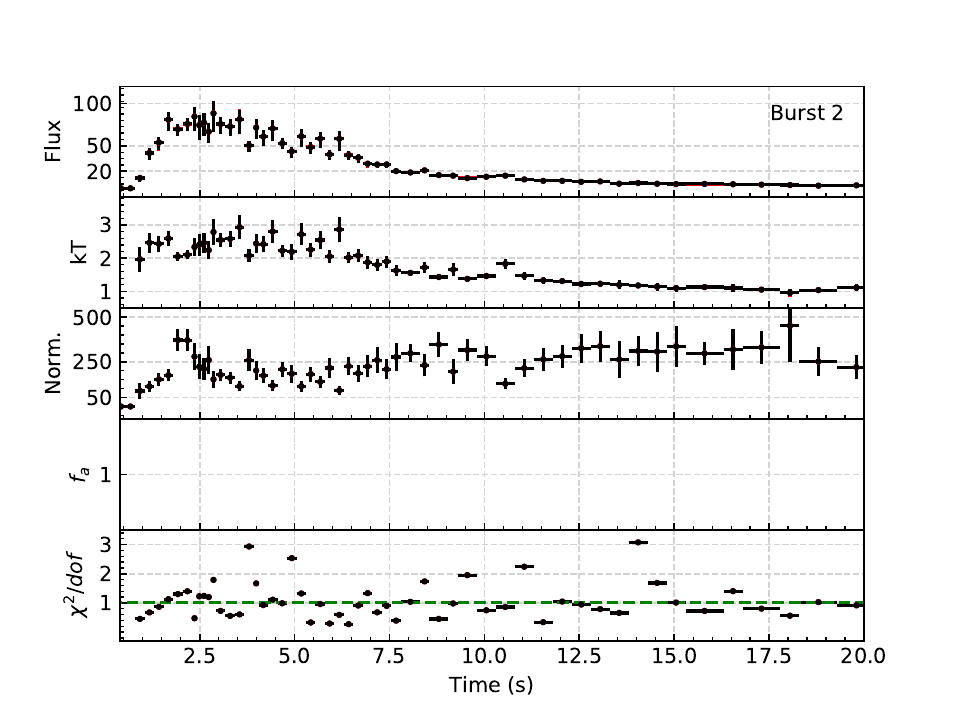}
    \includegraphics[scale=0.5]{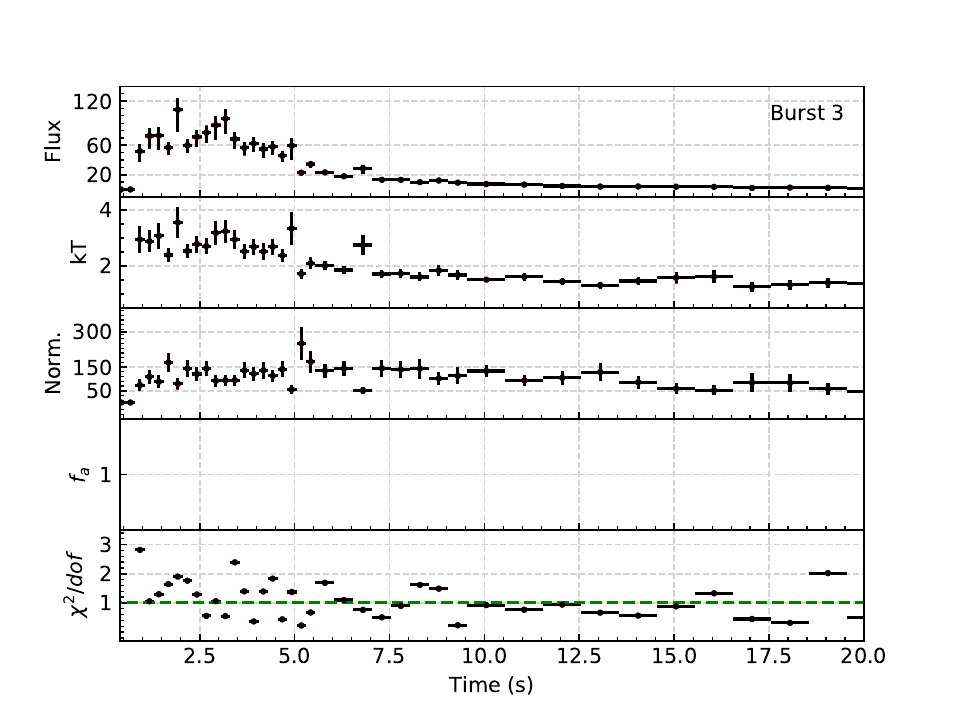}
    \includegraphics[scale=0.5]{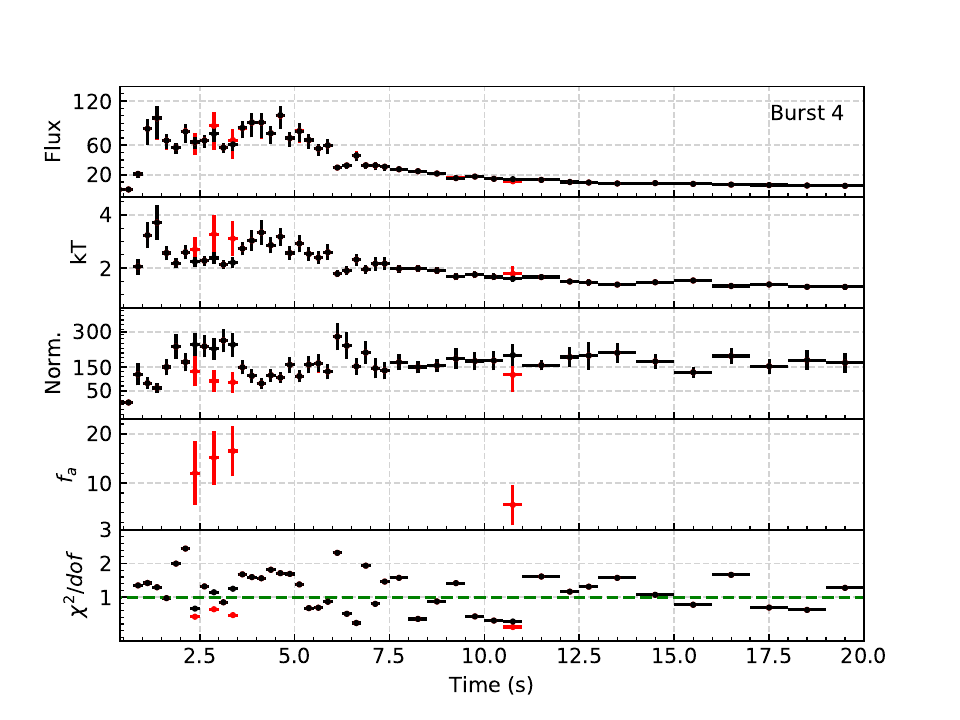}
    \includegraphics[scale=0.5]{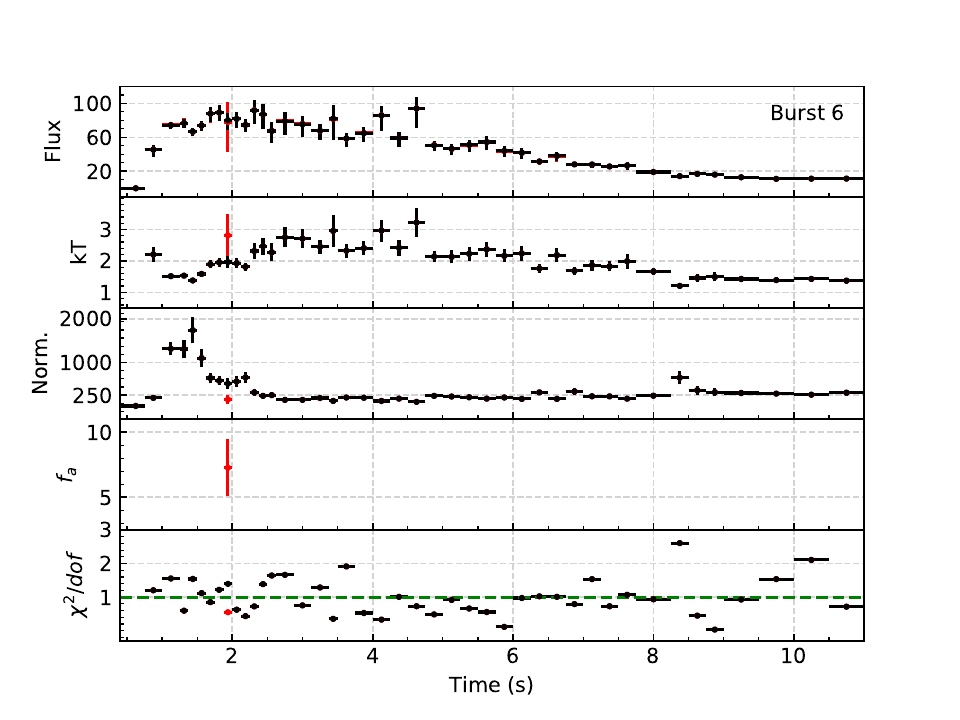}
    \includegraphics[scale=0.5]{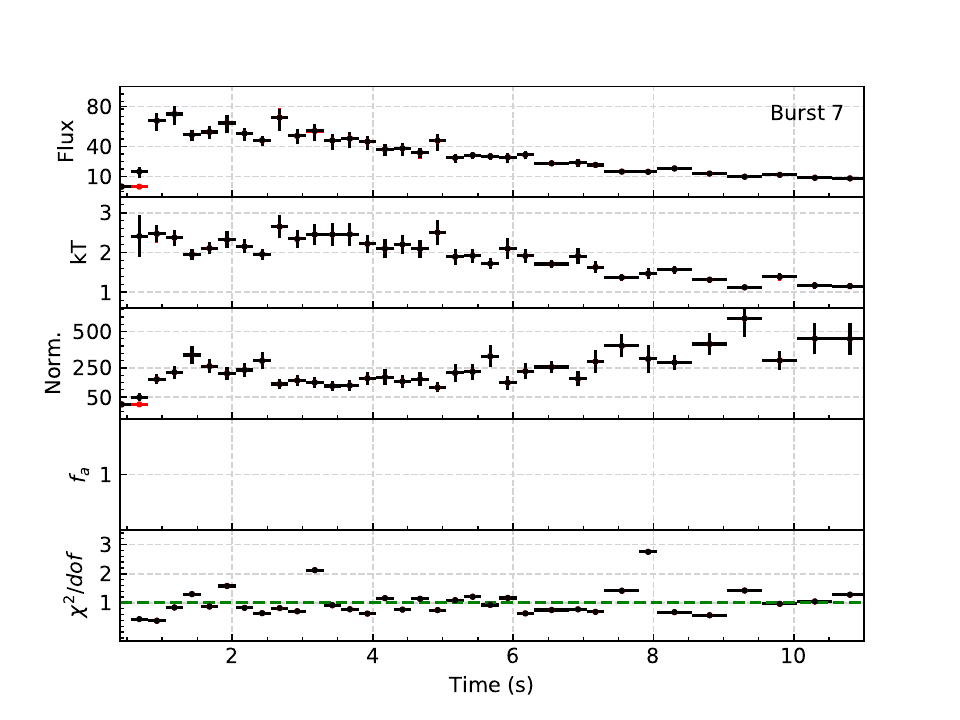}

    \caption{Time evolution of spectral parameters are shown. Red symbols show the results of the \fa method when applicable and black symbols show the results for constant background emission. In each panel, we show, from top to bottom, bolometric flux (in units of 10$^{-9}$~\fluxcgs), temperature (keV), blackbody normalization ($\rm{R}^{2}_{km}/\rm{D}^{2}_{10kpc}$), $f_a$, and finally the fit statistic, respectively.}
    \label{fig:bursts_plts1} 
\end{figure*}

\begin{figure*}
    \centering
    \includegraphics[scale=0.5]{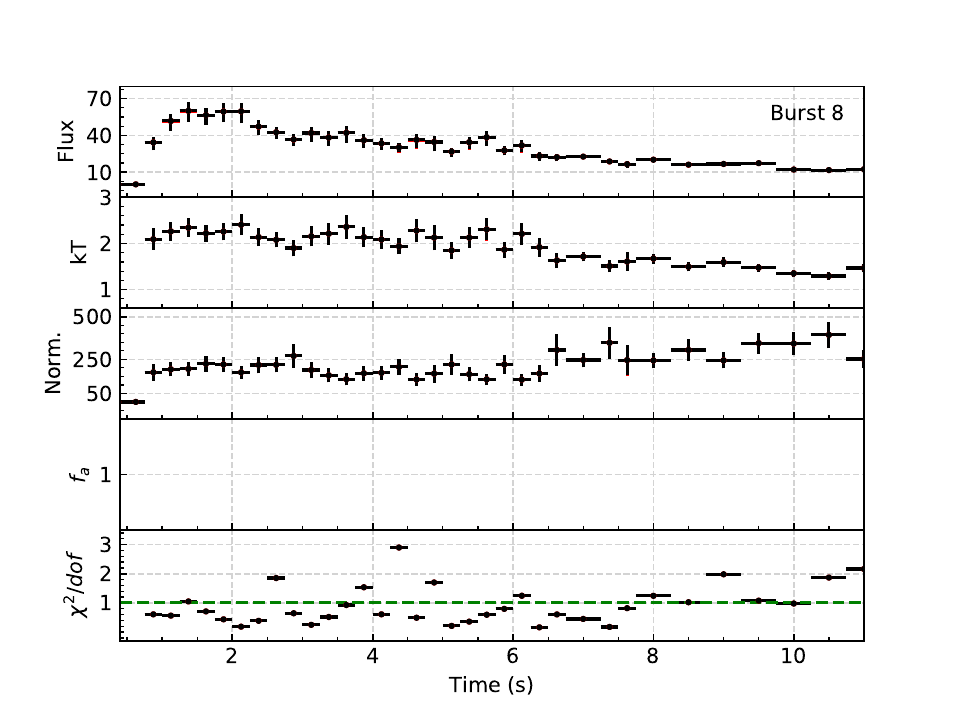}
    \includegraphics[scale=0.5]{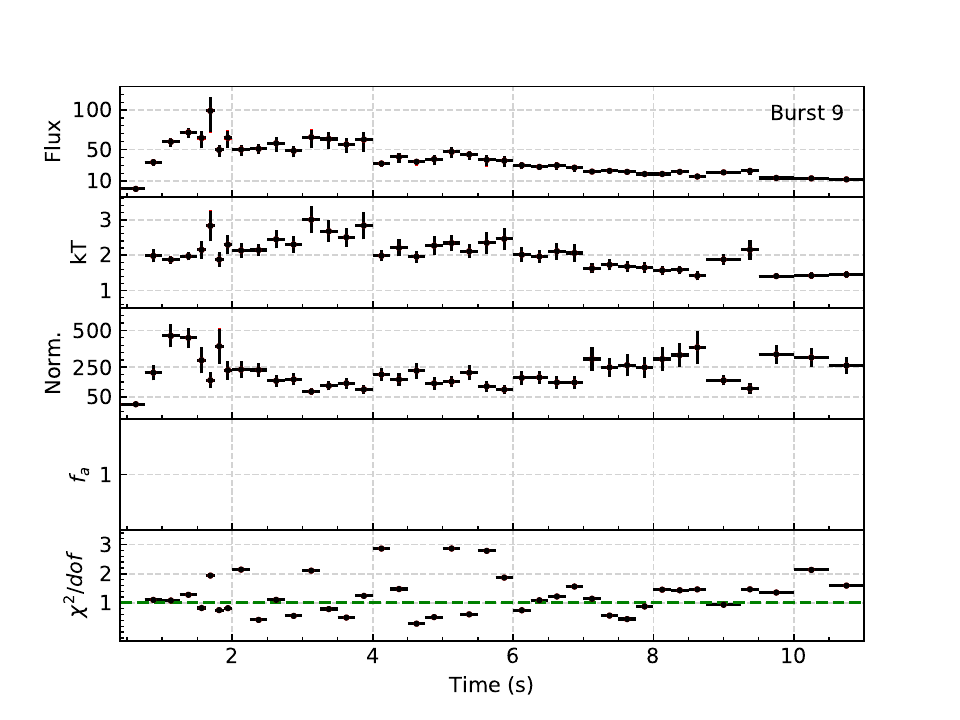}
    \includegraphics[scale=0.5]{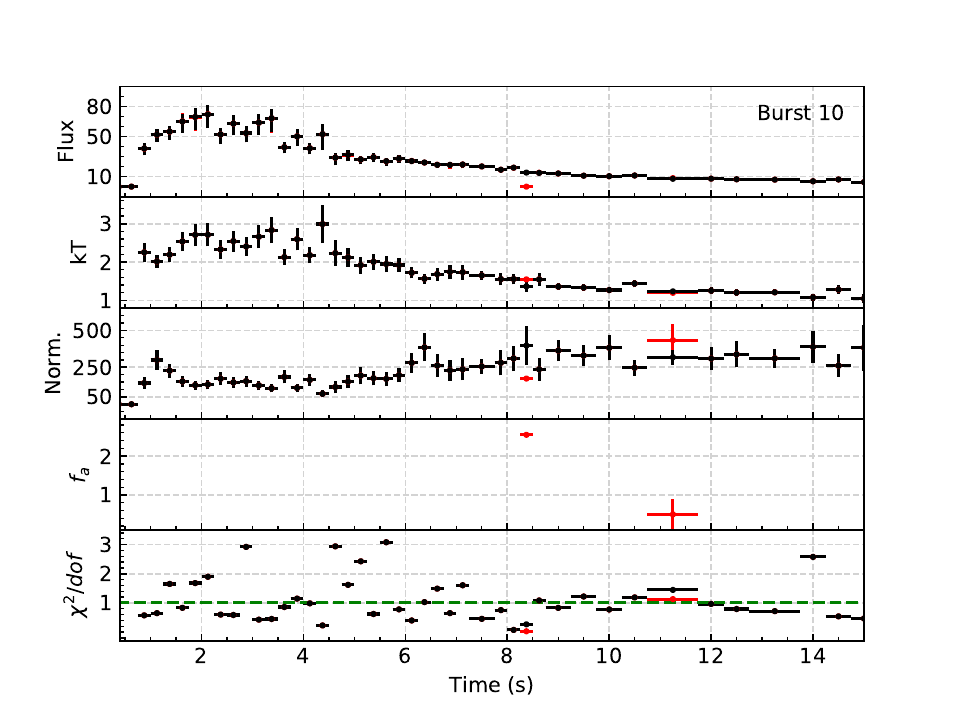}
    \includegraphics[scale=0.5]{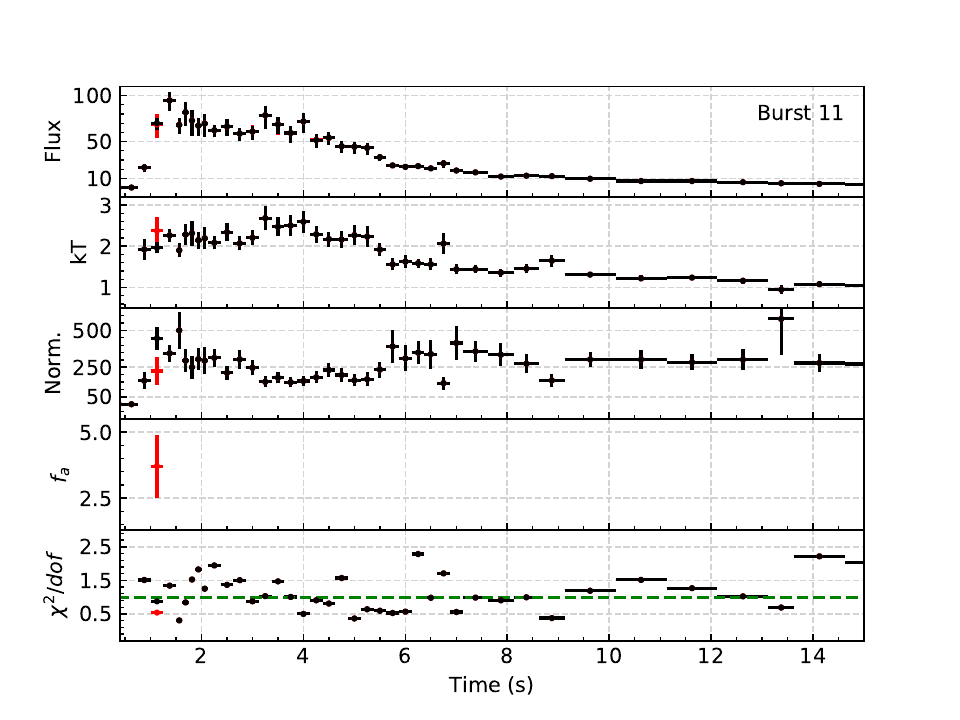}
    \caption{Same as Figure \ref{fig:bursts_plts1}.}
    \label{fig:bursts_plts2}
\end{figure*}

\begin{figure}
	%\centering
   	\includegraphics[scale=0.5]{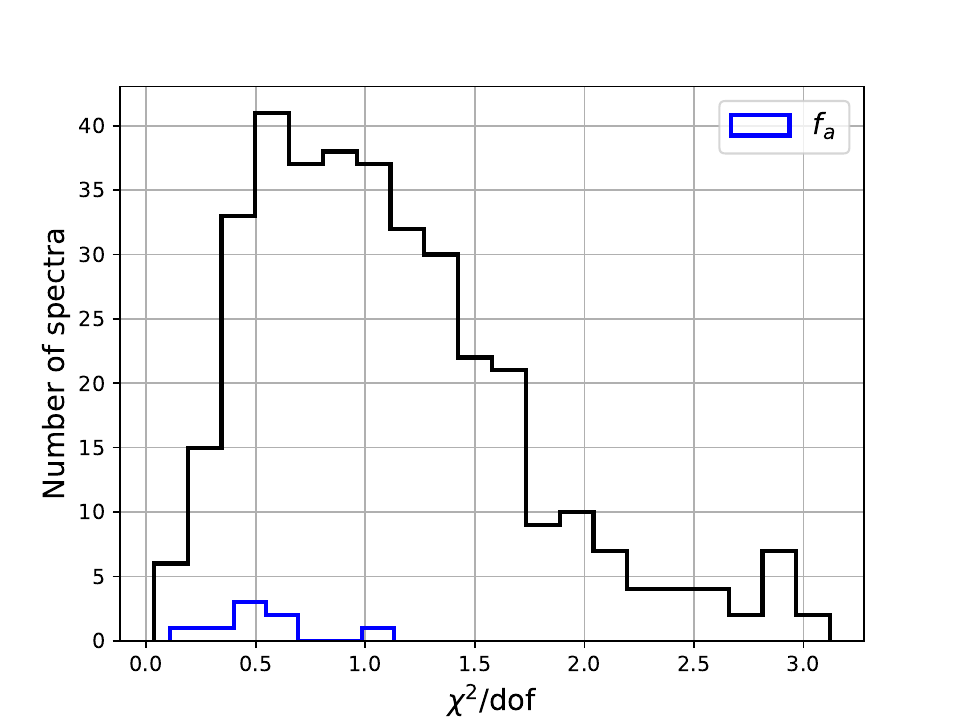}
    \caption{Histogram of the $\chi^2$ values with or without the application of the \fa parameter. Only in a very small fraction of the cases the use of \fa is statistically required.}
    \label{fig:burst_chi2}
\end{figure}

We present the resulting best fit parameters for the persistent emission preceding the detected bursts in \autoref{tab:pers}. In \autoref{fig:bursts_plts1} and \ref{fig:bursts_plts2} we present the observed spectral evolution in each burst. The inferred best fit parameters at the peak flux moment along with the fluences of each burst are summarized in \autoref{tab:burst_peaks}. In the calculation of the fluence, we integrated the bolometric fluxes starting from the onset of a burst till it declines to 10\% of the peak flux. Contrary to the earlier findings from \nicer \citep[see, e.g.,][]{2018ApJ...855L...4K,2018ApJ...856L..37K,2019ApJ...885L...1B,2020MNRAS.499..793B,2022MNRAS.510.1577G,2022ApJ...935..154G,2022ApJ...940...81B}, the spectral results reveal that in the case of \source~the multiplication of the persistent emission with a scaling factor does not improve the fits. In most cases this is because the fits are already statistically acceptable when we just use the persistent emission as a fixed model plus a blackbody for the burst emission, as shown in \autoref{fig:burst_chi2}. In rare cases, only about 2\% of the total spectra within the flux limits,  the application of a scaling factor is statistically favorable (f-test yields a chance probability smaller than 5\%). However, in these cases we see that the reduced $\chi^2$ values are mostly below unity indicating an overfitting issue. A simple explanation of this issue may be related to the fact that we use only the 1$-$10~keV band as opposed to most of the earlier studies where the authors used 0.5$-$10~keV range. We tested this by running our fits in the 0.5$-$10~keV range as well. We saw that in this case the fraction of X-ray spectra where the addition of the \fa improves the fit increases to 9\% of the total. However, this is still much smaller compared for example to 4U~1636$-$536 where in 63\% of the spectra a scaling factor is needed \citep{2022ApJ...935..154G}. This discrepancy may be attributed, in part, to the substantial hydrogen column density inferred along the line of sight towards \source~ N$_{\rm{H}}=4.18\times10^{22}$. Similarly, \cite{2021ApJ...910...37G} and \cite{2021ApJ...907...79B} also found that a scaling factor is not necessary for 4U~1608$-$52 and XTE~J1739$-$285. In both cases the hydrogen column densities in the line of sight to these sources were significantly high with N$_{\rm{H}}=1.4\times10^{22}$ and N$_{\rm{H}}=1.73\times10^{22}$, for 4U~1608$-$52 and XTE~J1739$-$285, respectively \citep{2021ApJ...910...37G,2021ApJ...907...79B}. These findings further confirm that the excess emission detected during X-ray bursts is mostly observed in the soft X-ray band, below 2.5~keV, irrespective of the observed low mass X-ray binary and does not really contribute significantly in the 3--10~keV band.

\begin{figure*}
    \includegraphics[scale=0.36]{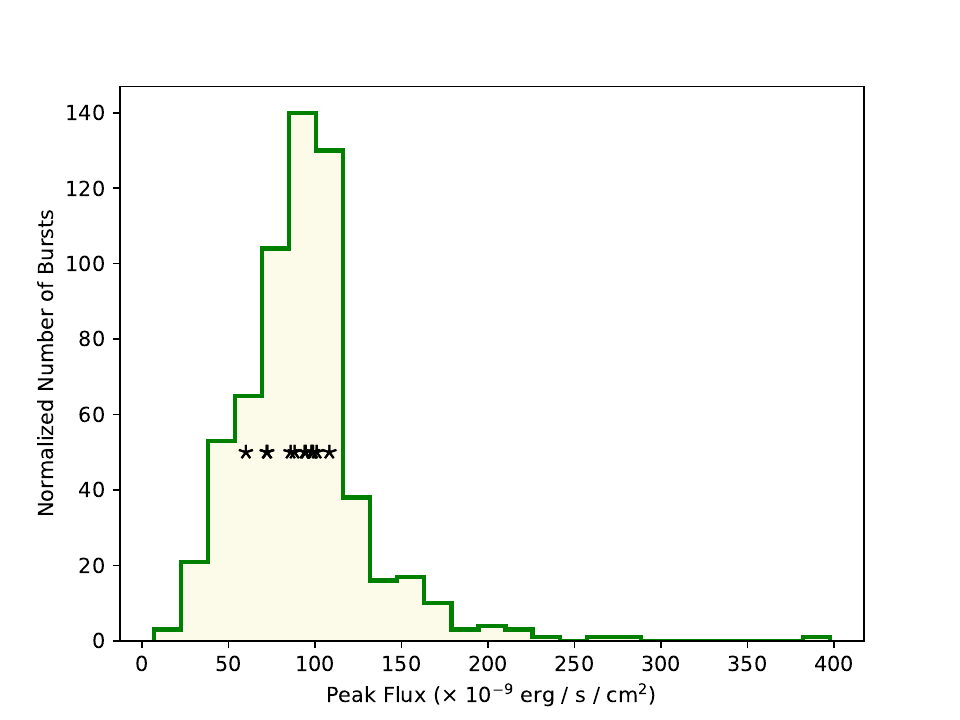}
    \includegraphics[scale=0.36]{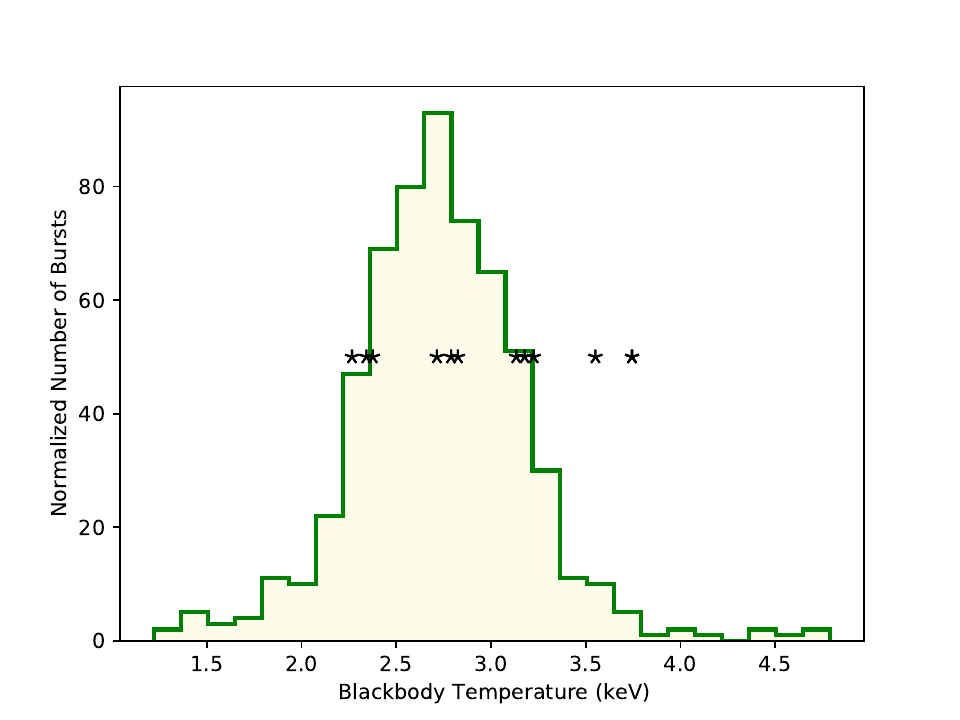}
    \includegraphics[scale=0.36]{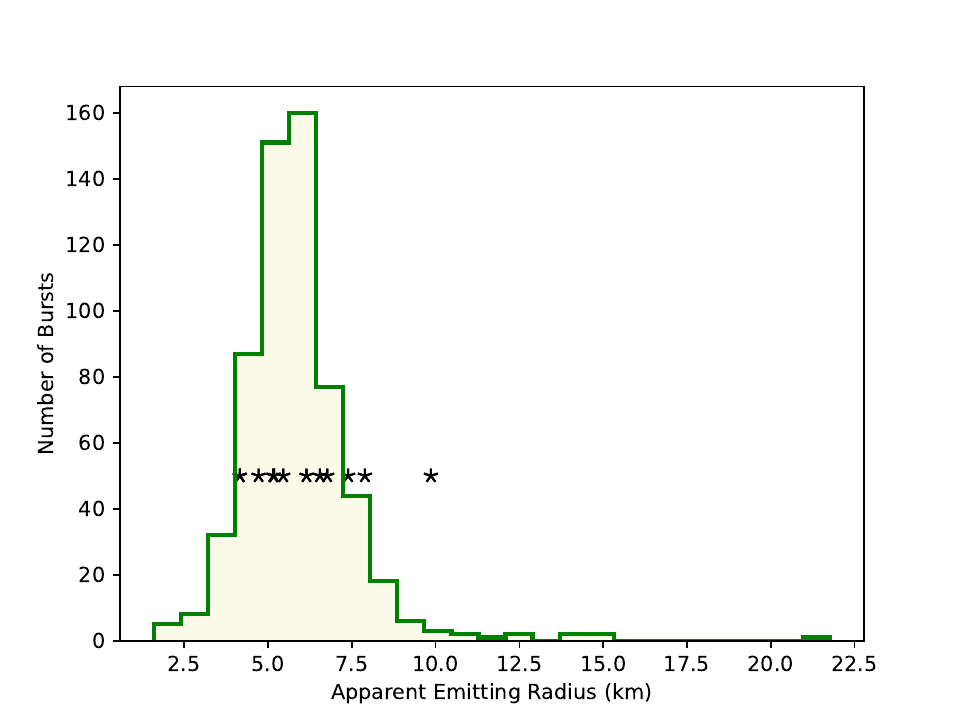}
    \caption{Histograms of peak flux, blackbody temperature and blackbody radius values assuming a source distance of 5.31~kpc in the MINBAR catalog together with the same values measured with \nicer for 10 bursts~(black stars) reported here.}
    \label{fig:burst_his_minbar}
\end{figure*}

In \autoref{fig:burst_his_minbar}, we compare our spectral parameters obtained at the peaks of each burst with those from the MINBAR catalog \citep{2020ApJS..249...32G}, which includes 611 bursts detected from \source. Our results seem to be in very good agreement with the range obtained from the MINBAR sample.

%%%%%%%%%%%%%%%%%%%%%%%%%%%%%%%%%%%%%%%%%%%%%%%%%%%%%%%%%%%%%%%%%%%%%%%%

\startlongtable
\begin{deluxetable*}{cccccc}
\tablehead{
\colhead{BID} & \colhead{Peak Flux\tablenotemark{a}} & \colhead{Peak kT} & 
\colhead{Peak Radius}\tablenotemark{b}& \colhead{Fluence\tablenotemark{c}} & \colhead{PRE}\tablenotemark{d}\\
\colhead{} & \colhead{} & \colhead{(keV)} & \colhead{(km)}& \colhead{} & \colhead{}
\tablecaption{Spectral parameters obtained at the peak flux moment for each burst. The fluence of each burst is also presented.}
}
\decimals
\startdata
1 & 9.81$\pm$2.43&3.74$\pm$0.71&4.16$\pm$0.69&36.8$\pm$1.43 &Y \\
2 &8.82$\pm$1.87&2.79$\pm$0.40&6.54$\pm$0.96&46.6$\pm$1.13 &M \\
3 &10.83$\pm$2.35&3.55$\pm$0.57&4.72$\pm$0.67&32.3$\pm$1.41&N \\
4 & 10.10$\pm$1.58&3.18$\pm$0.34&5.45$\pm$0.55&46.5$\pm$1.68&M \\
6 &9.40$\pm$1.82&3.23$\pm$0.45&5.16$\pm$0.67&55.4$\pm$1.17&Y \\
7 &7.24$\pm$0.96&2.38$\pm$0.20&7.89$\pm$0.75&29.4$\pm$0.85&N \\
8 &6.01$\pm$0.89&2.34$\pm$0.20&7.39$\pm$0.72&28.7$\pm$0.75&N \\
9 &9.89$\pm$2.21&2.83$\pm$0.44&6.77$\pm$1.05&38.6$\pm$0.97&M \\
10 &7.21$\pm$1.15&2.71$\pm$0.28&6.15$\pm$0.71&30.9$\pm$0.97&Y \\
11 &9.47$\pm$1.01&2.27$\pm$0.16&9.86$\pm$0.77&38.2$\pm$0.97&M \\
\enddata
\tablenotetext{a}{Unabsorbed bolometric flux in units of $\times10^{-8}$~\fluxcgs.}
\tablenotetext{b}{Apparent blackbody radius assuming a distance of 5.31~kpc.}
\tablenotetext{c}{In units of $\times10^{-8}$~\flencecgs.}
\tablenotetext{d}{The “PRE” column indicates whether the burst exhibited a photospheric radius expansion (Y) or not (N), or whether this is not clear (M).}
\label{tab:burst_peaks}
\end{deluxetable*}
%%%%%%%%%%%%%%%%%%%%%%%%%%%%%%%%%%%%%%%%

%%%%%%%%%%%%%%%%%%%%%%%%%%%%%%%%
%%%%%%%%%%%%%%%%%%%%%%%%%%%%%%%%%%%%

%%%%%%%%%%%%%%%%%%%%%%%%%%%%%%%%%%%%
%%%%%%%%%%%%%%%%%%%%%%%%%%%%%%%%%%%%%%%%%%%%

%%%%%%%%%%%%%%%%%%%%%%%%%%%%%%%%%%%%
%%%%%%%%%%%%%%%%%%%%%%%%%%%%%%%%%%%%%%%%%%%%
%%%%%%%%%%%%%%%%%%%%%%%%%%%%%%%%%%%%%%%%%%%%

\begin{figure}
	%\centering
   	\includegraphics[scale=0.5]{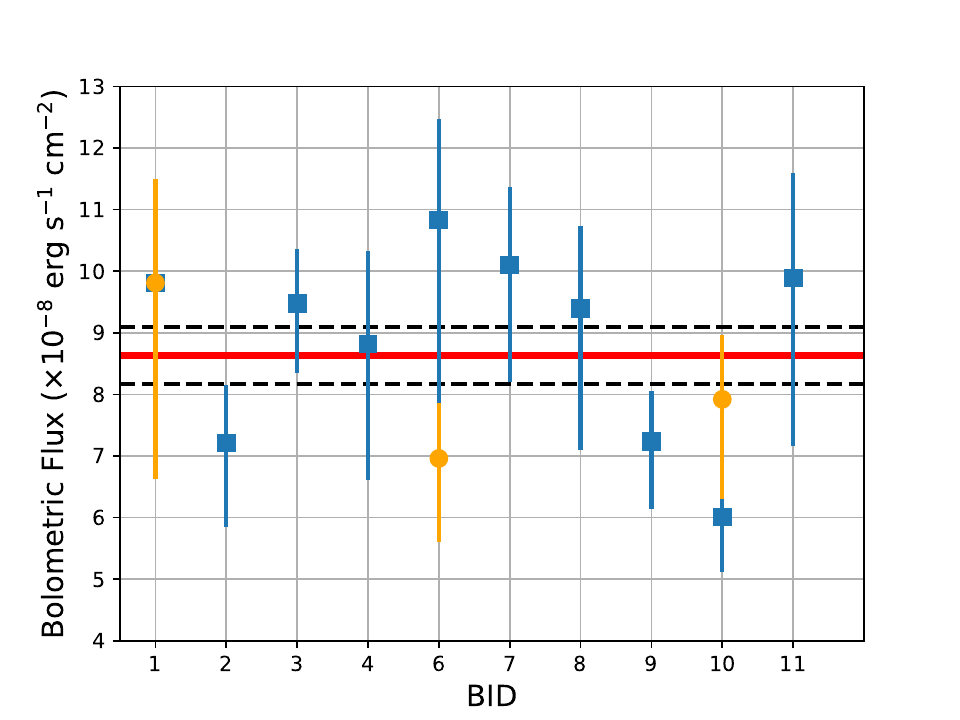}
    \caption{Touchdown (orange dots) and/or peak flux (blue squares) values measured here compared with the average touchdown flux value presented by \cite{2012ApJ...747...77G}, which is shown with the red solid line together with the systematic uncertainties shown with black dashed lines.}
    \label{fig:burst_td_flux}
\end{figure}

As shown by~\cite{2008ApJS..179..360G,2012ApJ...747...77G,2020ApJS..249...32G} \source~is one of the rare sources, together with 4U~1820-30 and 4U~1636-536, which show frequent photospheric radius expansion bursts. The combined effects of high hydrogen column density towards \source~ which decreases the observed count rate in the \nicer band and the fast evolution of the bursts limit our capability to infer much from the spectral analysis. However, following the criteria proposed by \cite{2008ApJS..179..360G} and \cite{2012ApJ...747...77G}, we identified three bursts that show evidence for photospheric radius expansion and three more candidates. These bursts are indicated in \autoref{tab:burst_peaks}. In \autoref{fig:burst_td_flux} we compare the touchdown fluxes and peak fluxes of the bursts as inferred with \nicer with the touchdown flux value inferred using 16 bursts by \cite{2012ApJ...747...77G}. Although with much larger error bars, our results remain consistent with previous results. Note that the average peak flux for \source~reported as ${F}_{peak}$=(9.4$\pm$3.6)$\times$10$^{-8}$~\fluxcgs~in MINBAR \citep{2020ApJS..249...32G}, also aligns well with the peak flux measurements presented in this study.
%%%%%%%%%%%%%%%%%%%%%%%%%%%%%%%%%%%%%%%%%%%%
  
%%%%%%%%%%%%%%%%%%%%%%%%%%%%%%%%%%%%%%%%%%%

Photospheric radius expansion bursts can be used to infer the Eddington Limits \cite[see, e.g.,][]{1979ApJ...234..609V,1990A&A...237..103D,2016ARA&A..54..401O}, as well as calculating the distances \citep{Basinska, 2003A&A...399..663K}. In order to calculate a distance for \source, we took the weighted average of the touchdown fluxes of bursts 1, 6 and 10 showing photospheric radius expansion $F_{TD}=(7.88_{-0.70}^{+1.18})\times10^{-8}$~\fluxcgs. Using the observed flux value and assuming 10~km and 1.4~$M_{\odot}$ for the radius and the mass of the neutron star as well as taking into account that the accreted matter can be hydrogen-rich~(X=0.7) or hydrogen-poor~(X=0), we estimate the distance as $d_{\rm H} = 4.09_{-0.17}^{+0.34}$~kpc, $d_{\rm He} = 5.31_{-0.22}^{+0.45}$~kpc, respectively. These values are in very good agreement with the distance estimations presented in \citep{2008ApJS..179..360G, 2003ApJ...593L..35S} for \source.

In \autoref{tab:burst_peaks} we presented apparent emitting radii of the blackbody at the peak of each burst for a distance of 5.31~kpc, assuming \source~is accreting pure-He. Considering the burst timescales inferred in this study and in \cite{2008ApJS..179..360G, 2020ApJS..249...32G} it is a fair assumption that \source~is a pure He accretor \citep{2003ApJ...593L..35S} and may be an ultracompact binary. We note however that recent simultaneous infrared and X-ray observations support that the companion is a helium star, not an ultra-compact \citep{2020MNRAS.495L..37V,Vincentelli2023}.

Finally, using the touchdown flux derived above one can also calculate the $\gamma$ value \citep{1988MNRAS.233..437V,2008ApJS..179..360G}, which is defined as the ratio of the persistent bolometric flux to the Eddington limit (derived from the touchdown or peak fluxes of the bursts) of a source and is expected to be related to the mass accretion rate. Due to the nature of the power-law model we used to fit the persistent emission, it is not reliable to calculate unabsorbed bolometric flux of the source by just extrapolating the function with the best fit parameters. We therefore calculated the unabsorbed touchdown flux of the source limiting to only 1$-$10~keV range as in the persistent flux measurements. This way we find $F_{TD}=(4.04_{-0.30}^{+0.2})\times10^{-8}$~\fluxcgs in the 1$-$10~keV. We used this value and the persistent state fluxes of the source to calculate the $\gamma$ values and present them in \autoref{tab:pers}. We here made the assumption that in the persistent emission there is no additional contribution from the accretion disc below or above the 1$-$10~keV range that would significantly change the ratio. Overall the inferred $\gamma$ values (see \autoref{tab:pers}) show that during \nicer observations the system was at about 10\% Eddington, with the exception of bursts 4 and 8.

\subsection{Timing Analysis and Detected Burst Oscillations}
\label{Sec:burst_osc_res}
We consider signals as candidates when a single-trial chance probability is calculated to be $<$ 10$^{-4}$ and a confidence level $>$ 99.7$\%$ is reached in either one of the 2 or 4~s search intervals. With these criteria we identified candidate burst oscillation signals in 8 of the 11 bursts. All of the bursts and the resulting Z$^{2}_{1}$ contours are shown in \autoref{fig:burst_lc2}. Properties of these signals are listed in Table~\ref{tab:bursts_oscil}, including the energy range in which the signal is detected, frequency, power of the signal, single-trial chance probability, confidence level, fractional rms amplitude, the time it is detected with respect to the burst peak and finally the search window in which the signal is found.  In our list, three candidate signals observed from bursts 4, 7, and 8 were also reported by \cite{2019ApJ...878..145M}, where the authors searched for oscillations in seven bursts covering the 360 and 365 Hz frequency range. Our findings for these three bursts are in agreement with the results presented in \cite{2019ApJ...878..145M}. Since the frequency range and the time interval we considered are wider, we found more candidate signals in the first seven bursts.

%%%%%%%%%%%%%%%%%%%%%%%%%%%%%%%%%%%%
\begin{deluxetable*}{ccccccrrc}
\tablehead{ \colhead{BID}&\colhead{Energy Range}&\colhead{Freq.}&\colhead{Z$^2_1$}&\colhead{Single Trial.}&\colhead{Conf. Level}&\colhead{A$_{rms}$}&\colhead{Time\tablenotemark{a}}&\colhead{Window Size}\\
\colhead{}&\colhead{keV}&\colhead{Hz}&\colhead{}&\colhead{}&\colhead{\%}&\colhead{\%}&\colhead{s}&\colhead{s}}
\tablecaption{Characteristic properties of all the candidate burst oscillations. The values shown in bold indicate the search interval where the signal is more significantly detected.
\label{tab:bursts_oscil}}
\startdata
1&0.5$-$12\tablenotemark{b}&356.1&30.50&2.0$\times10^{-7}$ &99.996&26.4$\pm$3.3&$-2.31$&2\\
3&0.5$-$6\tablenotemark{c}&363.2&25.64&	2.1$\times10^{-6}$ &99.960&9.7$\pm$1.4&2.35& 2, 4\\
3&0.5$-$6\tablenotemark{b}&367.0&23.01 &1.0$\times10^{-5}$ &99.849 & 15.5$\pm$2.3 &$-4.27$&4\\
4&0.5$-$6\tablenotemark{c}& 367.5&23.77 &6.9$\times10^{-6}$ &99.897& 19.9$\pm$3.2& 25.29&2, 4 \\
4\tablenotemark{d}&6$-$12& 362.5&30.48 &2.0$\times10^{-7}$ 	&99.996 &40.1$\pm$5.8&7.66& 2, 4 \\
5&0.5$-$12\tablenotemark{b} & 355.5&24.22 &5.5$\times10^{-6}$ & 99.918 & 13.0$\pm$1.9 & 9.23  & 2, 4  \\
7\tablenotemark{d}&0.5$-$12\tablenotemark{c}&363.1&28.09&8.0$\times10^{-7}$&99.988 & 9.2$\pm$1.3 &3.97& 2, 4  \\
8\tablenotemark{d}&6$-$12&363.6 &30.75 &2.0$\times10^{-7}$	&99.996& 45.7$\pm$6.7&13.45 & 2,  4\\
10&0.5$-$12\tablenotemark{b}&357.9 &21.87 &1.7$\times10^{-5}$ &99.733&11.0$\pm$1.6 & 9.31& 2, 4\\
10& 6$-$12&359.0&27.62&	1.0$\times10^{-6}$ &99.985&63.7$\pm$10.7 & $-1.28$  &  2, 4\\
11& 0.5$-$12\tablenotemark{b}&366.4&28.34&7.0$\times10^{-7}$&99.989&7.6$\pm$1.0&1.80& 2, 4\\
\enddata
\tablenotetext{a}{Time is given with respect to the peak moment of each burst.}
\tablenotetext{b}{Also detected in 0.5$-$6~keV band.}
\tablenotetext{c}{Also detected in 0.5$-$12~keV band.}
\tablenotetext{d}{Already reported in \cite{2019ApJ...878..145M}.}
\end{deluxetable*}    
%%%%%%%%%%%%%%%%%%%%%%%%%%%%%%%%%%%%%%%%%%%%%%%%%%%%%%%%%%%%%%%%%%%%%%%%

In three bursts (3, 7 and 11) we detected candidate oscillations from the peaks to the e-folding times, in both 0.5$-$6 and 0.5$-$12~keV bands, as well as in both 2 and 4~s search interval windows. Bursts 3 and 7 show signals at around 363~Hz while during burst 11 an oscillation at 366~Hz is observed, which is well beyond our uncertainty in frequency (0.1~Hz).  As seen in Table~\ref{tab:bursts_oscil}, the fractional rms amplitudes of these oscillations are in the range of 7 to 10\%, which are consistent with burst oscillation rms amplitudes around peaks reported in previous studies \citep{1997ApJ...487L..77S, 2001ApJ...551..907V, 2019ApJ...878..145M}. Errors in rms amplitudes show 1$\sigma$ confidence levels and are calculated from the best-fit parameters and their associated statistical uncertainties. 

Bursts 4, 5, 8 and 10 show oscillations during the burst tail. In bursts 5 and 10, we detected signals at frequencies of 355.5 and 357.9 Hz in both 0.5$-$6 and 0.5$-$12~keV and in both 2 and 4~s search interval windows with maximum Z$^{2}$ values of 22 and 24, respectively. The fractional rms amplitudes for these oscillations range from 11$-$13\% in the 0.5$-$6 and 0.5$-$12~keV bands. For bursts 4 and 8, we found signals around 363~Hz with the maximum power of just over 30 in the 6$-$12~keV band. The fractional rms amplitude of the signals in 6$-$12~keV band is very large, over 40\%. Our finding of these two bursts is consistent with the results reported by \cite{2019ApJ...878..145M}. We also noticed a tentative signal at 367.5~Hz after the decay time of burst 4. This signal is observed in both 0.5$-$6 and 0.5$-$12~keV bands and in both 2 and 4~s search interval windows with an rms amplitude of about 20\%$\pm$3\% (see, Table~\ref{tab:bursts_oscil}).

There are three bursts (1, 3, and 10) that deserve particular attention. As it can be seen in \autoref{tab:bursts_oscil} and Figure~\ref{fig:burst_lc2} oscillations are observed just prior to the X-ray bursts and they fade away when bursts start to rise. In the case of burst 1, the signal at 356~Hz is detected in 0.5$-$6 and 0.5$-$12~keV bands with an rms amplitude of about 26\%$\pm$3\%. The oscillation seems to reach a maximum Z$^2$ of slightly over 30 just 2.3$\pm$1.0~s before the burst peak time. Although the 356~Hz signal is also seen in 0.5$-$12~keV in 4~s search interval window it does not achieve statistical significance requisite for initial selection criteria. In the case of burst 3, a tentative oscillation at 367~Hz is seen in 0.5$-$6~keV band with an rms amplitude of about 16\%$\pm$2\%. The signal reaches a maximum power of 23 at 4.3$\pm$2.0~s before the burst peak time. In the case of burst 10, a potential candidate oscillation is detected at 359~Hz in the 6$-$12~keV band in both 2~s and 4~s search interval windows. The oscillation seems to reach a maximum Z$^2$ of just over 27 at 1.3$\pm$1.0~s before the burst peak time. The rms amplitude of the signal computed within the search time interval is 64\%$\pm$11\%.

\begin{figure*}
 	\includegraphics[scale=0.27]{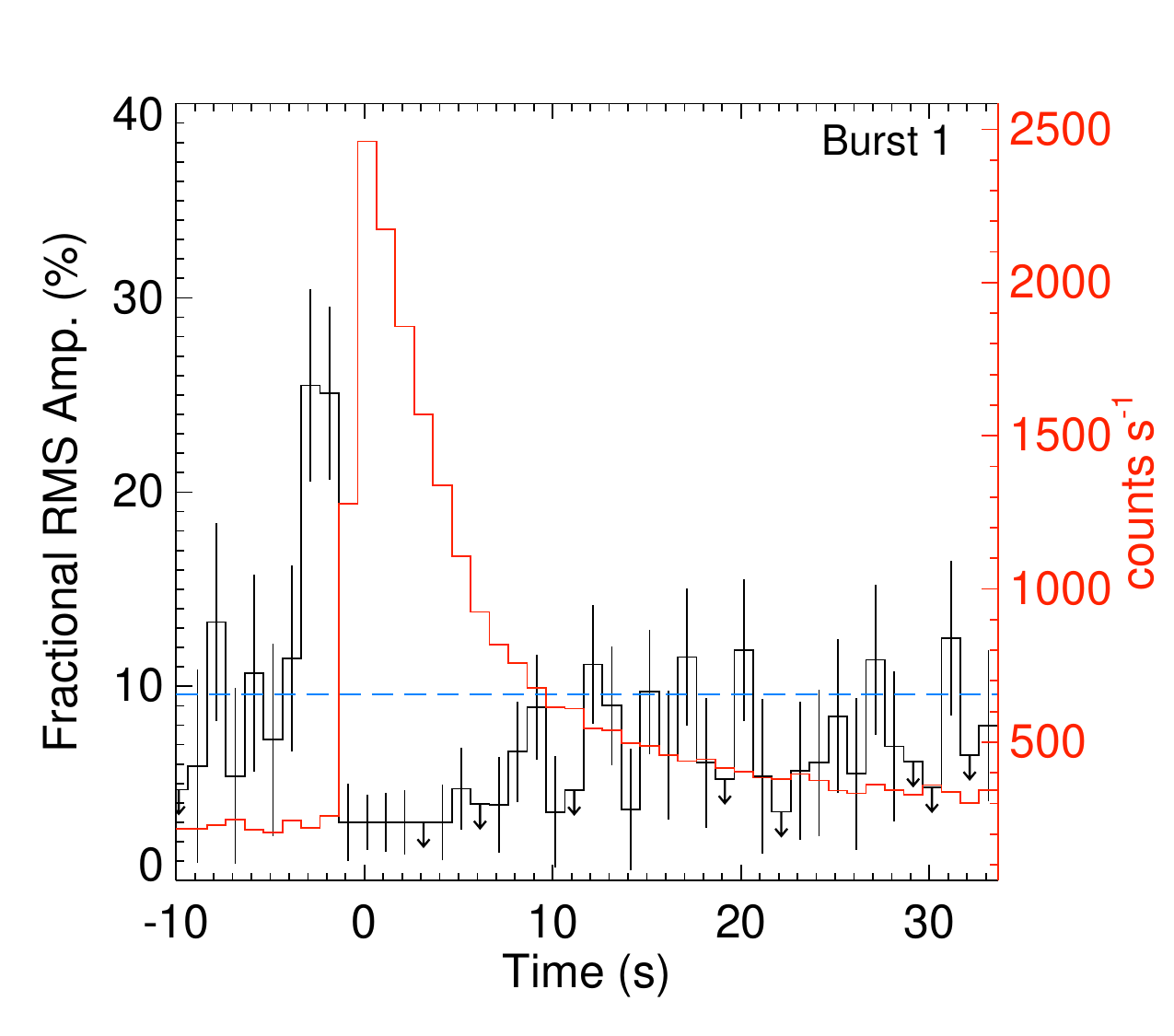}
	\includegraphics[scale=0.27]{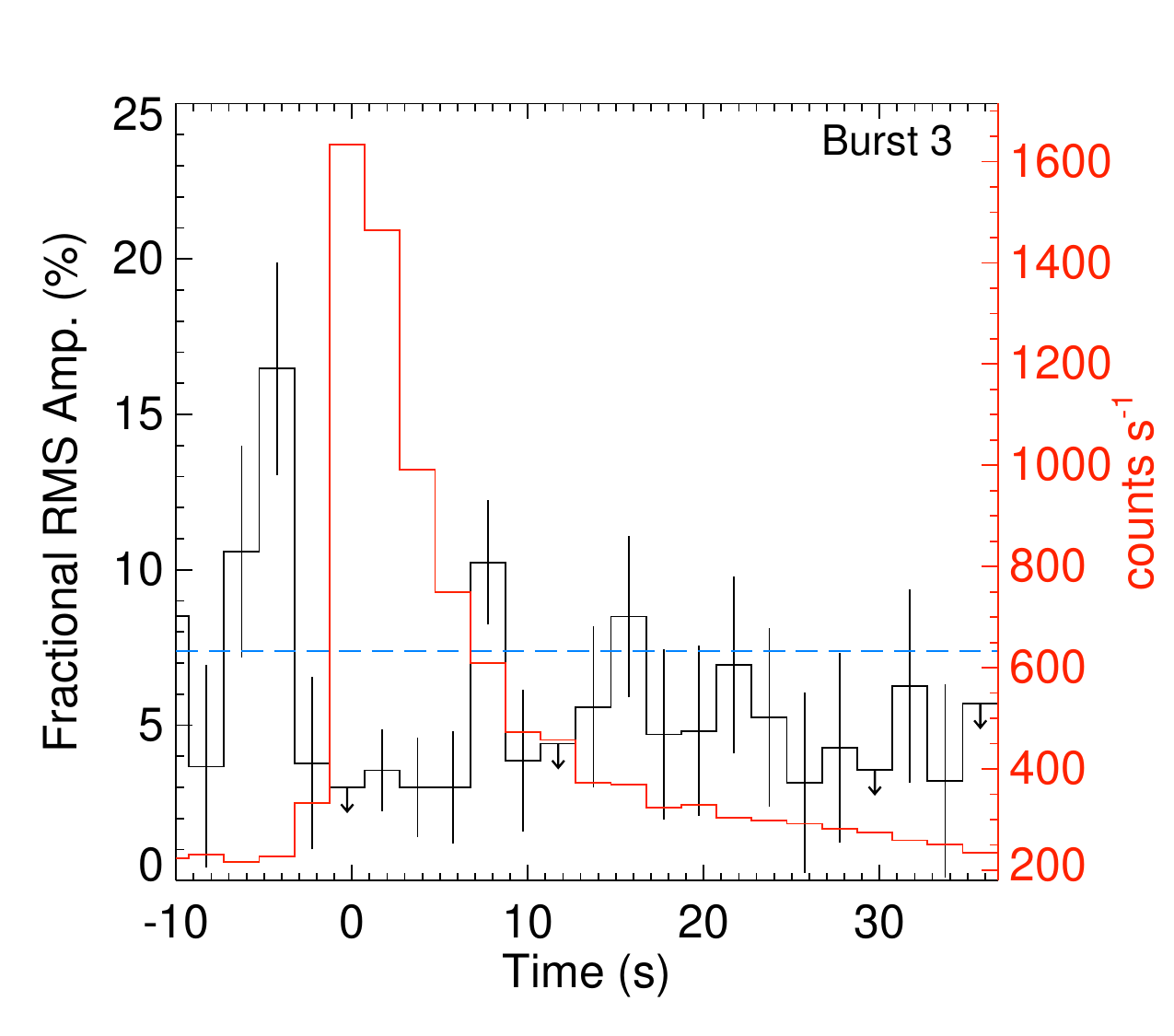}
    \includegraphics[scale=0.27]{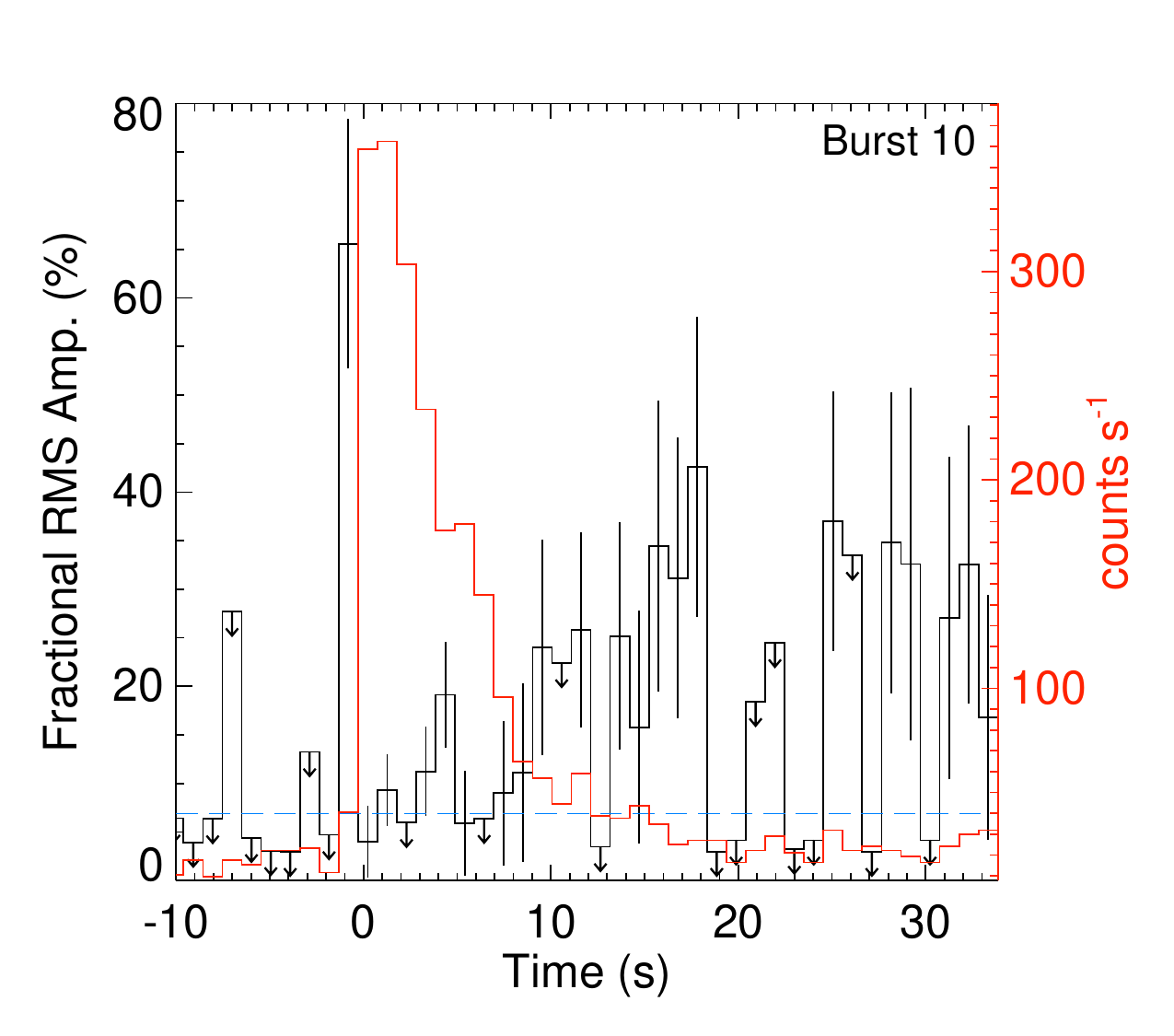} \\
    \includegraphics[scale=0.27]{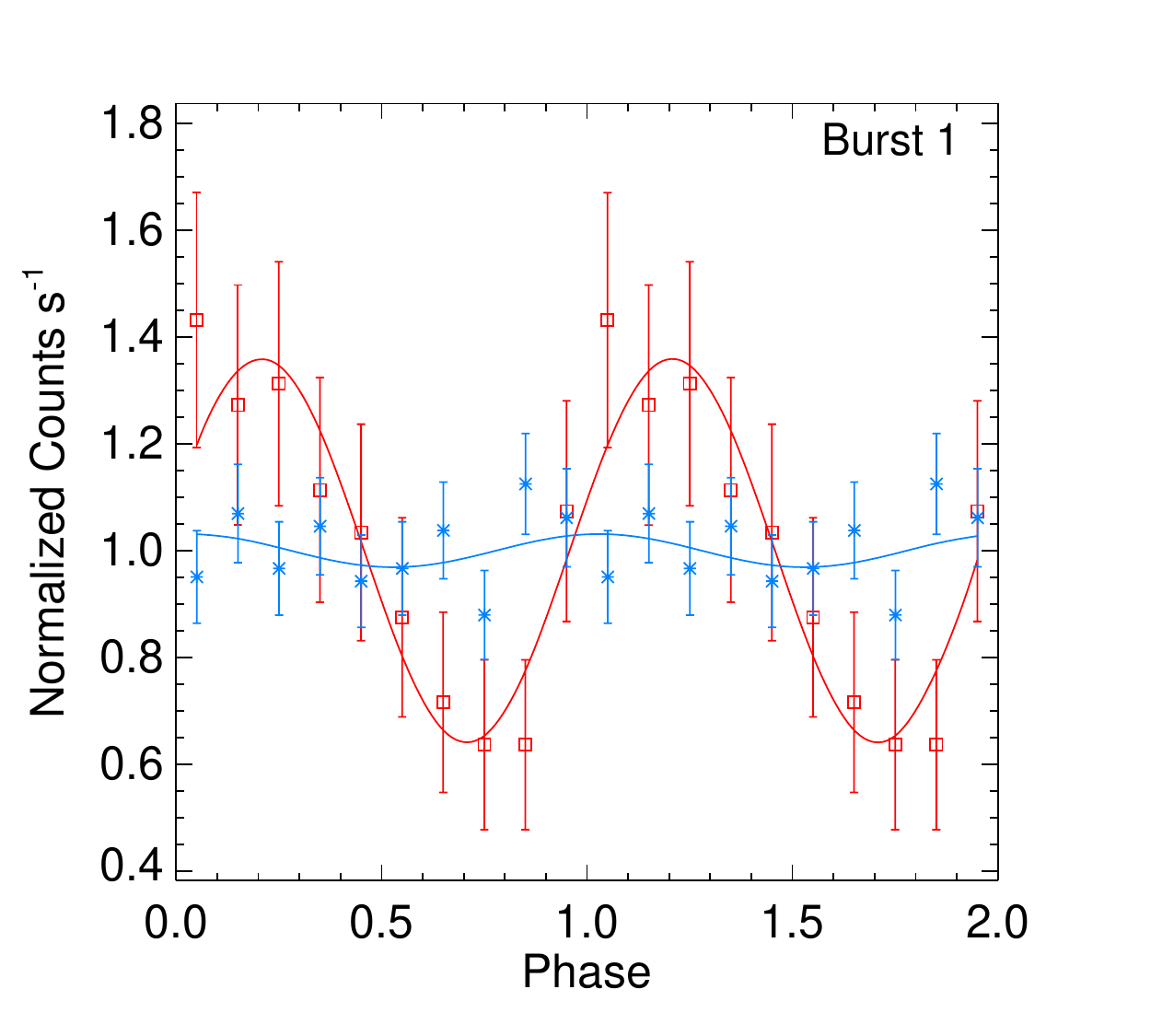}
   	\includegraphics[scale=0.27]{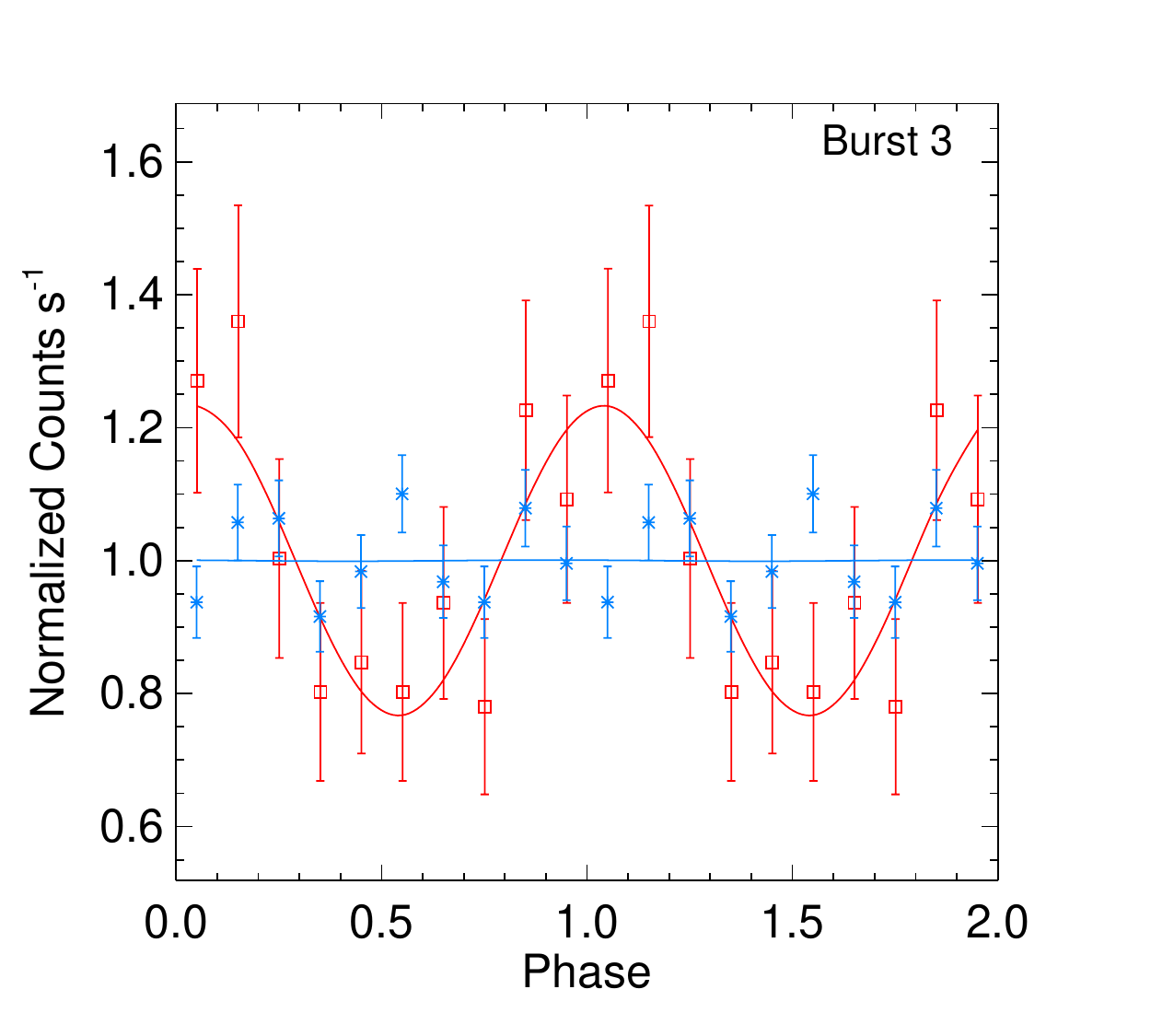} 
   	\includegraphics[scale=0.27]{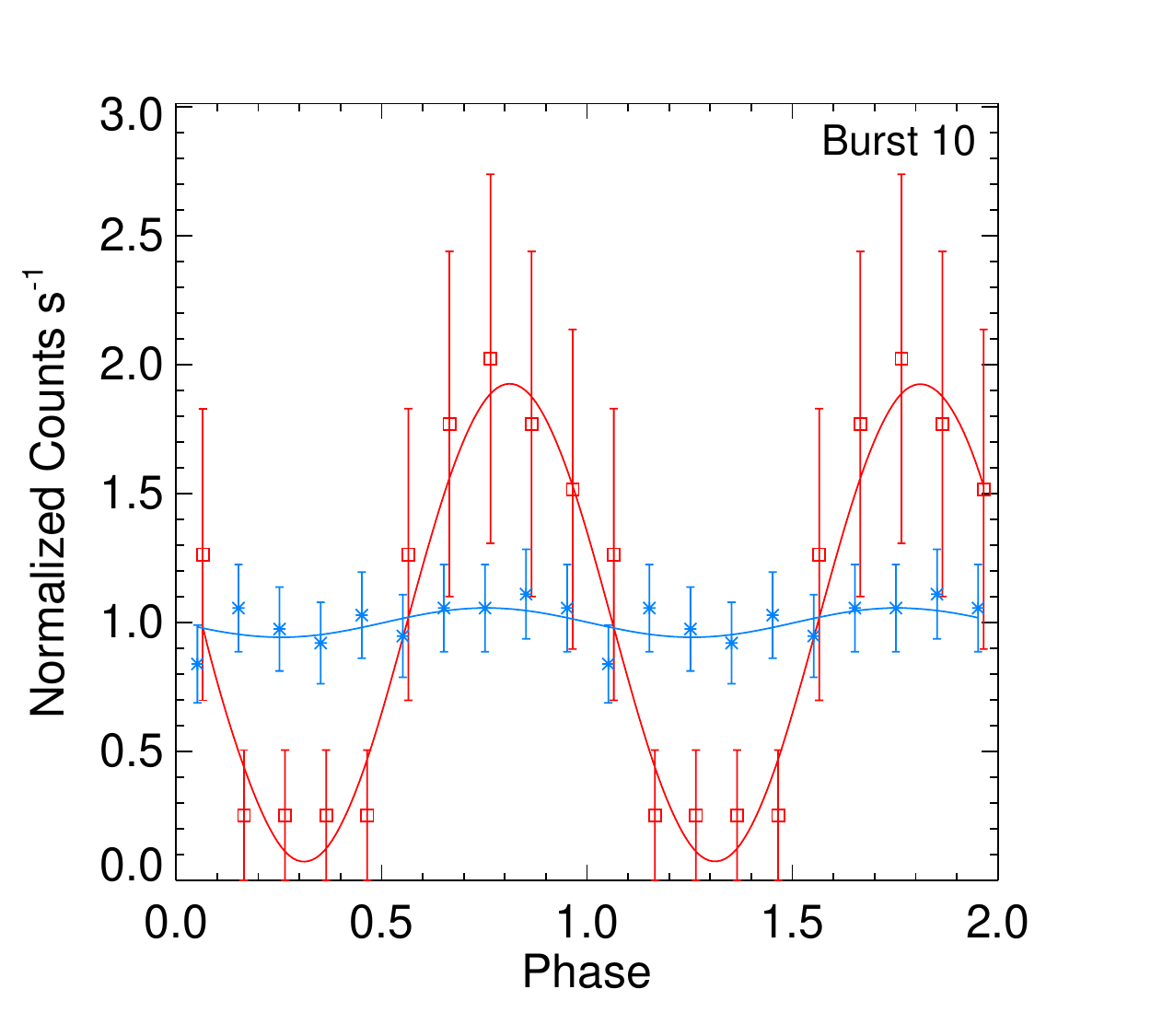}  
        \caption{{\it Upper Panel:} Under the assumption that there are oscillations at 356.1 Hz, 367.0 Hz and 359.0 Hz, the evolution of the fractional rms amplitudes (black lines) together with the burst light curves (red lines) in the 0.5$-$12, 0.5$-$6, and 6$-$12~keV for bursts 1, 3, and 10 (from left to right panels), respectively. Amplitudes are calculated in 1~s intervals for bursts 1 and 10 and in 2~s intervals for burst 3. Arrows indicate upper limits. The amplitudes are high in two bins (2 or 4~s for burst 1 and 3) just before burst onset, then drop below the detection level during the burst peak. The horizontal blue dashed lines show the average rms amplitude values calculated using the 100~s interval before the onset for bursts 1 and 3, and 30~s interval before the onset for burst 10. {\it Lower Panel:} Phase-folded light curves (squares and stars with the error bars) obtained by folding 1~s intervals for bursts 1 and 10, and 2~s intervals for burst 3. The best-fitting sinusoidal model (solid lines) for the bins just before the onset (red) and the peak (blue) are also shown. Phase-folded light curves were normalized by constant values from the best-fitting sinusoidal models, for clarity.}
    \label{fig:burst_one_ampl_ev1_3}
\end{figure*}

To investigate the temporal behavior of the oscillations seen immediately before the bursts, we divided the light curve from 100~s (30~s for burst 10) before the burst to the end into 1~s (or 2~s for burst 3) time intervals and determine fractional rms amplitudes of the oscillation signal for each interval. Results for bursts 1, 3, and 10 are presented in \autoref{fig:burst_one_ampl_ev1_3}, where, in the upper panels, we show the time dependent variation of the rms amplitudes determined in each interval and also the light curves in the energy range where the oscillation is observed. It is clearly seen that the rms amplitudes are high in the intervals where the power is maximum and then decrease as the burst rises. Similar to the fractional rms amplitude evolutions seen here, \cite{2014ApJ...792....4C} reported that burst oscillations detected during the rises of the bursts show a decreasing trend of fractional rms  amplitude with time. They infer a typical timescale for the oscillations to be undetectable as  2.5~s and attribute this time to the flame spreading. However, in the case of the oscillations reported here, the decrease in the rms amplitude happens instantaneously instead of showing a similar decreasing trend. Lower panels of Figure \ref{fig:burst_one_ampl_ev1_3} show pulse profiles obtained by folding the interval with the oscillation frequency where the measured rms is high (red curve) and at peak of the burst interval (blue curve).  

%%%%%%%%%%%%%%%%%%%%%%%%%%%%%%%%%%%%

%%%%%%%%%%%%%%%%%%%%%%%%%%%%%%%
\subsubsection{Monte Carlo simulations}
\label{sec:mc}
In order to assess the significance of the candidate oscillations observed during the X-ray bursts, we followed two different methods. As an initial step, we generated $10^5$ simulations of the null hypothesis (no oscillations) for each of the eight bursts including the entire time window we initially looked for. We randomized the arrival times of events in each time window (2s or 4s) and also allowed for variation of the observed count rates assuming a Poisson distribution. We followed exactly the same procedures for the timing analysis of simulated burst profiles for each burst and evaluated how frequently a maximum Z$^2$ value equal to or greater than the observed value in the real data is obtained in the simulated data. The resulting maximum $Z^2$ values are distributed almost homogeneously around the burst times although there is a minor trend following the count rate, especially in the 6$-$12~keV band. We determined the p-value from the distribution of simulated maximum Z$^2$ values for each burst and transformed them to $\sigma$-values to establish their significance. We found from the simulation that six out of eight bursts show significances between 2.5$-$3.2$\sigma$ while two bursts are insignificant ($\leqslant$2.1$\sigma$). This further supports the conclusion that these six bursts have real signals. This study presents the discovery of oscillations in bursts 1, 10, and 11 for the first time, whereas oscillations observed during bursts 4, 7 and 8 have already been reported by \citet{2019ApJ...878..145M}. 

For bursts 1 and 10 showing oscillations prior to the burst, we found 301 and 662 cases out of $10^5$ simulations, where the maximum $Z^2$ is greater than the value we report, indicating 3$\sigma$ and 2.7$\sigma$ significance, respectively. We note that 30 and 61 cases ($\simeq$10\%) out of them are seen prior to simulated bursts for bursts 1 and 10, respectively. However, we found 3592 cases with the maximum $Z^2$ greater than the reported value for burst 3. This shows that the oscillation seen prior to burst 3 is not significant enough, with only 2.1$\sigma$.

\begin{figure*}
	\includegraphics[scale=0.27]{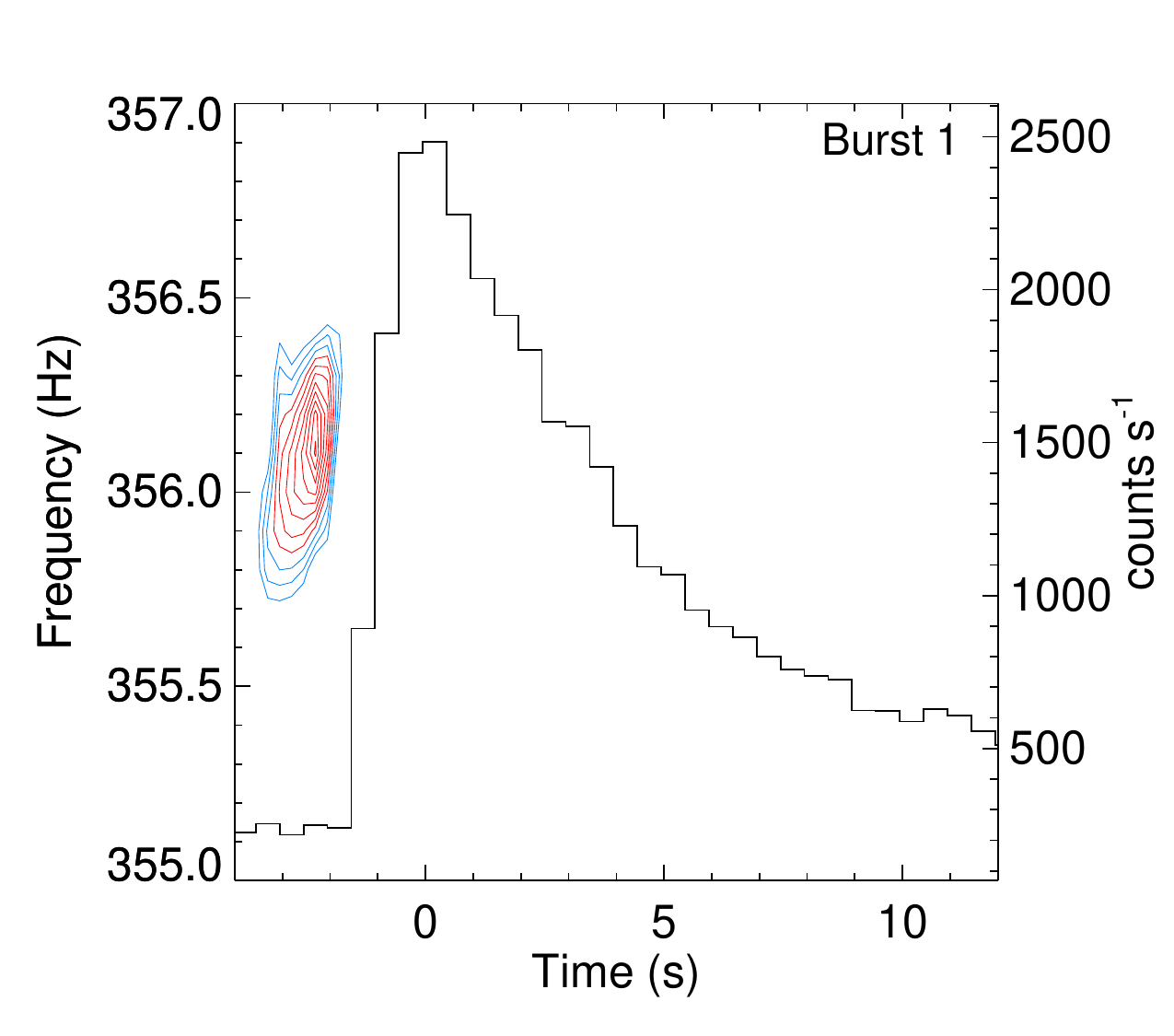} 
    \includegraphics[scale=0.27]{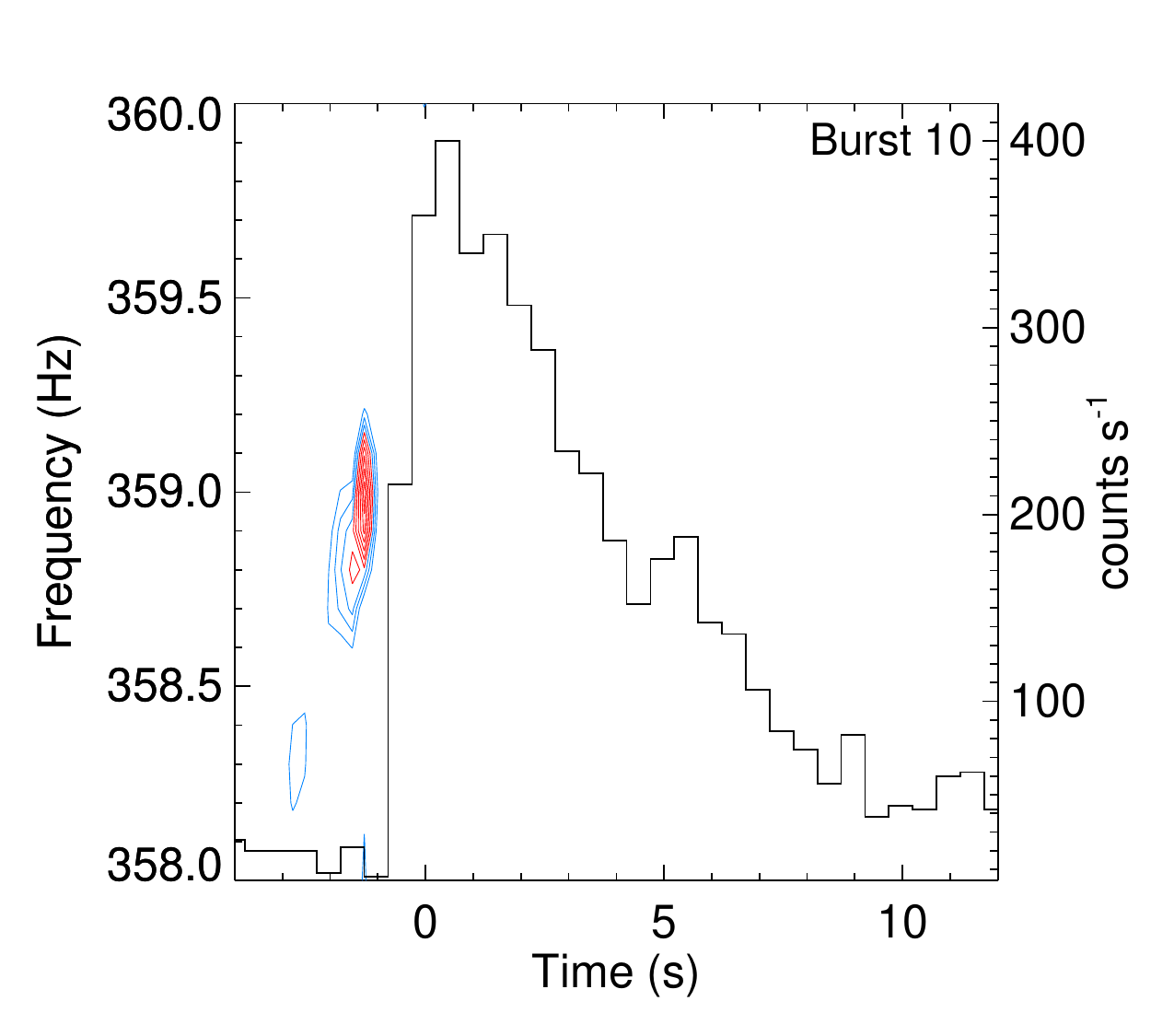}   
    \includegraphics[scale=0.27]{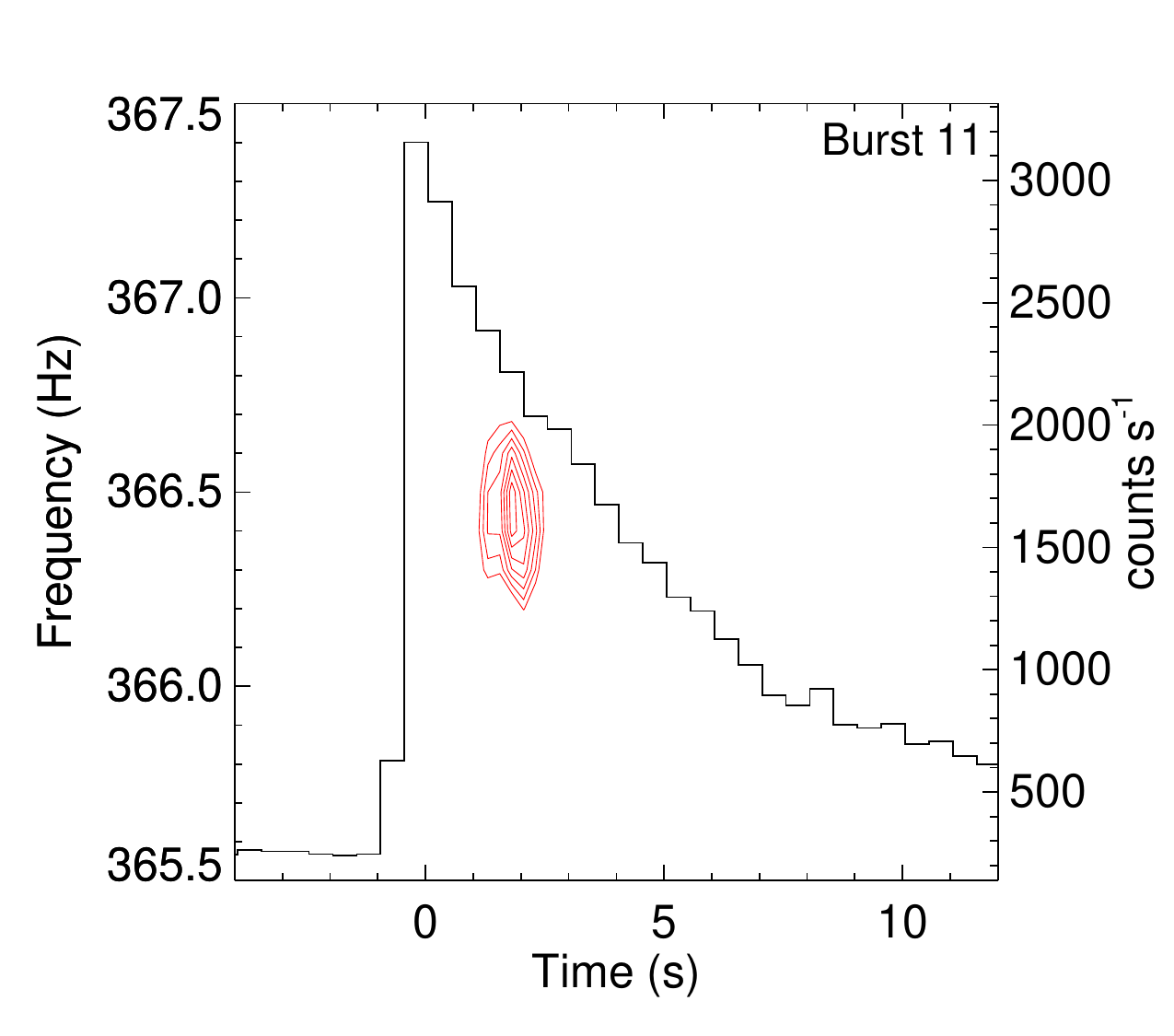}\\
    \includegraphics[scale=0.27]{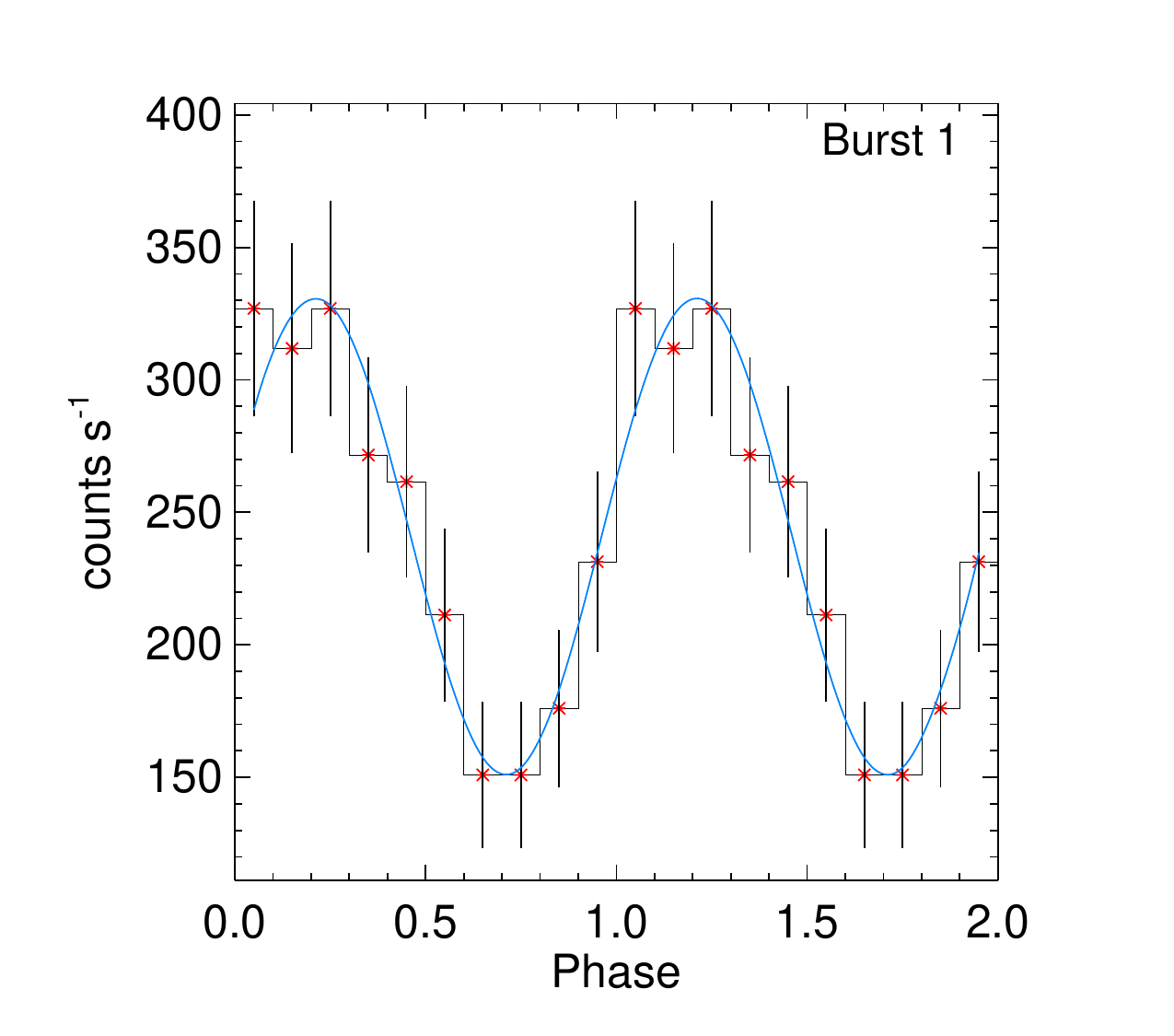}
    \includegraphics[scale=0.27]{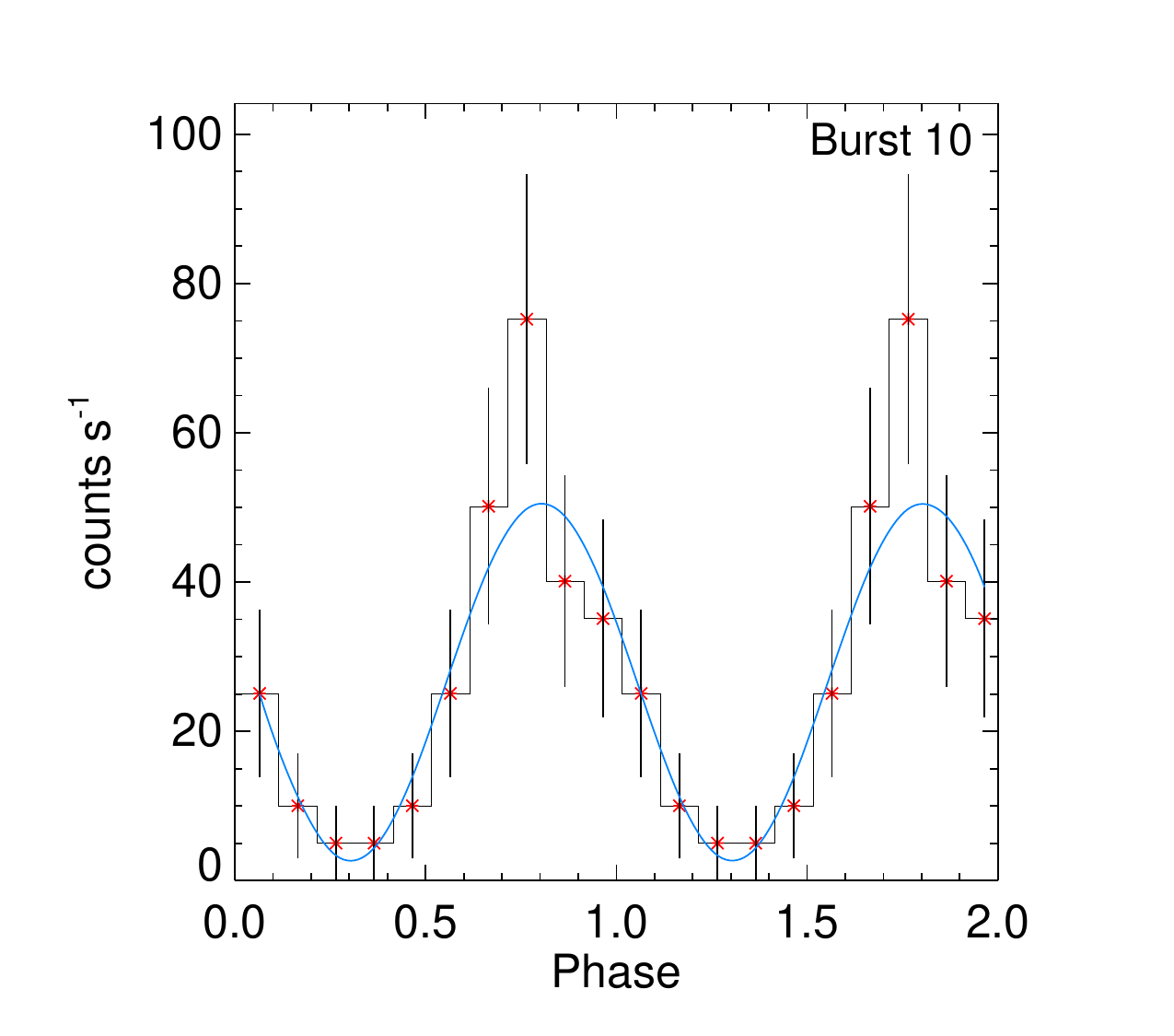}
    \includegraphics[scale=0.27]{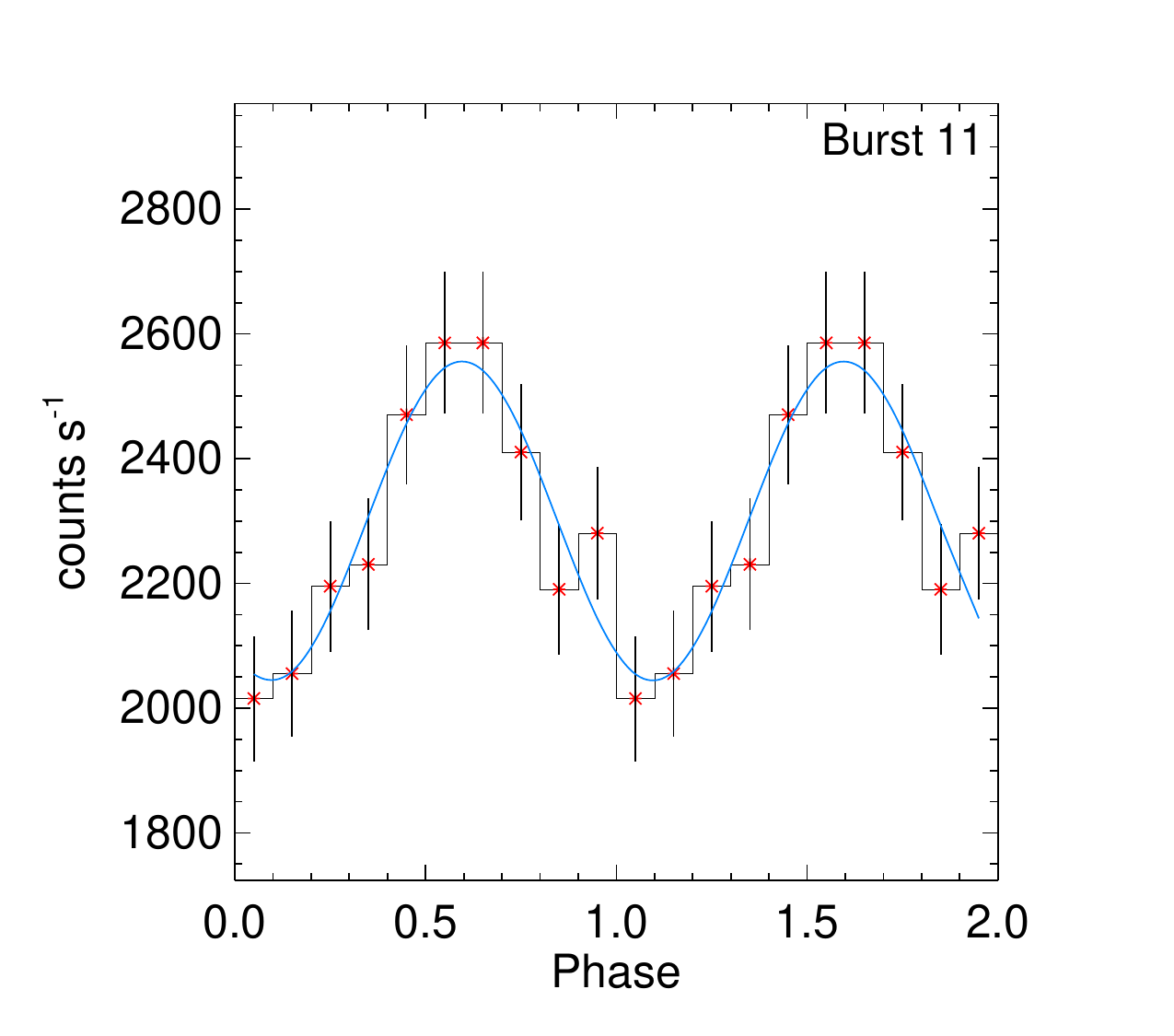}
     \caption{{\it Upper Panels:} Light curves of the X-ray bursts with a bin size of 0.5~s (black), where we modified the time axis of the plots to begin at the peak and contours of dynamical power spectra showing burst oscillations for bursts 1, 10 and 11 from \source\, (from left to right). Contours refer to Z$^2$ $\geq 10$ (blue) and  $\geq15$ (red) up to the maximum with steps of 2. {\it Lower Panels:} Pulse profiles calculated using the 2~s interval together with the best-fitting sinusoidal model (blue line).}
    \label{fig:burst_oscillations}
\end{figure*}

 In Figure~\ref{fig:burst_oscillations}, we only present resulting contour maps together with light curves of bursts 1, 10 and 11 since the remaining three bursts have been reported previously in \citet{2019ApJ...878..145M}. Power spectra are reconstructed using windows being shifted by 0.25~s and contours are plotted for Z$^{2}_{1}$ values of 10 and 15 to the maximum value, in steps of 2, in blue and red, respectively. We adjusted the frequency range in the figure according to the frequency of the significant signal. The lower panels in  Figure~\ref{fig:burst_oscillations} show phase-folded light curves calculated in the search interval window and in the energy band in which the signal with maximum power is identified.

%%%%%%%%%%%%%%%%%%%%%%%%%%%%%%%%%%%%%%%%%%%%

\begin{figure}	
   	\includegraphics[scale=0.22]{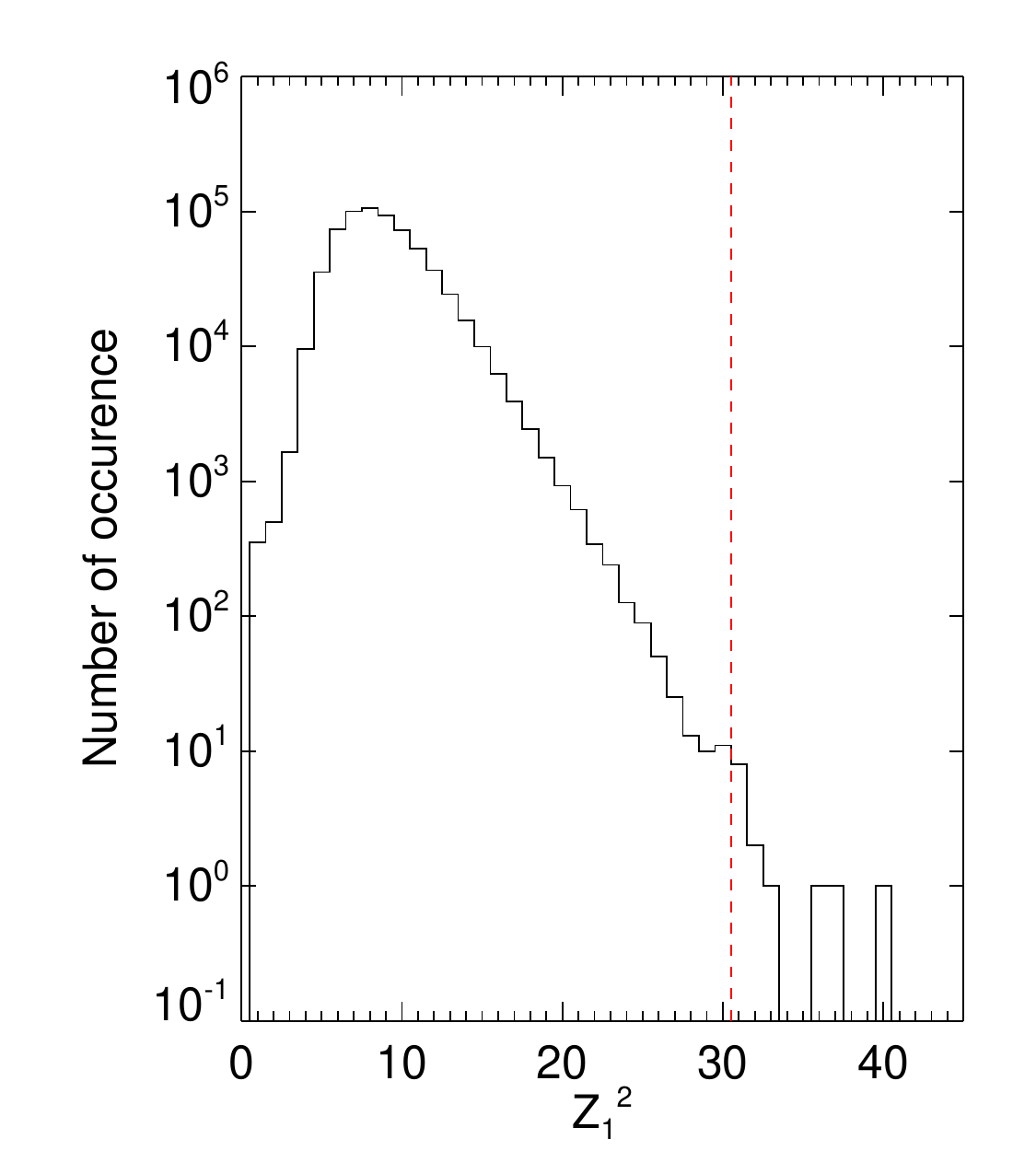}
    \includegraphics[scale=0.22]{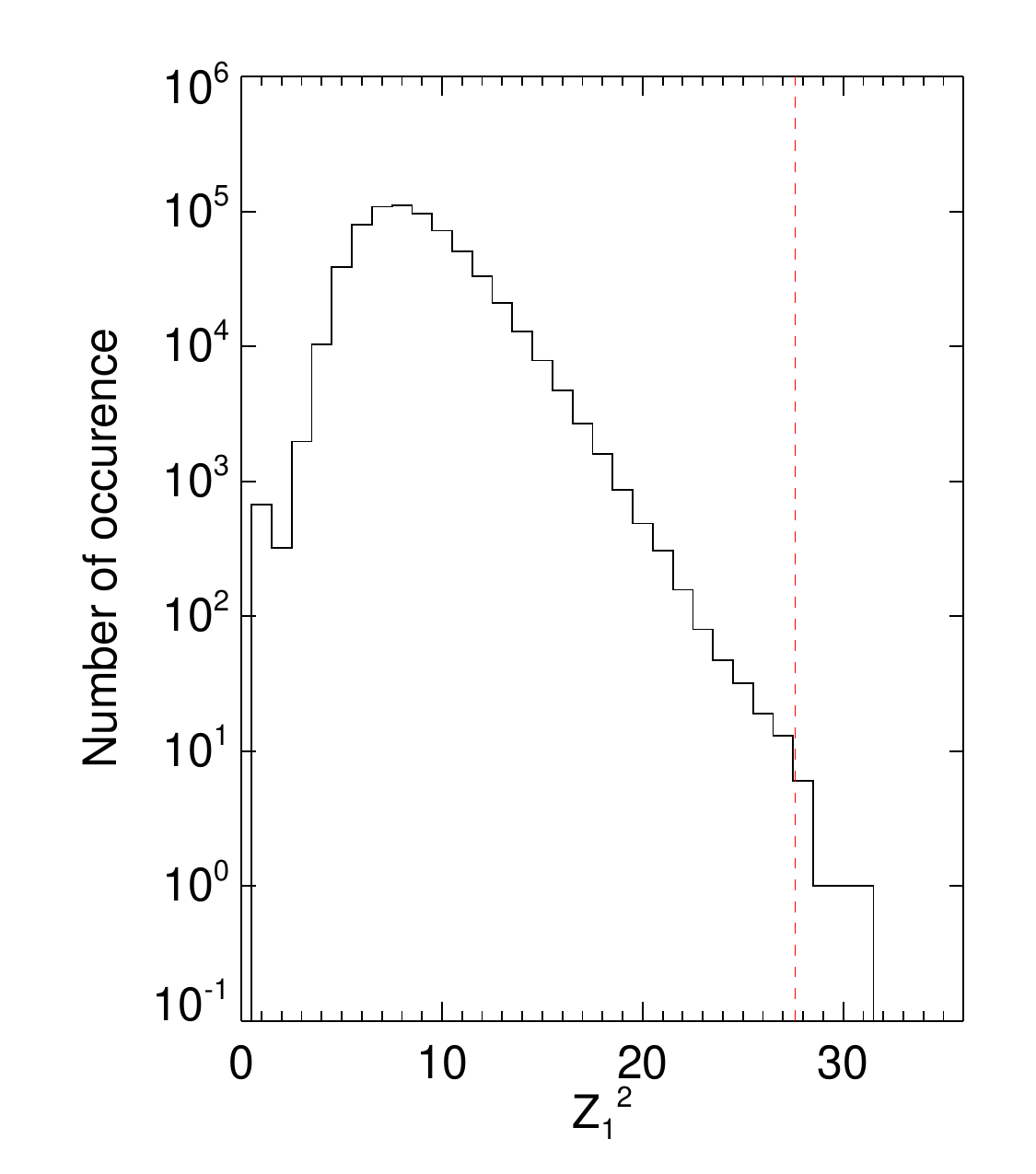}
        \caption{ Distribution of maximum $Z^2$ values obtained from whole 73 \nicer observations of \source, in the energy range of 0.5$-$12~keV (left), and 6$-$12~keV (right). The red dashed lines show the maximum $Z^2$ values measured from the oscillations prior to the bursts in the related energy band.}
    \label{fig:whole_obs_hist}
\end{figure}

As another method to check the chance occurrence of the oscillatory signals prior to the bursts we also performed the same timing analysis procedures (but time windows being shifted by 0.25~s) for all 73 \nicer observations of \source,\ only excluding the burst times, already used to search for burst oscillations in Subsection \ref{Sec:burst_osc_res}. In \autoref{fig:whole_obs_hist} we present histograms of maximum $Z^2$ values from the total 656,685 and 659,792 time steps of 2~s and 4~s time windows, respectively, in the three energy bands. First of all, we could not find any time interval within the existing clean event files of the analyzed observations here where the $Z^2$ is systematically larger for a time interval longer than the size of the search window (2 or 4 seconds). This indicates that no intermittent pulsation or oscillation behaviour is observed from \source~within any of the \nicer observations in the 355$-$370~Hz range. 

The analysis of non-bursting times across all observations revealed that the chance probability of obtaining $Z^2$ values as high as 30.5 and 27.6 (detected in the 0.5$-$12 and 6$-$12 keV bands, respectively, similar to bursts 1 and 10) was remarkably low, at 0.0026\% (17 cases) and 0.0023\% (15 cases). We note that for the burst 3 the same probability is found to be 0.12\% (817 cases). This test also supports the conclusion that the detection significance of the oscillations are highly unlikely to be obtained by chance and likely related to bursts 1 and 10.

%%%%%%%%%%%%%%%%%%%%%%%%%%%%%%%%%%%%

%%%%%%%%%%%%%%%%%%%%%%%%%%%%%%%%%%%%
\subsubsection{The origin of the pulsations before bursts 1 and 10}

To the best of our knowledge this is the first time from any bursting low mass X-ray binary that burst oscillations are detected just prior to  bursts and end with their rise. We must here note that usually an increase in the X-ray count rate is taken as an indication that the burst has started, but most likely the thermonuclear runaway starts before the observed rise as there should be a finite time for the heat/radiation to diffuse from the burning layer depth to the photosphere. This time difference likely depends on the ignition depth and the dominating transport mechanism \citep[dependent, in turn, on composition and accretion rate; see, e.g.,][and \citealt{2000ApJ...544..453C} for analytical estimates]{tugba_boztepe_2022_7600607,2008ApJS..174..261F,2010ApJS..189..204J}. We looked for any statistically significant deviation in the observed count rate when the oscillations are detected compared to the source count rates before. Unfortunately we could not obtain any significant deviation. 

To examine potential frequency evolution of the oscillations, in \autoref{fig:burst_oscillations} we lowered our limit on the Z$^2$ to 10 and showed in blue some additional contours. Although such a low Z$^2$ value is not statistically significant these additional contours show that in both cases there may be an increase in the oscillation frequency by about 0.5$-$1~Hz within about one and two seconds prior to the bursts in bursts 1 and 10, respectively.

\begin{figure}
    \centering
    \includegraphics[scale=0.4]{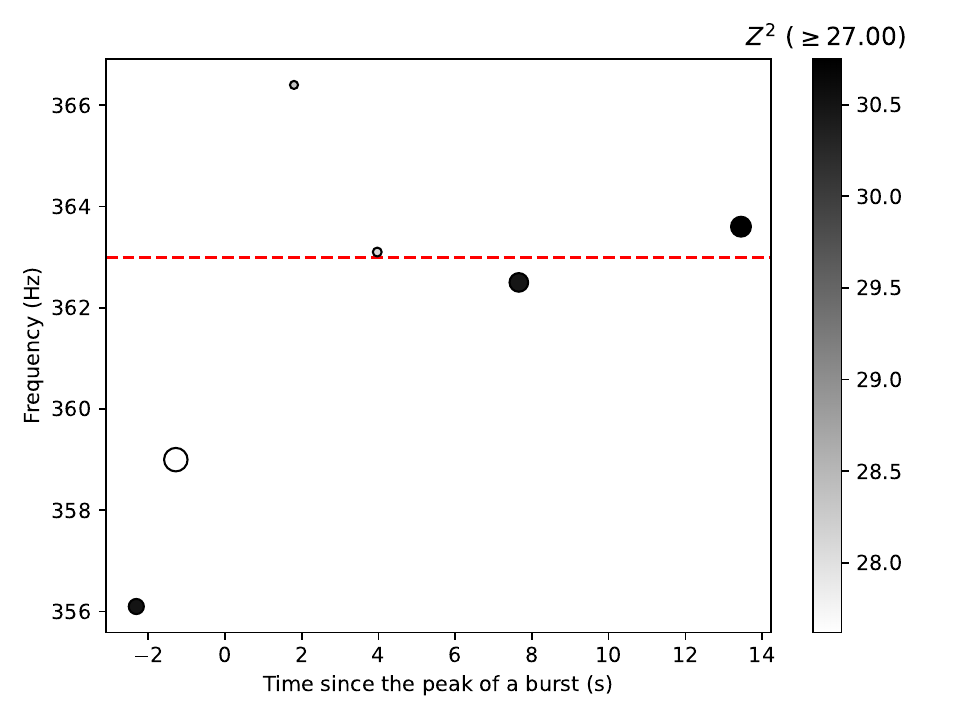}
    \caption{Frequencies of the detected oscillations as a function of time since the peak of a burst. Only the oscillations which were selected in Subsection \ref{sec:mc} are shown. Red dashed line shows the 363~Hz level. The color scale shows Z$^2$ values, while the size of the symbols changes with increasing fractional rms amplitude}.
    \label{fig:osc_drift}
\end{figure}

Frequency drifts in burst oscillation detections have been observed in various sources \citep{2002ApJ...580.1048M, 2012ARA&A..50..609W}. Often as a burst progresses, the detected oscillation frequency drifts upwards by a few Hz \citep{2012ARA&A..50..609W}. A notable instance involves  \cite{2001ApJ...549L..71W}, who reported a 5~Hz frequency drift during a burst from the X-ray binary X1658--298. A similar, upward drift is also observed here in burst 4, where the frequency increases by 5~Hz, although the oscillations are detected at different energy bands and with varying significances (as in the case of \cite{2001ApJ...549L..71W}; see \autoref{tab:bursts_oscil}). The oscillations detected here are not strong enough to be followed individually in terms of frequency drifts. However, we plot the frequencies of all the significant oscillations as a function of the time they are detected before or after the peak in \autoref{fig:osc_drift}, where the asymptotic drift towards 363~Hz can be seen especially when considering the oscillations detected prior to the bursts. This suggests that the nature of these oscillations may be similar to what is observed during the bursts. Based on all these findings we now discuss possible origins for our detections.

One plausible explanation for the oscillations before the rise is related to a hot spot and its time evolution. For this scenario to work, the initial burst rise may have a weak, slowly increasing part, that is not significant enough to discern in the count rate, but might be showing up as pulsations. This is partly at odds with the expectation that hydrogen-poor bursts would manifest relatively quickly after ignition \citep{tugba_boztepe_2022_7600607,2000ApJ...544..453C}, but it could be related to multi-D effects such as confinement or finite time flame spreading. As shown by \cite{2002ApJ...566.1018S} and \cite{2019ApJ...882..142C} such a flame can then quickly cover the surface of the neutron star causing the disappearance of the detected oscillations. A hot spot would be more justified by an off-equator ignition and this is more likely for slow rotators such as \source{} \citep{2002ApJ...566.1018S}. 
Furthermore, \cite{2007ApJ...657L..29C} and \cite{2020MNRAS.499.2148C} suggest that as the mass accretion rate increases the ignition latitude may also rise, due to the fact that burning on the equator should approach stability, which fits nicely with the fact that the rms of the detections before the bursts increases with the persistent count rate (as a proxy to the accretion rate; see \autoref{tab:bursts} and \autoref{tab:bursts_oscil}), since ignition at the equator should lead quickly to a ring around the star, more than a hot spot. However, in such a case one would expect to see similar oscillations associated to bursts 2, 9 and 11 as these bursts seem to be happening at similar persistent fluxes. We note that these bursts are labelled as possible photospheric radius expansion events and, unlike bursts 1 and 10, not clear PREs, and this may have some bearing on the fact that we do or do not see any pulsation, although in general PREs weaken the pulsations near the peak. Regarding the hot spot propagation scenario it is worth highlighting that, given the time between the start of the bursts and the detection of the oscillations and using the speed for flame spread from simulations \citep{2019ApJ...882..142C} we find that for burst 1 and burst 10 the flame spread reaches 5.6 and 2.8~km, respectively.

Another effect may be that the energy input from the thermonuclear burning excites oscillation modes \citep[such as r-modes, initially proposed as an explanation for burst oscillations by][see also \citealt{1996ApJ...467..773S}]{2004ApJ...600..939H}, which could influence the surface emission pattern. However, we note that these modes take some time to grow, and they are global waves on the star, so one would expect that the burning flame should have expanded significantly in order to put enough energy into them (and are indeed expected more during the tail of the bursts; see \citealt{2004ApJ...600..939H} and also \citealt{2020MNRAS.491.6032C}). Other related mode instabilities, such as the shear instabilities proposed by \citet{2005ApJ...630..441C}, are similar and also only suitable to explain pulsations in the tails.

A final speculation  may be related to accretion powered pulsations (APPs) as also suggested by \cite{2019ApJ...878..145M} for oscillations observed in the tails of bursts 4 and 8 with a large fractional rms amplitude. Using RXTE observations of HETE J1900.1$-$2455, \cite{2007ApJ...654L..73G} reported that APPs are influenced by the bursts. They reported an increase in the amplitude more or less in coincidence with some bursts and then a decline afterwards. On the other hand, \cite{2009ApJ...690.1856P} reported similar findings from SAX J1748.9$-$2021, but concluded that there is not a clear trend, noting that some bursts appeared to strengthen the APPs and others did not. In both of these sources persistent or intermittent pulsations have been observed clearly. In the case of \source\, no such pulsation has been reported before and our search for all the non-bursting times in the \nicer data revealed no such significant pulsation.  One explanation could be that \source\, has extremely weak APPs, if any at all, and that the oscillations reported here are APPs which are enhanced by the burning/burst occurrence, similar to what \cite{2019ApJ...878..145M} suggested for the burst oscillations in the tail of bursts 4 and 8. The fact that we detect oscillations before the start of the bursts perhaps makes these detections more suggestive of that phenomenology. If this is the explanation, than the detections reported here are not the first burst oscillations detected before a burst, but perhaps the first APPs from \source.

\section{Conclusions}
We have searched two years of archival \nicer data of the low mass X-ray binary, \source. We detected 11 X-ray bursts, 3 of which show photospheric radius expansion. Our results show that, unlike some of the earlier results from \nicer, the use of a scaling factor is statistically not required to model the X-ray spectra extracted during the bursts. This result is most likely due to the significantly large hydrogen column density value in the line of sight towards \source. Similar results are obtained for 4U~1608$-$52 and XTE~1739$-$286 \citep{2021ApJ...907...79B, 2021ApJ...910...37G} for which the absorption due to interstellar medium is similarly large. We compared our spectral results at the peak times of each burst to the extensive MINBAR sample. The results seem to agree with earlier measurements in terms of peak flux and blackbody parameters. Similarly, \cite{2022MNRAS.510.1577G} compared the spectral parameters at the peaks of the bursts observed from Aql~X-1 with the MINBAR sample and showed that when the \fa method is not employed, the inferred parameters show a systematic trend of being lower than what is inferred from the MINBAR sample, which is based on data obtained in the 3$-$25~keV band. The fact that we do not see such a systematic trend here, as well as the much better fits compared to the results from other sources \citep{2022MNRAS.510.1577G,2022ApJ...935..154G}, together with the $N_{H}$, further imply that the excess observed in some of the bursters is probably only limited to the soft X-ray band (below 2.5~keV). This is similar to the findings by \cite{2022MNRAS.510.1577G}, where fitting only the 3$-$10~keV data without an \fa factor resulted in similar spectral parameters for the bursts when using the full band of the \nicer but using the \fa factor. 

We also reported our search for burst oscillations during the 11 bursts detected. In 6 of these 11 events, we detected significant oscillations at around 363~Hz, similar to previous reports from this source \citep[see, e.g.,][]{1996ApJ...469L...9S,2019ApJ...878..145M,2020ApJS..249...32G}. We found that two bursts featured oscillations between their peak and e-folding time, while another two displayed oscillations during burst tails. Although, previous reports of burst oscillations from \source~were confined to the 363$\pm$5~Hz interval we here focused on a broader frequency range. Such an analysis enabled us to detect several similarly significant signals below or above the previous frequency limits.

Most remarkably, in two bursts we detect significant oscillations just preceding the observed X-ray bursts. To the best of our knowledge this is the first time from any bursting low mass X-ray binary that burst oscillations are detected just prior to  bursts and end with their rise. In burst 1 we detect oscillations prior to the burst in the 0.5$-$12~keV (Z$^2$ = 30.5) and 0.5$-$6~keV (Z$^2$ = 27) bands, while they are absent in the 6$-$12~keV range (Z$^2$ $< 15$). On the other hand in burst 10 while the oscillations are detected in the 6$-$12~keV band (Z$^2$ = 27.6) they are not detected at lower energies. The difference may at least be partly related to the observed number of counts in each case, since during burst 10 the source is brighter by about 15\%. Given the strong interstellar absorption towards the source which compensates for the large effective area of \nicer at lower energies it would be expected that such oscillations may be detected in the archival Rossi X-ray Timing Explorer (RXTE) data. Finding more examples of oscillations outside the bursts of \source{} will certainly help understanding their nature and their connections to the thermonuclear burning. A separate analysis on the search for similar events in the RXTE archive is currently underway and will be presented elsewhere.

%%%%%%%%%%%%%%%%%%%%%%%%%%%%%%%%%%%%%%%%%%%%%%%%%
% moved text was here
%%%%%%%%%%%%%%%%%%%%%%%%%%%%%%%%%%%%%%%%%%%%%%%%%

%%%%%%%%%%%%%%%%%%%%%%%%%%%%%%%%%%%%%%%%%%%%%%%%%%
\section*{Acknowledgements}
We thank the referee for valuable comments and suggestions that improved the manuscript.
This work is supported by the Scientific Research Projects Coordination Unit of Istanbul University (ADEP Project No: FBA-2023-39409) and the Turkish Republic, Presidency of Strategy and Budget project, 2016K121370.~YC acknowledges support from the grant RYC2021-032718-I, financed by MCIN/AEI/10.13039/501100011033 and the European Union NextGenerationEU/PRTR. This work was supported by NASA through the \nicer mission and the Astrophysics Explorers Program. 

%%%%%%%%%%%%%%%%%%%%%%%%%%%%%%%%%%%%%%%%%%%%%%%%%%
\section*{Data Availability}
All the data used in this publication is publicly available through NASA/HEASARC archives.

\bibliography{sample631}{}

\begin{thebibliography}{}
\expandafter\ifx\csname natexlab\endcsname\relax\def\natexlab#1{#1}\fi
\providecommand{\url}[1]{\href{#1}{#1}}
\providecommand{\dodoi}[1]{doi:~\href{http://doi.org/#1}{\nolinkurl{#1}}}
\providecommand{\doeprint}[1]{\href{http://ascl.net/#1}{\nolinkurl{http://ascl.net/#1}}}
\providecommand{\doarXiv}[1]{\href{https://arxiv.org/abs/#1}{\nolinkurl{https://arxiv.org/abs/#1}}}

\bibitem[{{Astropy Collaboration} {et~al.}(2018){Astropy Collaboration},
  {Price-Whelan}, {Sip{\H{o}}cz}, {G{\"u}nther}, {Lim}, {Crawford}, {Conseil},
  {Shupe}, {Craig}, {Dencheva}, {Ginsburg}, {Vand erPlas}, {Bradley},
  {P{\'e}rez-Su{\'a}rez}, {de Val-Borro}, {Aldcroft}, {Cruz}, {Robitaille},
  {Tollerud}, {Ardelean}, {Babej}, {Bach}, {Bachetti}, {Bakanov}, {Bamford},
  {Barentsen}, {Barmby}, {Baumbach}, {Berry}, {Biscani}, {Boquien}, {Bostroem},
  {Bouma}, {Brammer}, {Bray}, {Breytenbach}, {Buddelmeijer}, {Burke},
  {Calderone}, {Cano Rodr{\'\i}guez}, {Cara}, {Cardoso}, {Cheedella}, {Copin},
  {Corrales}, {Crichton}, {D'Avella}, {Deil}, {Depagne}, {Dietrich}, {Donath},
  {Droettboom}, {Earl}, {Erben}, {Fabbro}, {Ferreira}, {Finethy}, {Fox},
  {Garrison}, {Gibbons}, {Goldstein}, {Gommers}, {Greco}, {Greenfield},
  {Groener}, {Grollier}, {Hagen}, {Hirst}, {Homeier}, {Horton}, {Hosseinzadeh},
  {Hu}, {Hunkeler}, {Ivezi{\'c}}, {Jain}, {Jenness}, {Kanarek}, {Kendrew},
  {Kern}, {Kerzendorf}, {Khvalko}, {King}, {Kirkby}, {Kulkarni}, {Kumar},
  {Lee}, {Lenz}, {Littlefair}, {Ma}, {Macleod}, {Mastropietro}, {McCully},
  {Montagnac}, {Morris}, {Mueller}, {Mumford}, {Muna}, {Murphy}, {Nelson},
  {Nguyen}, {Ninan}, {N{\"o}the}, {Ogaz}, {Oh}, {Parejko}, {Parley}, {Pascual},
  {Patil}, {Patil}, {Plunkett}, {Prochaska}, {Rastogi}, {Reddy Janga},
  {Sabater}, {Sakurikar}, {Seifert}, {Sherbert}, {Sherwood-Taylor}, {Shih},
  {Sick}, {Silbiger}, {Singanamalla}, {Singer}, {Sladen}, {Sooley},
  {Sornarajah}, {Streicher}, {Teuben}, {Thomas}, {Tremblay}, {Turner},
  {Terr{\'o}n}, {van Kerkwijk}, {de la Vega}, {Watkins}, {Weaver}, {Whitmore},
  {Woillez}, {Zabalza}, \& {Astropy Contributors}}]{2018AJ....156..123A}
{Astropy Collaboration}, {Price-Whelan}, A.~M., {Sip{\H{o}}cz}, B.~M., {et~al.}
  2018, \aj, 156, 123, \dodoi{10.3847/1538-3881/aabc4f}

\bibitem[{{Basinska} {et~al.}(1984{\natexlab{a}}){Basinska}, {Lewin},
  {Sztajno}, {Cominsky}, \& {Marshall}}]{1984ApJ...281..337B}
{Basinska}, E.~M., {Lewin}, W.~H.~G., {Sztajno}, M., {Cominsky}, L.~R., \&
  {Marshall}, F.~J. 1984{\natexlab{a}}, \apj, 281, 337, \dodoi{10.1086/162103}

\bibitem[{{Basinska} {et~al.}(1984{\natexlab{b}}){Basinska}, {Lewin},
  {Sztajno}, {Cominsky}, \& {Marshall}}]{Basinska}
---. 1984{\natexlab{b}}, \apj, 281, 337, \dodoi{10.1086/162103}

\bibitem[{{Bhattacharyya}(2010)}]{2010AdSpR..45..949B}
{Bhattacharyya}, S. 2010, Advances in Space Research, 45, 949,
  \dodoi{10.1016/j.asr.2010.01.010}

\bibitem[{{Bhattacharyya} {et~al.}(2018){Bhattacharyya}, {Yadav}, {Sridhar},
  {Verdhan Chauhan}, {Agrawal}, {Antia}, {Pahari}, {Misra}, {Katoch},
  {Manchanda}, \& {Paul}}]{2018ApJ...860...88B}
{Bhattacharyya}, S., {Yadav}, J.~S., {Sridhar}, N., {et~al.} 2018, \apj, 860,
  88, \dodoi{10.3847/1538-4357/aac495}

\bibitem[{{Bilous} \& {Watts}(2019)}]{2019ApJS..245...19B}
{Bilous}, A.~V., \& {Watts}, A.~L. 2019, \apjs, 245, 19,
  \dodoi{10.3847/1538-4365/ab2fe1}

\bibitem[{{Bogdanov} {et~al.}(2019){Bogdanov}, {Lamb}, {Mahmoodifar}, {Miller},
  {Morsink}, {Riley}, {Strohmayer}, {Tung}, {Watts}, {Dittmann}, {Chakrabarty},
  {Guillot}, {Arzoumanian}, \& {Gendreau}}]{2019ApJ...887L..26B}
{Bogdanov}, S., {Lamb}, F.~K., {Mahmoodifar}, S., {et~al.} 2019, \apjl, 887,
  L26, \dodoi{10.3847/2041-8213/ab5968}

\bibitem[{{Buccheri} {et~al.}(1983){Buccheri}, {Bennett}, {Bignami}, {Bloemen},
  {Boriakoff}, {Caraveo}, {Hermsen}, {Kanbach}, {Manchester}, {Masnou},
  {Mayer-Hasselwander}, {{\"O}zel}, {Paul}, {Sacco}, {Scarsi}, \&
  {Strong}}]{buccheri}
{Buccheri}, R., {Bennett}, K., {Bignami}, G.~F., {et~al.} 1983, \aap, 128, 245

\bibitem[{{Buisson} {et~al.}(2020){Buisson}, {Altamirano}, {Bult}, {Mancuso},
  {G{\"u}ver}, {Jaisawal}, {Hare}, {Albayati}, {Arzoumanian}, {Castro Segura},
  {Chakrabarty}, {Gandhi}, {Guillot}, {Homan}, {Gendreau}, {Jiang},
  {Malacaria}, {Miller}, {{\"O}zbey Arabac{\i}}, {Remillard}, {Strohmayer},
  {Tombesi}, {Tomsick}, {Vincentelli}, \& {Walton}}]{2020MNRAS.499..793B}
{Buisson}, D.~J.~K., {Altamirano}, D., {Bult}, P., {et~al.} 2020, \mnras, 499,
  793, \dodoi{10.1093/mnras/staa2749}

\bibitem[{{Bult} {et~al.}(2019){Bult}, {Jaisawal}, {G{\"u}ver}, {Strohmayer},
  {Altamirano}, {Arzoumanian}, {Ballantyne}, {Chakrabarty}, {Chenevez},
  {Gendreau}, {Guillot}, \& {Ludlam}}]{2019ApJ...885L...1B}
{Bult}, P., {Jaisawal}, G.~K., {G{\"u}ver}, T., {et~al.} 2019, \apjl, 885, L1,
  \dodoi{10.3847/2041-8213/ab4ae1}

\bibitem[{{Bult} {et~al.}(2021){Bult}, {Altamirano}, {Arzoumanian}, {Bilous},
  {Chakrabarty}, {Gendreau}, {G{\"u}ver}, {Jaisawal}, {Kuulkers}, {Malacaria},
  {Ng}, {Sanna}, \& {Strohmayer}}]{2021ApJ...907...79B}
{Bult}, P., {Altamirano}, D., {Arzoumanian}, Z., {et~al.} 2021, \apj, 907, 79,
  \dodoi{10.3847/1538-4357/abd54b}

\bibitem[{{Bult} {et~al.}(2022){Bult}, {Mancuso}, {Strohmayer}, {Albayati},
  {Altamirano}, {Buisson}, {Chenevez}, {Guillot}, {G{\"u}ver}, {Iwakiri},
  {Jaisawal}, {Ng}, {Sanna}, \& {Swank}}]{2022ApJ...940...81B}
{Bult}, P., {Mancuso}, G.~C., {Strohmayer}, T.~E., {et~al.} 2022, \apj, 940,
  81, \dodoi{10.3847/1538-4357/ac9b26}

\bibitem[{{Cavecchi} {et~al.}(2020){Cavecchi}, {Galloway}, {Goodwin},
  {Johnston}, \& {Heger}}]{2020MNRAS.499.2148C}
{Cavecchi}, Y., {Galloway}, D.~K., {Goodwin}, A.~J., {Johnston}, Z., \&
  {Heger}, A. 2020, \mnras, 499, 2148, \dodoi{10.1093/mnras/staa2858}

\bibitem[{{Cavecchi} \& {Spitkovsky}(2019)}]{2019ApJ...882..142C}
{Cavecchi}, Y., \& {Spitkovsky}, A. 2019, \apj, 882, 142,
  \dodoi{10.3847/1538-4357/ab3650}

\bibitem[{{Chakrabarty} {et~al.}(2003){Chakrabarty}, {Morgan}, {Muno},
  {Galloway}, {Wijnands}, {van der Klis}, \& {Markwardt}}]{2003Natur.424...42C}
{Chakrabarty}, D., {Morgan}, E.~H., {Muno}, M.~P., {et~al.} 2003, \nat, 424,
  42, \dodoi{10.1038/nature01732}

\bibitem[{{Chakraborty} \& {Bhattacharyya}(2014)}]{2014ApJ...792....4C}
{Chakraborty}, M., \& {Bhattacharyya}, S. 2014, \apj, 792, 4,
  \dodoi{10.1088/0004-637X/792/1/4}

\bibitem[{{Chambers} \& {Watts}(2020)}]{2020MNRAS.491.6032C}
{Chambers}, F.~R.~N., \& {Watts}, A.~L. 2020, \mnras, 491, 6032,
  \dodoi{10.1093/mnras/stz3449}

\bibitem[{{Cooper} \& {Narayan}(2007)}]{2007ApJ...657L..29C}
{Cooper}, R.~L., \& {Narayan}, R. 2007, \apjl, 657, L29, \dodoi{10.1086/513077}

\bibitem[{{Cumming}(2005)}]{2005ApJ...630..441C}
{Cumming}, A. 2005, \apj, 630, 441, \dodoi{10.1086/431731}

\bibitem[{{Cumming} \& {Bildsten}(2000)}]{2000ApJ...544..453C}
{Cumming}, A., \& {Bildsten}, L. 2000, \apj, 544, 453, \dodoi{10.1086/317191}

\bibitem[{{D'A{\'\i}} {et~al.}(2006){D'A{\'\i}}, {di Salvo}, {Iaria},
  {M{\'e}ndez}, {Burderi}, {Lavagetto}, {Lewin}, {Robba}, {Stella}, \& {van der
  Klis}}]{2006A&A...448..817D}
{D'A{\'\i}}, A., {di Salvo}, T., {Iaria}, R., {et~al.} 2006, \aap, 448, 817,
  \dodoi{10.1051/0004-6361:20053228}

\bibitem[{{Damen} {et~al.}(1990){Damen}, {Magnier}, {Lewin}, {Tan}, {Penninx},
  \& {van Paradijs}}]{1990A&A...237..103D}
{Damen}, E., {Magnier}, E., {Lewin}, W.~H.~G., {et~al.} 1990, \aap, 237, 103

\bibitem[{{Di Salvo} {et~al.}(2000){Di Salvo}, {Iaria}, {Burderi}, \&
  {Robba}}]{2000ApJ...542.1034D}
{Di Salvo}, T., {Iaria}, R., {Burderi}, L., \& {Robba}, N.~R. 2000, \apj, 542,
  1034, \dodoi{10.1086/317029}

\bibitem[{{Egron} {et~al.}(2011){Egron}, {di Salvo}, {Burderi}, {Papitto},
  {Barrag{\'a}n}, {Dauser}, {Wilms}, {D'A{\`\i}}, {Riggio}, {Iaria}, \&
  {Robba}}]{2011A&A...530A..99E}
{Egron}, E., {di Salvo}, T., {Burderi}, L., {et~al.} 2011, \aap, 530, A99,
  \dodoi{10.1051/0004-6361/201016093}

\bibitem[{{Fisker} {et~al.}(2008){Fisker}, {Schatz}, \&
  {Thielemann}}]{2008ApJS..174..261F}
{Fisker}, J.~L., {Schatz}, H., \& {Thielemann}, F.-K. 2008, \apjs, 174, 261,
  \dodoi{10.1086/521104}

\bibitem[{{Fragile} {et~al.}(2020){Fragile}, {Ballantyne}, \&
  {Blankenship}}]{2020NatAs...4..541F}
{Fragile}, P.~C., {Ballantyne}, D.~R., \& {Blankenship}, A. 2020, Nature
  Astronomy, 4, 541, \dodoi{10.1038/s41550-019-0987-5}

\bibitem[{{Fragile} {et~al.}(2018){Fragile}, {Ballantyne}, {Maccarone}, \&
  {Witry}}]{2018ApJ...867L..28F}
{Fragile}, P.~C., {Ballantyne}, D.~R., {Maccarone}, T.~J., \& {Witry}, J. W.~L.
  2018, \apjl, 867, L28, \dodoi{10.3847/2041-8213/aaeb99}

\bibitem[{{Franco}(2001)}]{2001ApJ...554..340F}
{Franco}, L.~M. 2001, \apj, 554, 340, \dodoi{10.1086/321341}

\bibitem[{{Freeman} {et~al.}(2001){Freeman}, {Doe}, \&
  {Siemiginowska}}]{2001SPIE.4477...76F}
{Freeman}, P., {Doe}, S., \& {Siemiginowska}, A. 2001, in Society of
  Photo-Optical Instrumentation Engineers (SPIE) Conference Series, Vol. 4477,
  Astronomical Data Analysis, ed. J.-L. {Starck} \& F.~D. {Murtagh}, 76--87,
  \dodoi{10.1117/12.447161}

\bibitem[{{Galloway} {et~al.}(2007){Galloway}, {Morgan}, {Krauss}, {Kaaret}, \&
  {Chakrabarty}}]{2007ApJ...654L..73G}
{Galloway}, D.~K., {Morgan}, E.~H., {Krauss}, M.~I., {Kaaret}, P., \&
  {Chakrabarty}, D. 2007, \apjl, 654, L73, \dodoi{10.1086/510741}

\bibitem[{{Galloway} {et~al.}(2008){Galloway}, {Muno}, {Hartman}, {Psaltis}, \&
  {Chakrabarty}}]{2008ApJS..179..360G}
{Galloway}, D.~K., {Muno}, M.~P., {Hartman}, J.~M., {Psaltis}, D., \&
  {Chakrabarty}, D. 2008, \apjs, 179, 360, \dodoi{10.1086/592044}

\bibitem[{{Galloway} {et~al.}(2003){Galloway}, {Psaltis}, {Chakrabarty}, \&
  {Muno}}]{2003ApJ...590..999G}
{Galloway}, D.~K., {Psaltis}, D., {Chakrabarty}, D., \& {Muno}, M.~P. 2003,
  \apj, 590, 999, \dodoi{10.1086/375049}

\bibitem[{{Galloway} {et~al.}(2020){Galloway}, {in't Zand}, {Chenevez},
  {W{\"o}rpel}, {Keek}, {Ootes}, {Watts}, {Gisler}, {Sanchez-Fernandez}, \&
  {Kuulkers}}]{2020ApJS..249...32G}
{Galloway}, D.~K., {in't Zand}, J., {Chenevez}, J., {et~al.} 2020, \apjs, 249,
  32, \dodoi{10.3847/1538-4365/ab9f2e}

\bibitem[{{Gendreau} {et~al.}(2016){Gendreau}, {Arzoumanian}, {Adkins},
  {Albert}, {Anders}, {Aylward}, {Baker}, {Balsamo}, {Bamford}, {Benegalrao},
  {Berry}, {Bhalwani}, {Black}, {Blaurock}, {Bronke}, {Brown}, {Budinoff},
  {Cantwell}, {Cazeau}, {Chen}, {Clement}, {Colangelo}, {Coleman},
  {Coopersmith}, {Dehaven}, {Doty}, {Egan}, {Enoto}, {Fan}, {Ferro}, {Foster},
  {Galassi}, {Gallo}, {Green}, {Grosh}, {Ha}, {Hasouneh}, {Heefner}, {Hestnes},
  {Hoge}, {Jacobs}, {J{\o}rgensen}, {Kaiser}, {Kellogg}, {Kenyon}, {Koenecke},
  {Kozon}, {LaMarr}, {Lambertson}, {Larson}, {Lentine}, {Lewis}, {Lilly},
  {Liu}, {Malonis}, {Manthripragada}, {Markwardt},
  {2020ApJS..249...32GMatonak}, {Mcginnis}, {Miller}, {Mitchell}, {Mitchell},
  {Mohammed}, {Monroe}, {Montt de Garcia}, {Mul{\'e}}, {Nagao}, {Ngo},
  {Norris}, {Norwood}, {Novotka}, {Okajima}, {Olsen}, {Onyeachu}, {Orosco},
  {Peterson}, {Pevear}, {Pham}, {Pollard}, {Pope}, {Powers}, {Powers}, {Price},
  {Prigozhin}, {Ramirez}, {Reid}, {Remillard}, {Rogstad}, {Rosecrans}, {Rowe},
  {Sager}, {Sanders}, {Savadkin}, {Saylor}, {Schaeffer}, {Schweiss}, {Semper},
  {Serlemitsos}, {Shackelford}, {Soong}, {Struebel}, {Vezie}, {Villasenor},
  {Winternitz}, {Wofford}, {Wright}, {Yang}, \& {Yu}}]{2016SPIE.9905E..1HG}
{Gendreau}, K.~C., {Arzoumanian}, Z., {Adkins}, P.~W., {et~al.} 2016, in
  Society of Photo-Optical Instrumentation Engineers (SPIE) Conference Series,
  Vol. 9905, Space Telescopes and Instrumentation 2016: Ultraviolet to Gamma
  Ray, ed. J.-W.~A. {den Herder}, T.~{Takahashi}, \& M.~{Bautz}, 99051H,
  \dodoi{10.1117/12.2231304}

\bibitem[{{G{\"u}ver} {et~al.}(2012{\natexlab{a}}){G{\"u}ver}, {{\"O}zel}, \&
  {Psaltis}}]{2012ApJ...747...77G}
{G{\"u}ver}, T., {{\"O}zel}, F., \& {Psaltis}, D. 2012{\natexlab{a}}, \apj,
  747, 77, \dodoi{10.1088/0004-637X/747/1/77}

\bibitem[{{G{\"u}ver} {et~al.}(2012{\natexlab{b}}){G{\"u}ver}, {Psaltis}, \&
  {{\"O}zel}}]{2012ApJ...747...76G}
{G{\"u}ver}, T., {Psaltis}, D., \& {{\"O}zel}, F. 2012{\natexlab{b}}, \apj,
  747, 76, \dodoi{10.1088/0004-637X/747/1/76}

\bibitem[{{G{\"u}ver} {et~al.}(2021){G{\"u}ver}, {Boztepe},
  {G{\"o}{\u{g}}{\"u}{\c{s}}}, {Chakraborty}, {Strohmayer}, {Bult},
  {Altamirano}, {Jaisawal}, {Kocab{\i}y{\i}k}, {Malacaria}, {Kashyap},
  {Gendreau}, {Arzoumanian}, \& {Chakrabarty}}]{2021ApJ...910...37G}
{G{\"u}ver}, T., {Boztepe}, T., {G{\"o}{\u{g}}{\"u}{\c{s}}}, E., {et~al.} 2021,
  \apj, 910, 37, \dodoi{10.3847/1538-4357/abe1ae}

\bibitem[{{G{\"u}ver} {et~al.}(2022{\natexlab{a}}){G{\"u}ver}, {Boztepe},
  {Ballantyne}, {Bostanc{\i}}, {Bult}, {Jaisawal}, {G{\"o}{\u{g}}{\"u}{\c{s}}},
  {Strohmayer}, {Altamirano}, {Guillot}, \&
  {Chakrabarty}}]{2022MNRAS.510.1577G}
{G{\"u}ver}, T., {Boztepe}, T., {Ballantyne}, D.~R., {et~al.}
  2022{\natexlab{a}}, \mnras, 510, 1577, \dodoi{10.1093/mnras/stab3422}

\bibitem[{{G{\"u}ver} {et~al.}(2022{\natexlab{b}}){G{\"u}ver}, {Bostanc{\i}},
  {Boztepe}, {G{\"o}{\u{g}}{\"u}{\c{s}}}, {Bult}, {Kashyap}, {Chakraborty},
  {Ballantyne}, {Ludlam}, {Malacaria}, {Jaisawal}, {Strohmayer}, {Guillot}, \&
  {Ng}}]{2022ApJ...935..154G}
{G{\"u}ver}, T., {Bostanc{\i}}, Z.~F., {Boztepe}, T., {et~al.}
  2022{\natexlab{b}}, \apj, 935, 154, \dodoi{10.3847/1538-4357/ac8106}

\bibitem[{{Hansen} \& {van Horn}(1975)}]{1975ApJ...195..735H}
{Hansen}, C.~J., \& {van Horn}, H.~M. 1975, \apj, 195, 735,
  \dodoi{10.1086/153375}

\bibitem[{{Hasinger} \& {van der Klis}(1989)}]{1989A&A...225...79H}
{Hasinger}, G., \& {van der Klis}, M. 1989, \aap, 225, 79

\bibitem[{{Heyl}(2004)}]{2004ApJ...600..939H}
{Heyl}, J.~S. 2004, \apj, 600, 939, \dodoi{10.1086/379966}

\bibitem[{{Hoffman} {et~al.}(1976){Hoffman}, {Lewin}, {Doty}, {Hearn}, {Clark},
  {Jernigan}, \& {Li}}]{1976ApJ...210L..13H}
{Hoffman}, J.~A., {Lewin}, W.~H.~G., {Doty}, J., {et~al.} 1976, \apjl, 210,
  L13, \dodoi{10.1086/182292}

\bibitem[{Hunter(2007)}]{Hunter:2007}
Hunter, J.~D. 2007, Computing in Science \& Engineering, 9, 90,
  \dodoi{10.1109/MCSE.2007.55}

\bibitem[{{Jos{\'e}} {et~al.}(2010){Jos{\'e}}, {Moreno}, {Parikh}, \&
  {Iliadis}}]{2010ApJS..189..204J}
{Jos{\'e}}, J., {Moreno}, F., {Parikh}, A., \& {Iliadis}, C. 2010, \apjs, 189,
  204, \dodoi{10.1088/0067-0049/189/1/204}

\bibitem[{{Kajava} {et~al.}(2017){Kajava}, {S{\'a}nchez-Fern{\'a}ndez},
  {Kuulkers}, \& {Poutanen}}]{2017A&A...599A..89K}
{Kajava}, J.~J.~E., {S{\'a}nchez-Fern{\'a}ndez}, C., {Kuulkers}, E., \&
  {Poutanen}, J. 2017, \aap, 599, A89, \dodoi{10.1051/0004-6361/201629542}

\bibitem[{{Keek} {et~al.}(2018{\natexlab{a}}){Keek}, {Arzoumanian}, {Bult},
  {Cackett}, {Chakrabarty}, {Chenevez}, {Fabian}, {Gendreau}, {Guillot},
  {G{\"u}ver}, {Homan}, {Jaisawal}, {Lamb}, {Ludlam}, {Mahmoodifar},
  {Markwardt}, {Miller}, {Prigozhin}, {Soong}, {Strohmayer}, \&
  {Wolff}}]{2018ApJ...855L...4K}
{Keek}, L., {Arzoumanian}, Z., {Bult}, P., {et~al.} 2018{\natexlab{a}}, \apjl,
  855, L4, \dodoi{10.3847/2041-8213/aab104}

\bibitem[{{Keek} {et~al.}(2018{\natexlab{b}}){Keek}, {Arzoumanian},
  {Chakrabarty}, {Chenevez}, {Gendreau}, {Guillot}, {G{\"u}ver}, {Homan},
  {Jaisawal}, {LaMarr}, {Lamb}, {Mahmoodifar}, {Markwardt}, {Okajima},
  {Strohmayer}, \& {in 't Zand}}]{2018ApJ...856L..37K}
{Keek}, L., {Arzoumanian}, Z., {Chakrabarty}, D., {et~al.} 2018{\natexlab{b}},
  \apjl, 856, L37, \dodoi{10.3847/2041-8213/aab904}

\bibitem[{{Kellogg} {et~al.}(1971){Kellogg}, {Gursky}, {Murray}, {Tananbaum},
  \& {Giacconi}}]{1971ApJ...169L..99K}
{Kellogg}, E., {Gursky}, H., {Murray}, S., {Tananbaum}, H., \& {Giacconi}, R.
  1971, \apjl, 169, L99, \dodoi{10.1086/180820}

\bibitem[{{Kuulkers} {et~al.}(2003){Kuulkers}, {den Hartog}, {in't Zand},
  {Verbunt}, {Harris}, \& {Cocchi}}]{2003A&A...399..663K}
{Kuulkers}, E., {den Hartog}, P.~R., {in't Zand}, J.~J.~M., {et~al.} 2003,
  \aap, 399, 663, \dodoi{10.1051/0004-6361:20021781}

\bibitem[{{Lamb} \& {Lamb}(1978)}]{1978ApJ...220..291L}
{Lamb}, D.~Q., \& {Lamb}, F.~K. 1978, \apj, 220, 291, \dodoi{10.1086/155905}

\bibitem[{{Lewin} {et~al.}(1976){Lewin}, {Clark}, \&
  {Doty}}]{1976IAUC.2922....1L}
{Lewin}, W.~H.~G., {Clark}, G., \& {Doty}, J. 1976, \iaucirc, 2922, 1

\bibitem[{{Mahmoodifar} {et~al.}(2019){Mahmoodifar}, {Strohmayer}, {Bult},
  {Altamirano}, {Arzoumanian}, {Chakrabarty}, {Gendreau}, {Guillot}, {Homan},
  {Jaisawal}, {Keek}, \& {Wolff}}]{2019ApJ...878..145M}
{Mahmoodifar}, S., {Strohmayer}, T.~E., {Bult}, P., {et~al.} 2019, \apj, 878,
  145, \dodoi{10.3847/1538-4357/ab20c4}

\bibitem[{{Mondal} {et~al.}(2017){Mondal}, {Pahari}, {Dewangan}, {Misra}, \&
  {Raychaudhuri}}]{2017MNRAS.466.4991M}
{Mondal}, A.~S., {Pahari}, M., {Dewangan}, G.~C., {Misra}, R., \&
  {Raychaudhuri}, B. 2017, \mnras, 466, 4991, \dodoi{10.1093/mnras/stx039}

\bibitem[{{Muno} {et~al.}(2002{\natexlab{a}}){Muno}, {Chakrabarty}, {Galloway},
  \& {Psaltis}}]{2002ApJ...580.1048M}
{Muno}, M.~P., {Chakrabarty}, D., {Galloway}, D.~K., \& {Psaltis}, D.
  2002{\natexlab{a}}, \apj, 580, 1048, \dodoi{10.1086/343793}

\bibitem[{{Muno} {et~al.}(2002{\natexlab{b}}){Muno}, {{\"O}zel}, \&
  {Chakrabarty}}]{2002ApJ...581..550M}
{Muno}, M.~P., {{\"O}zel}, F., \& {Chakrabarty}, D. 2002{\natexlab{b}}, \apj,
  581, 550, \dodoi{10.1086/344152}

\bibitem[{{Okajima} {et~al.}(2016){Okajima}, {Soong}, {Balsamo}, {Enoto},
  {Olsen}, {Koenecke}, {Lozipone}, {Kearney}, {Fitzsimmons}, {Numata},
  {Kenyon}, {Arzoumanian}, \& {Gendreau}}]{2016SPIE.9905E..4XO}
{Okajima}, T., {Soong}, Y., {Balsamo}, E.~R., {et~al.} 2016, in Society of
  Photo-Optical Instrumentation Engineers (SPIE) Conference Series, Vol. 9905,
  Space Telescopes and Instrumentation 2016: Ultraviolet to Gamma Ray, ed.
  J.-W.~A. {den Herder}, T.~{Takahashi}, \& M.~{Bautz}, 99054X,
  \dodoi{10.1117/12.2234436}

\bibitem[{{{\"O}zel} \& {Freire}(2016)}]{2016ARA&A..54..401O}
{{\"O}zel}, F., \& {Freire}, P. 2016, \araa, 54, 401,
  \dodoi{10.1146/annurev-astro-081915-023322}

\bibitem[{{{\"O}zel} {et~al.}(2016){{\"O}zel}, {Psaltis}, {G{\"u}ver}, {Baym},
  {Heinke}, \& {Guillot}}]{2016ApJ...820...28O}
{{\"O}zel}, F., {Psaltis}, D., {G{\"u}ver}, T., {et~al.} 2016, \apj, 820, 28,
  \dodoi{10.3847/0004-637X/820/1/28}

\bibitem[{{Patruno} {et~al.}(2009){Patruno}, {Altamirano}, {Hessels},
  {Casella}, {Wijnands}, \& {van der Klis}}]{2009ApJ...690.1856P}
{Patruno}, A., {Altamirano}, D., {Hessels}, J. W.~T., {et~al.} 2009, \apj, 690,
  1856, \dodoi{10.1088/0004-637X/690/2/1856}

\bibitem[{{Qiao} \& {Liu}(2019)}]{2019MNRAS.487.1626Q}
{Qiao}, E., \& {Liu}, B.~F. 2019, \mnras, 487, 1626,
  \dodoi{10.1093/mnras/stz1365}

\bibitem[{{Remillard} {et~al.}(2022){Remillard}, {Loewenstein}, {Steiner},
  {Prigozhin}, {LaMarr}, {Enoto}, {Gendreau}, {Arzoumanian}, {Markwardt},
  {Basak}, {Stevens}, {Ray}, {Altamirano}, \& {Buisson}}]{2022AJ....163..130R}
{Remillard}, R.~A., {Loewenstein}, M., {Steiner}, J.~F., {et~al.} 2022, \aj,
  163, 130, \dodoi{10.3847/1538-3881/ac4ae6}

\bibitem[{{Shaposhnikov} {et~al.}(2003){Shaposhnikov}, {Titarchuk}, \&
  {Haberl}}]{2003ApJ...593L..35S}
{Shaposhnikov}, N., {Titarchuk}, L., \& {Haberl}, F. 2003, \apjl, 593, L35,
  \dodoi{10.1086/378255}

\bibitem[{{Sleator} {et~al.}(2016){Sleator}, {Tomsick}, {King}, {Miller},
  {Boggs}, {Bachetti}, {Barret}, {Chenevez}, {Christensen}, {Craig}, {Hailey},
  {Harrison}, {Rahoui}, {Stern}, {Walton}, \& {Zhang}}]{2016ApJ...827..134S}
{Sleator}, C.~C., {Tomsick}, J.~A., {King}, A.~L., {et~al.} 2016, \apj, 827,
  134, \dodoi{10.3847/0004-637X/827/2/134}

\bibitem[{{Speicher} {et~al.}(2022){Speicher}, {Ballantyne}, \&
  {Fragile}}]{2022MNRAS.509.1736S}
{Speicher}, J., {Ballantyne}, D.~R., \& {Fragile}, P.~C. 2022, \mnras, 509,
  1736, \dodoi{10.1093/mnras/stab3087}

\bibitem[{{Spitkovsky} {et~al.}(2002){Spitkovsky}, {Levin}, \&
  {Ushomirsky}}]{2002ApJ...566.1018S}
{Spitkovsky}, A., {Levin}, Y., \& {Ushomirsky}, G. 2002, \apj, 566, 1018,
  \dodoi{10.1086/338040}

\bibitem[{Strohmayer \& Bildsten(2006)}]{2006csxs.book..113S}
Strohmayer, T., \& Bildsten, L. 2006, New views of thermonuclear bursts,
  Cambridge Astrophysics (Cambridge University Press), 113–156,
  \dodoi{10.1017/CBO9780511536281.004}

\bibitem[{{Strohmayer} \& {Lee}(1996)}]{1996ApJ...467..773S}
{Strohmayer}, T.~E., \& {Lee}, U. 1996, \apj, 467, 773, \dodoi{10.1086/177651}

\bibitem[{{Strohmayer} {et~al.}(1997){Strohmayer}, {Zhang}, \&
  {Swank}}]{1997ApJ...487L..77S}
{Strohmayer}, T.~E., {Zhang}, W., \& {Swank}, J.~H. 1997, \apjl, 487, L77,
  \dodoi{10.1086/310880}

\bibitem[{{Strohmayer} {et~al.}(1996){Strohmayer}, {Zhang}, {Swank}, {Smale},
  {Titarchuk}, {Day}, \& {Lee}}]{1996ApJ...469L...9S}
{Strohmayer}, T.~E., {Zhang}, W., {Swank}, J.~H., {et~al.} 1996, \apjl, 469,
  L9, \dodoi{10.1086/310261}

\bibitem[{Van Der~Walt {et~al.}(2011)Van Der~Walt, Colbert, \&
  Varoquaux}]{van2011numpy}
Van Der~Walt, S., Colbert, S.~C., \& Varoquaux, G. 2011, Computing in Science
  \& Engineering, 13, 22

\bibitem[{{van Paradijs}(1979)}]{1979ApJ...234..609V}
{van Paradijs}, J. 1979, \apj, 234, 609, \dodoi{10.1086/157535}

\bibitem[{{van Paradijs} {et~al.}(1988){van Paradijs}, {Penninx}, \&
  {Lewin}}]{1988MNRAS.233..437V}
{van Paradijs}, J., {Penninx}, W., \& {Lewin}, W.~H.~G. 1988, \mnras, 233, 437,
  \dodoi{10.1093/mnras/233.2.437}

\bibitem[{{van Straaten} {et~al.}(2001){van Straaten}, {van der Klis},
  {Kuulkers}, \& {M{\'e}ndez}}]{2001ApJ...551..907V}
{van Straaten}, S., {van der Klis}, M., {Kuulkers}, E., \& {M{\'e}ndez}, M.
  2001, \apj, 551, 907, \dodoi{10.1086/320234}

\bibitem[{{Verdhan Chauhan} {et~al.}(2017){Verdhan Chauhan}, {Yadav}, {Misra},
  {Agrawal}, {Antia}, {Pahari}, {Sridhar}, {Dedhia}, {Katoch}, {Madhwani},
  {Manchanda}, {Paul}, \& {Shah}}]{2017ApJ...841...41V}
{Verdhan Chauhan}, J., {Yadav}, J.~S., {Misra}, R., {et~al.} 2017, \apj, 841,
  41, \dodoi{10.3847/1538-4357/aa6d7e}

\bibitem[{{Verner} \& {Yakovlev}(1995)}]{1995A&AS..109..125V}
{Verner}, D.~A., \& {Yakovlev}, D.~G. 1995, \aaps, 109, 125

\bibitem[{{Vincentelli} {et~al.}(2020){Vincentelli}, {Cavecchi}, {Casella},
  {Migliari}, {Altamirano}, {Belloni}, \& {Diaz-Trigo}}]{2020MNRAS.495L..37V}
{Vincentelli}, F.~M., {Cavecchi}, Y., {Casella}, P., {et~al.} 2020, \mnras,
  495, L37, \dodoi{10.1093/mnrasl/slaa049}

\bibitem[{{Vincentelli} {et~al.}(2023){Vincentelli}, {Casella}, {Borghese},
  {Cavecchi}, {Mastroserio}, {Stella}, {Altamirano}, {Padilla}, {Baglio},
  {Belloni}, {Casares}, {C{\'u}neo}, {Degenaar}, {Trigo}, {Fender},
  {Maccarone}, {Malzac}, {S{\'a}nchez}, {Middleton}, {Migliari},
  {Mu{\~n}oz-Darias}, {O'Brien}, {Panizo-Espinar}, {S{\'a}nchez-Sierras},
  {Russell}, \& {Uttley}}]{Vincentelli2023}
{Vincentelli}, F.~M., {Casella}, P., {Borghese}, A., {et~al.} 2023, \mnras,
  525, 2509, \dodoi{10.1093/mnras/stad2414}

\bibitem[{{Wang} {et~al.}(2019){Wang}, {M{\'e}ndez}, {Altamirano}, {Zhang},
  {Belloni}, {Ribeiro}, {Linares}, {Sanna}, {Motta}, \&
  {Tomsick}}]{2019MNRAS.484.3004W}
{Wang}, Y., {M{\'e}ndez}, M., {Altamirano}, D., {et~al.} 2019, \mnras, 484,
  3004, \dodoi{10.1093/mnras/stz169}

\bibitem[{{Watts}(2012)}]{2012ARA&A..50..609W}
{Watts}, A.~L. 2012, \araa, 50, 609,
  \dodoi{10.1146/annurev-astro-040312-132617}

\bibitem[{Watts {et~al.}(2005)Watts, Strohmayer, \& Markwardt}]{Watts_2005}
Watts, A.~L., Strohmayer, T.~E., \& Markwardt, C.~B. 2005, The Astrophysical
  Journal, 634, 547, \dodoi{10.1086/496953}

\bibitem[{{W}es {M}c{K}inney(2010)}]{mckinney-proc-scipy-2010}
{W}es {M}c{K}inney. 2010, in {P}roceedings of the 9th {P}ython in {S}cience
  {C}onference, ed. {S}t\'efan van~der {W}alt \& {J}arrod {M}illman, 56 -- 61,
  \dodoi{10.25080/Majora-92bf1922-00a}

\bibitem[{{Wijnands} {et~al.}(2001){Wijnands}, {Strohmayer}, \&
  {Franco}}]{2001ApJ...549L..71W}
{Wijnands}, R., {Strohmayer}, T., \& {Franco}, L.~M. 2001, \apjl, 549, L71,
  \dodoi{10.1086/319128}

\bibitem[{{Wilms} {et~al.}(2000){Wilms}, {Allen}, \&
  {McCray}}]{2000ApJ...542..914W}
{Wilms}, J., {Allen}, A., \& {McCray}, R. 2000, \apj, 542, 914,
  \dodoi{10.1086/317016}

\bibitem[{{Woosley} {et~al.}(2004){Woosley}, {Heger}, {Cumming}, {Hoffman},
  {Pruet}, {Rauscher}, {Fisker}, {Schatz}, {Brown}, \&
  {Wiescher}}]{tugba_boztepe_2022_7600607}
{Woosley}, S.~E., {Heger}, A., {Cumming}, A., {et~al.} 2004, \apjs, 151, 75,
  \dodoi{10.1086/381533}

\bibitem[{{Worpel} {et~al.}(2013){Worpel}, {Galloway}, \&
  {Price}}]{2013ApJ...772...94W}
{Worpel}, H., {Galloway}, D.~K., \& {Price}, D.~J. 2013, \apj, 772, 94,
  \dodoi{10.1088/0004-637X/772/2/94}

\bibitem[{{Worpel} {et~al.}(2015){Worpel}, {Galloway}, \&
  {Price}}]{2015ApJ...801...60W}
---. 2015, \apj, 801, 60, \dodoi{10.1088/0004-637X/801/1/60}

\bibitem[{{Wroblewski} {et~al.}(2008){Wroblewski}, {Guver}, \&
  {Ozel}}]{2008arXiv0810.0007W}
{Wroblewski}, P., {Guver}, T., \& {Ozel}, F. 2008, arXiv e-prints,
  arXiv:0810.0007, \dodoi{10.48550/arXiv.0810.0007}

\bibitem[{{Zhang} {et~al.}(2016){Zhang}, {M{\'e}ndez}, {Zamfir}, \&
  {Cumming}}]{2016MNRAS.455.2004Z}
{Zhang}, G., {M{\'e}ndez}, M., {Zamfir}, M., \& {Cumming}, A. 2016, \mnras,
  455, 2004, \dodoi{10.1093/mnras/stv2482}

\end{thebibliography}
\bibliographystyle{aasjournal}

%% This command is needed to show the entire author+affiliation list when
%% the collaboration and author truncation commands are used.  It has to
%% go at the end of the manuscript.
%\allauthors

%% Include this line if you are using the \added, \replaced, \deleted
%% commands to see a summary list of all changes at the end of the article.
%\listofchanges

\appendix
\restartappendixnumbering
\section{Light curves of detected bursts}

Light curves of each burst as observed in the 0.5$-$10~keV range are given together with the burst start, decaying e-folding and decay times. Z$^2$ contours are also shown in cases where a significant detection is observed.

%%%%%%%%%%%%%%%%%%%%%%%%%%%%%%%%%%%%
\begin{figure*}
	\includegraphics[scale=0.26]{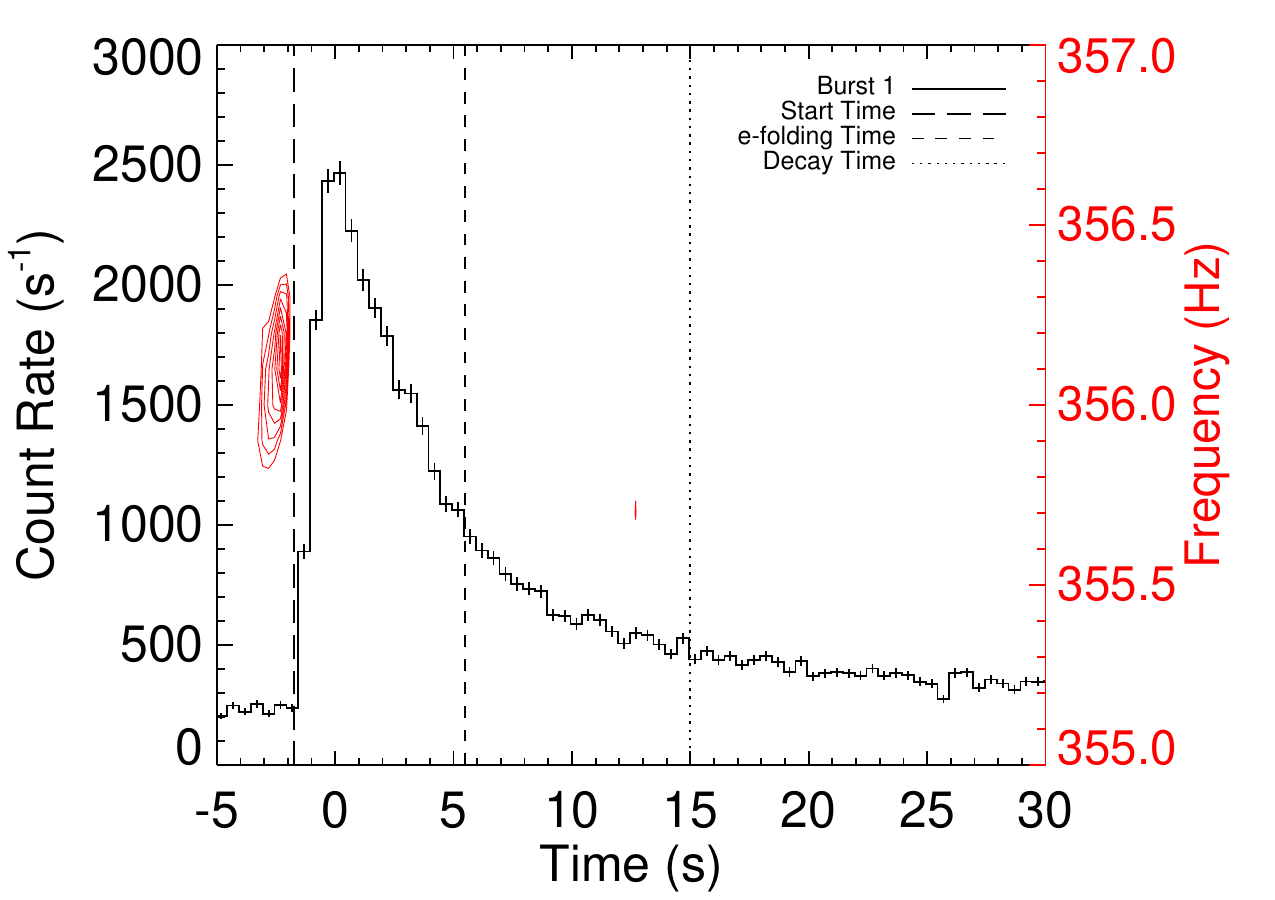}
    \includegraphics[scale=0.26]{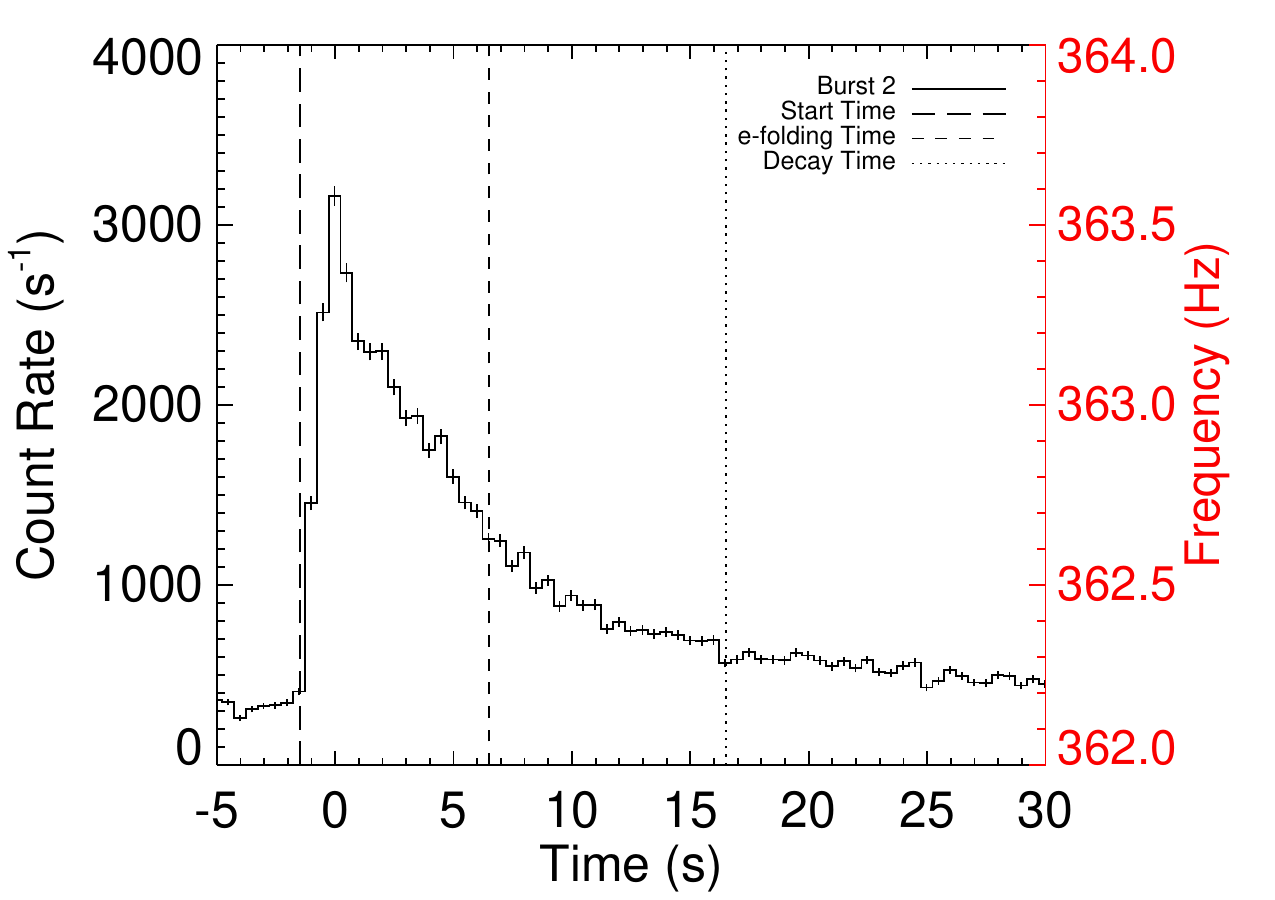}
    \includegraphics[scale=0.26]{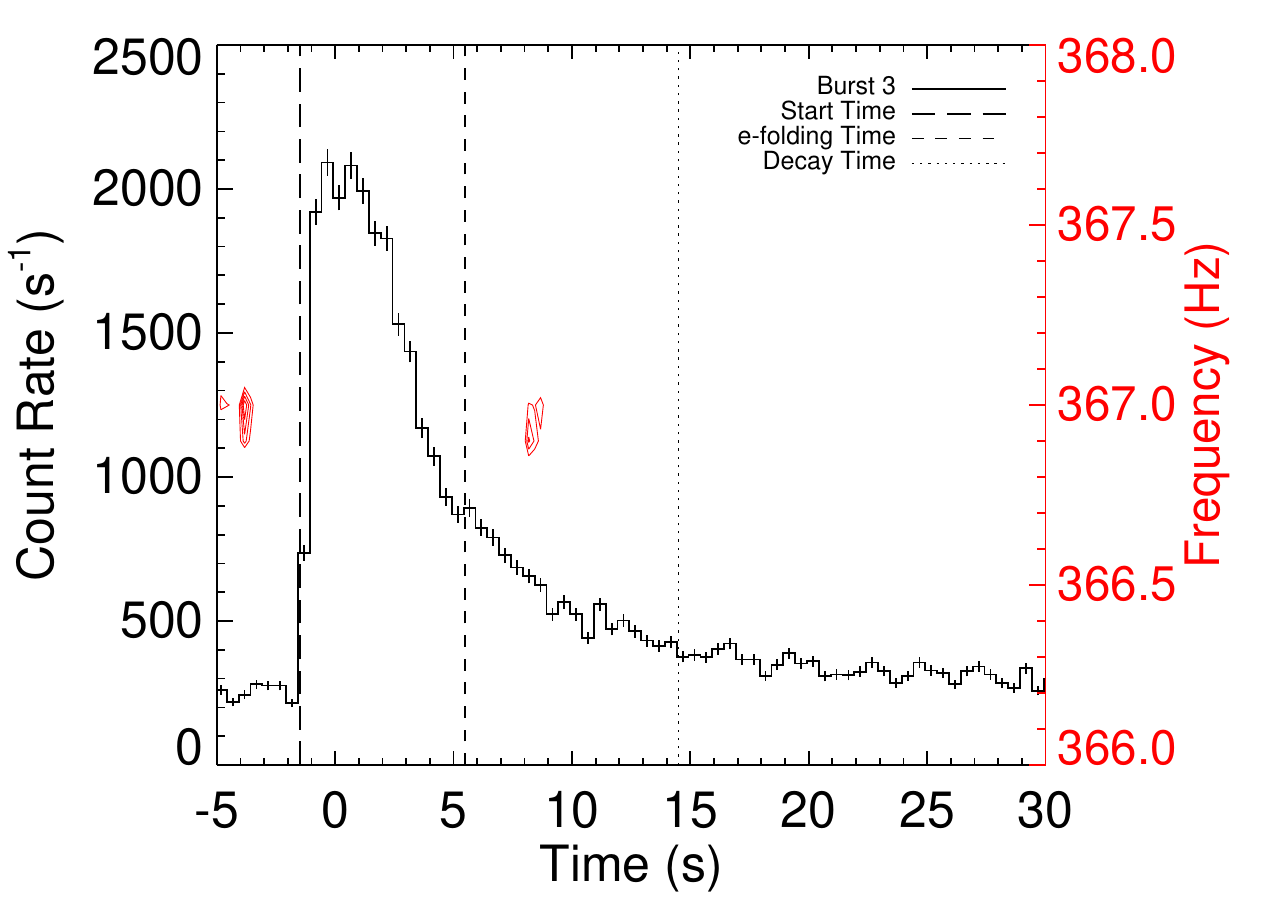}
    \includegraphics[scale=0.26]{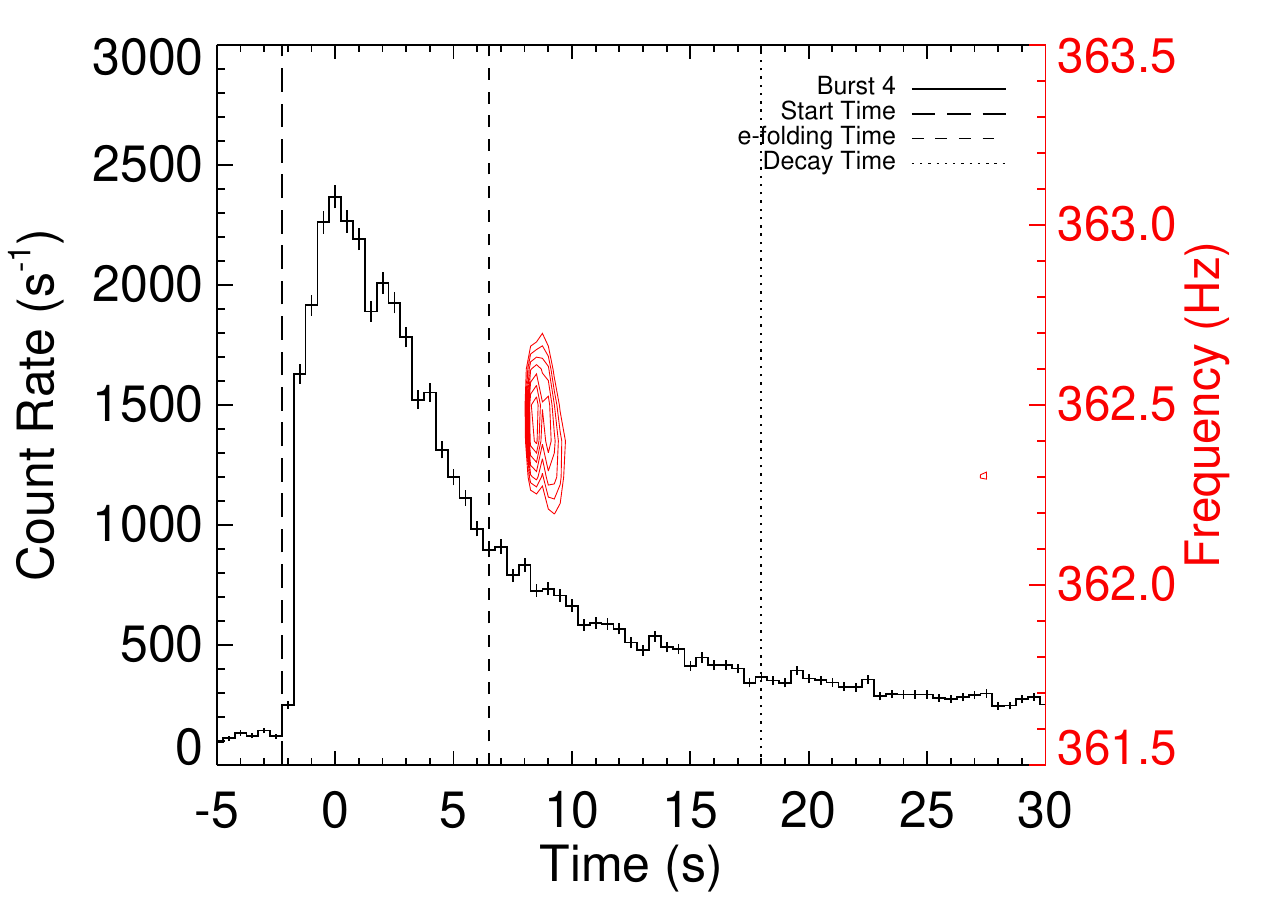}
    \includegraphics[scale=0.26]{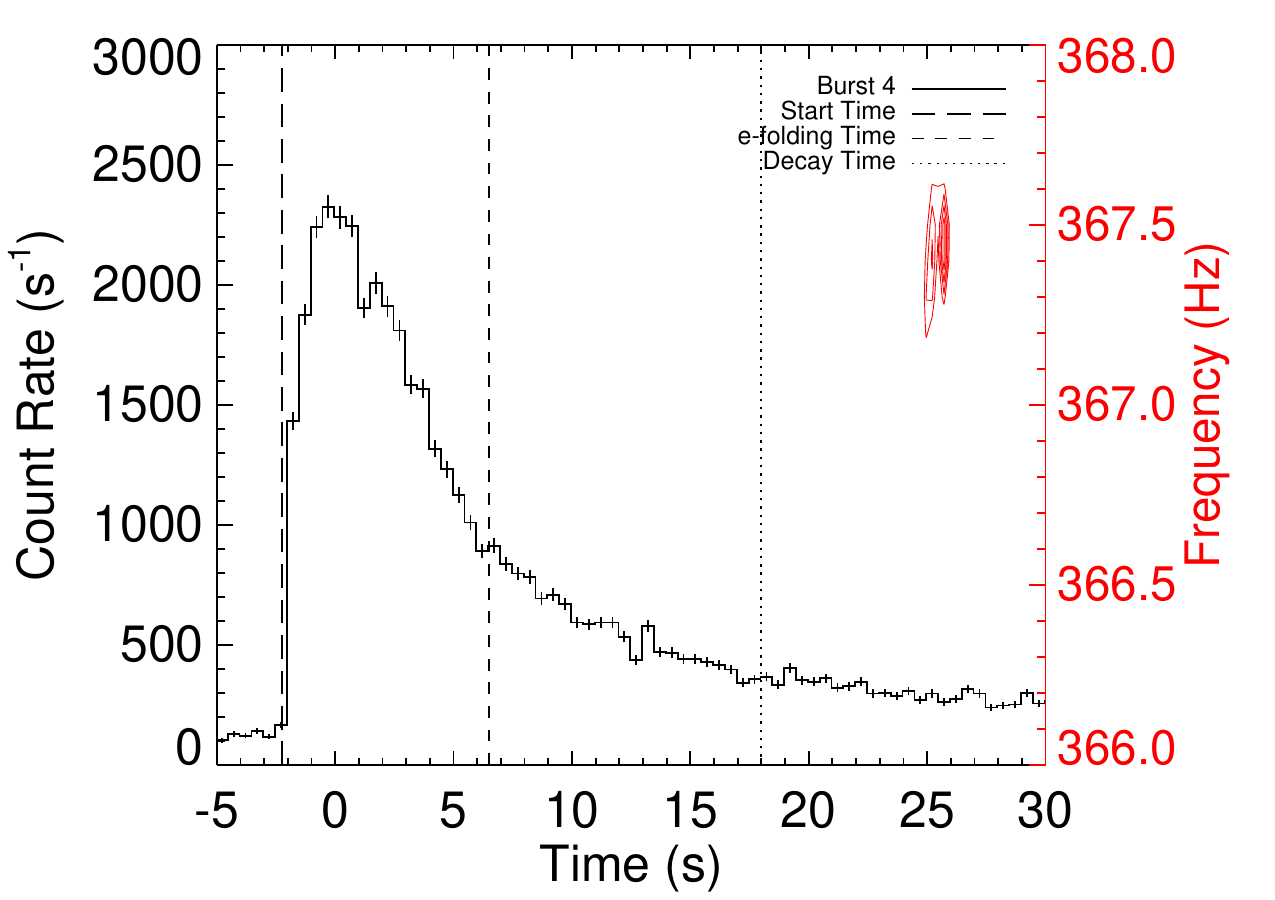}
    \includegraphics[scale=0.26]{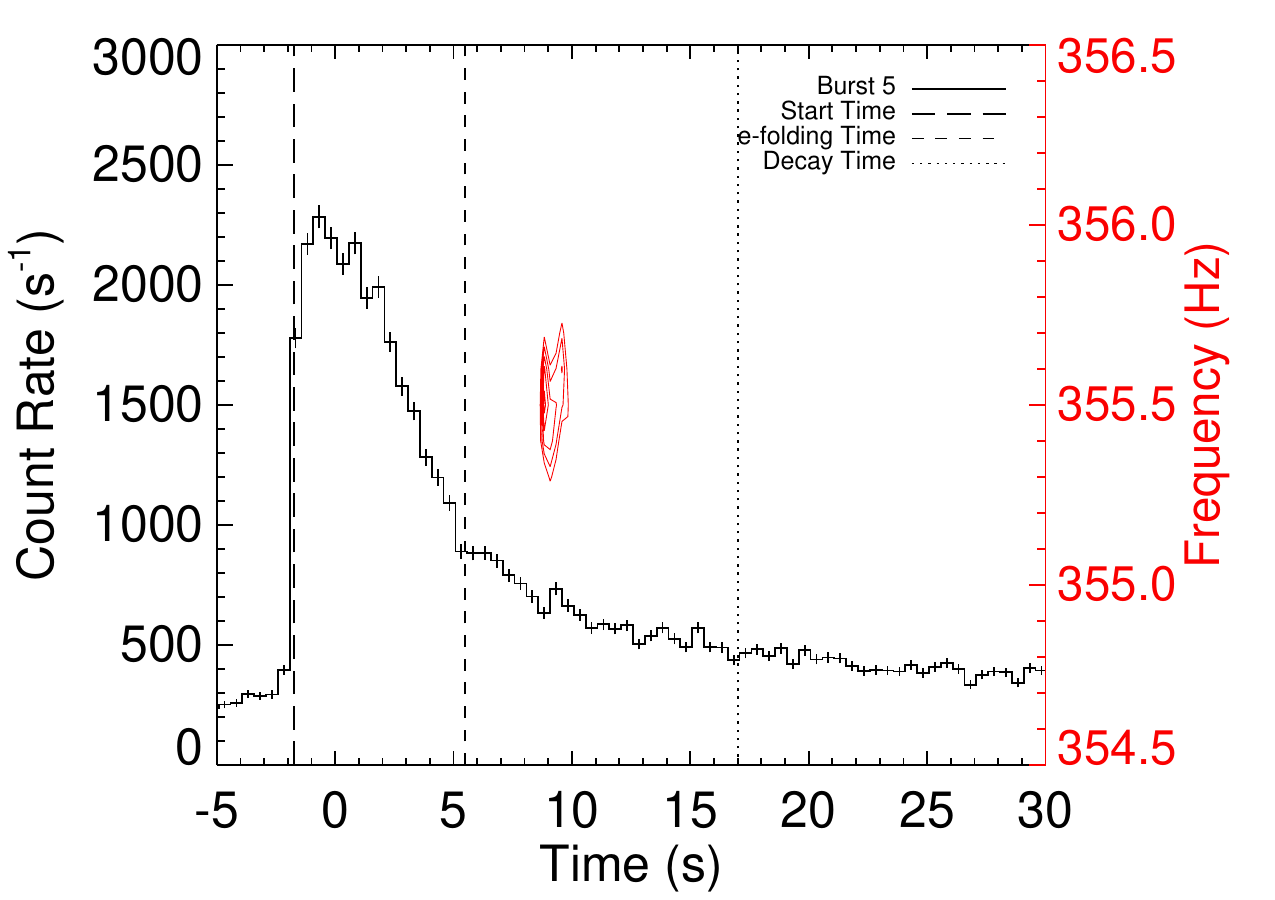}
    \includegraphics[scale=0.26]{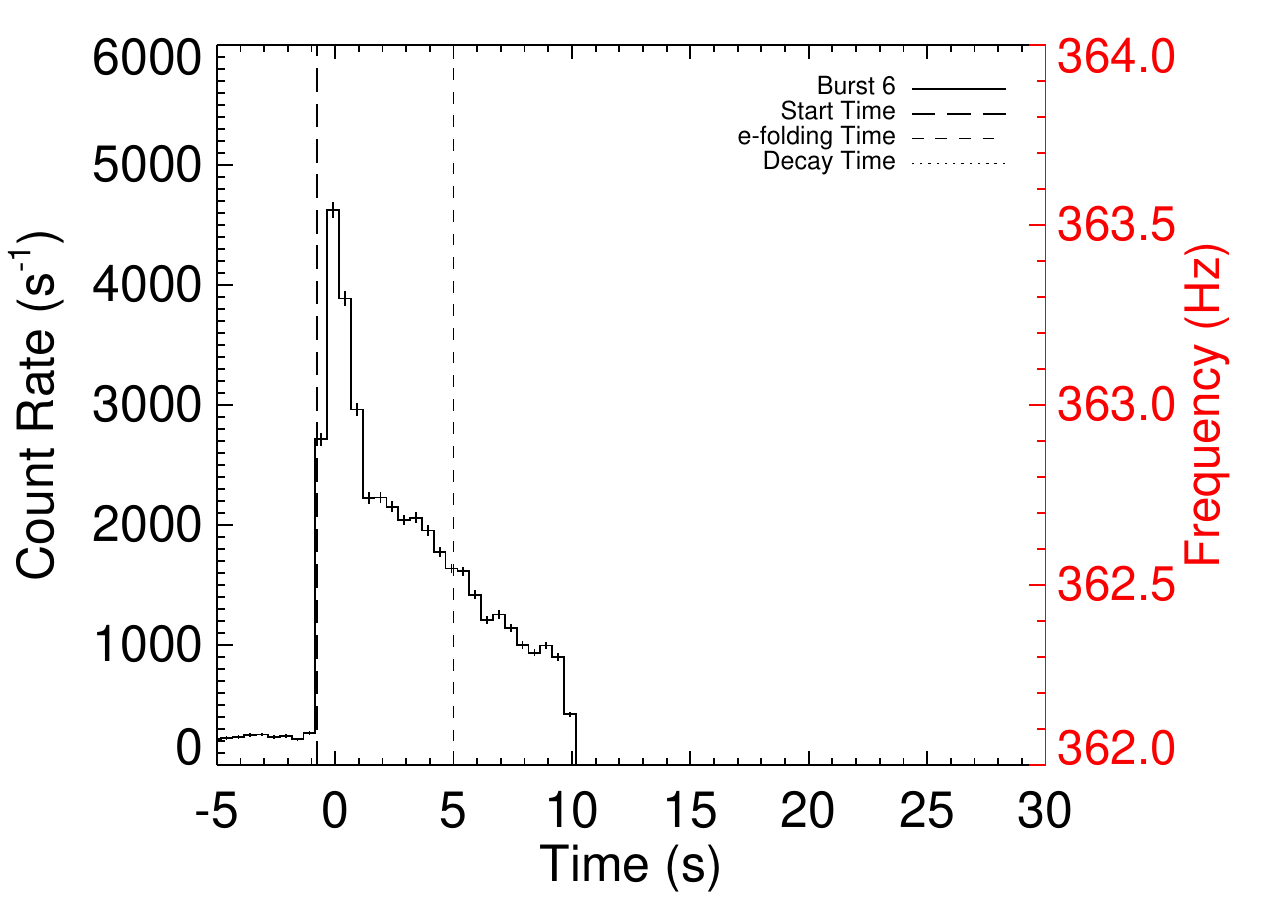}
    \includegraphics[scale=0.26]{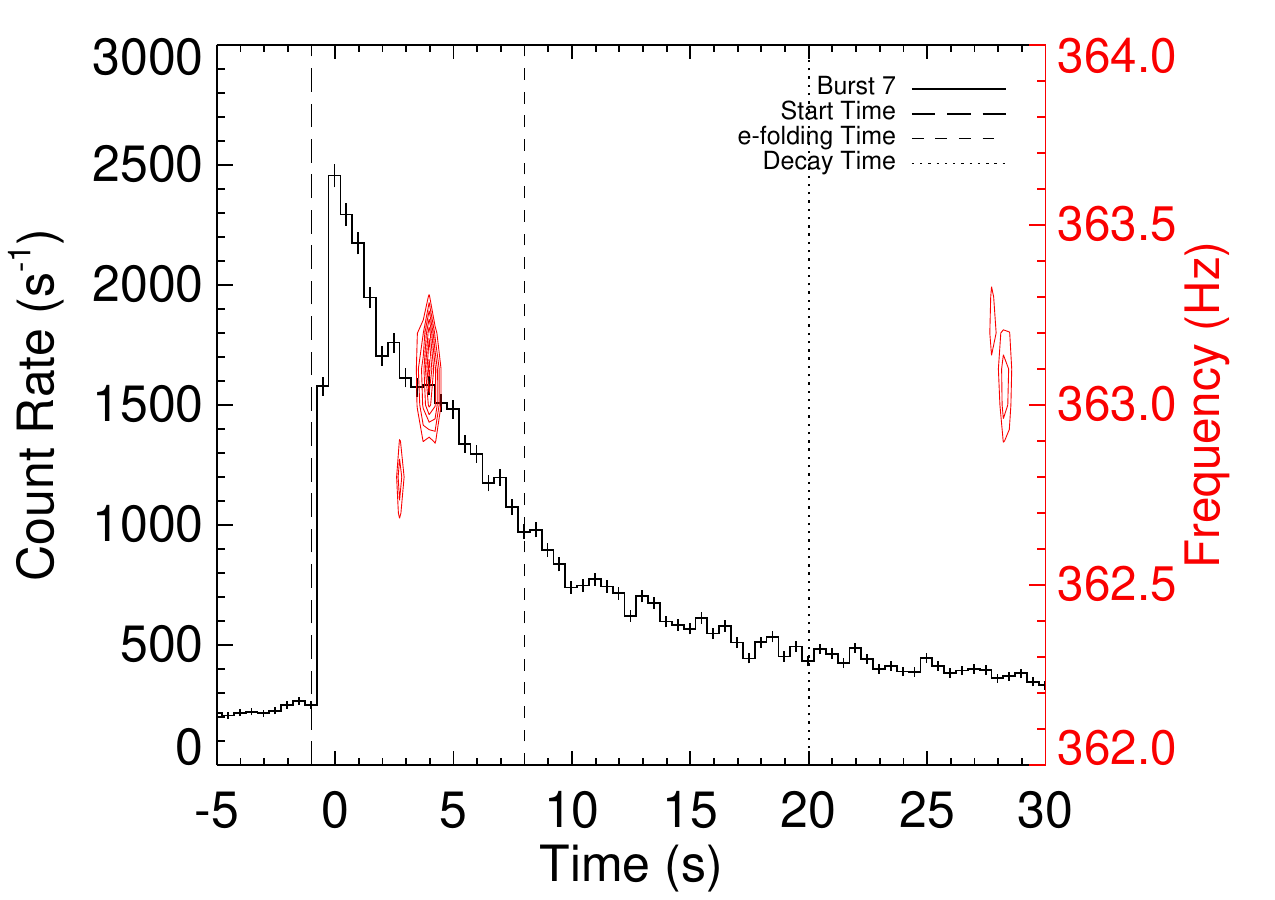}
    \includegraphics[scale=0.26]{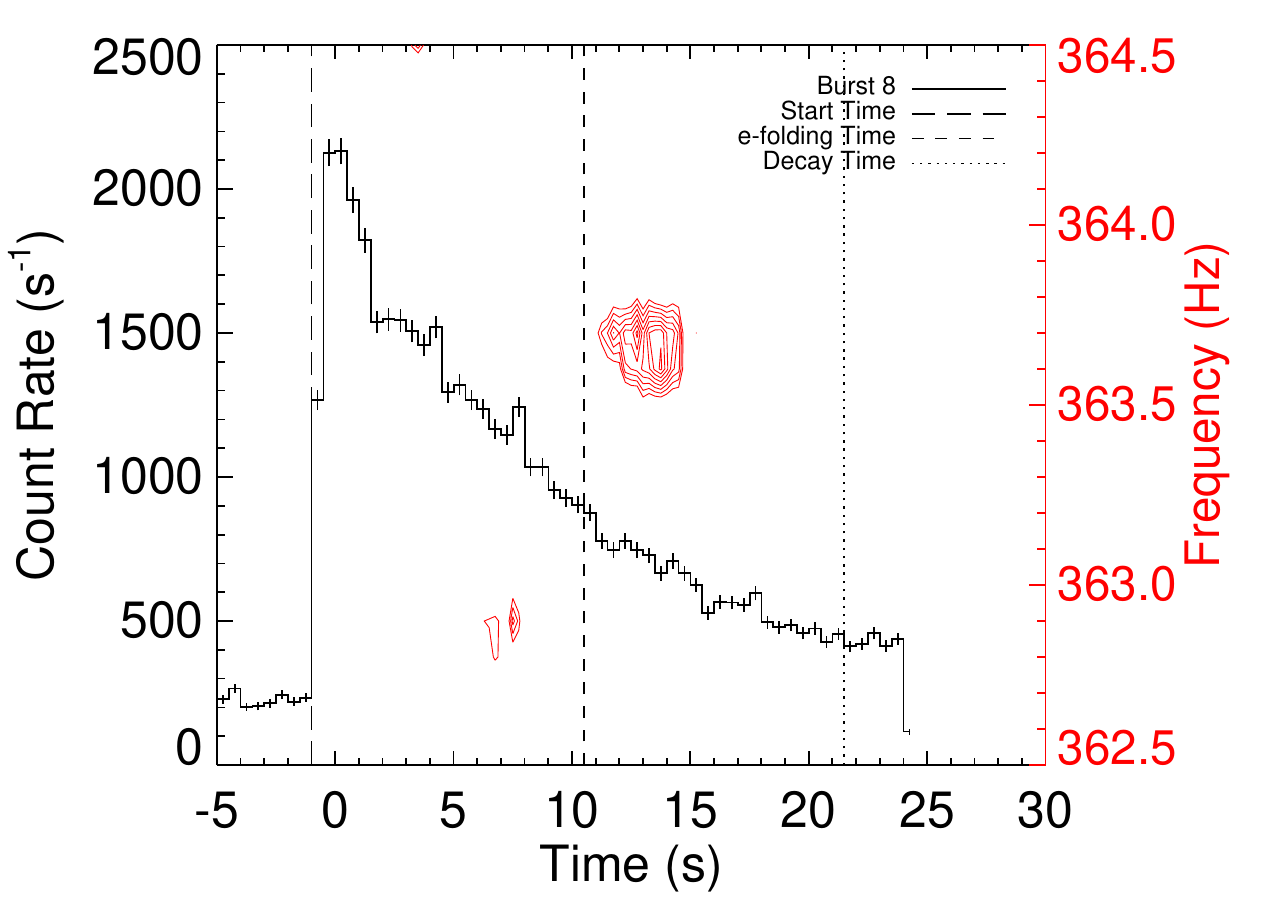}
    \includegraphics[scale=0.26]{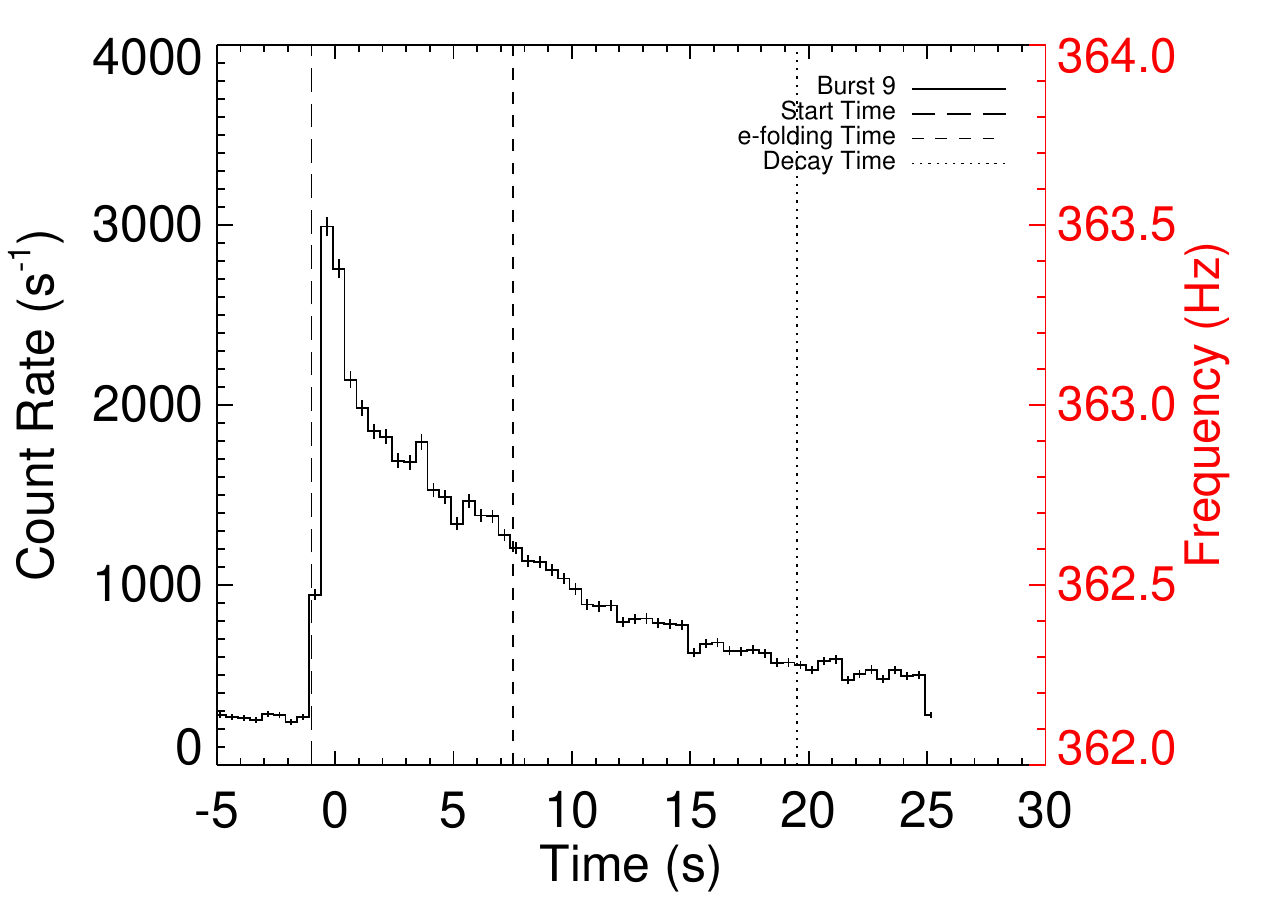}
    \includegraphics[scale=0.26]{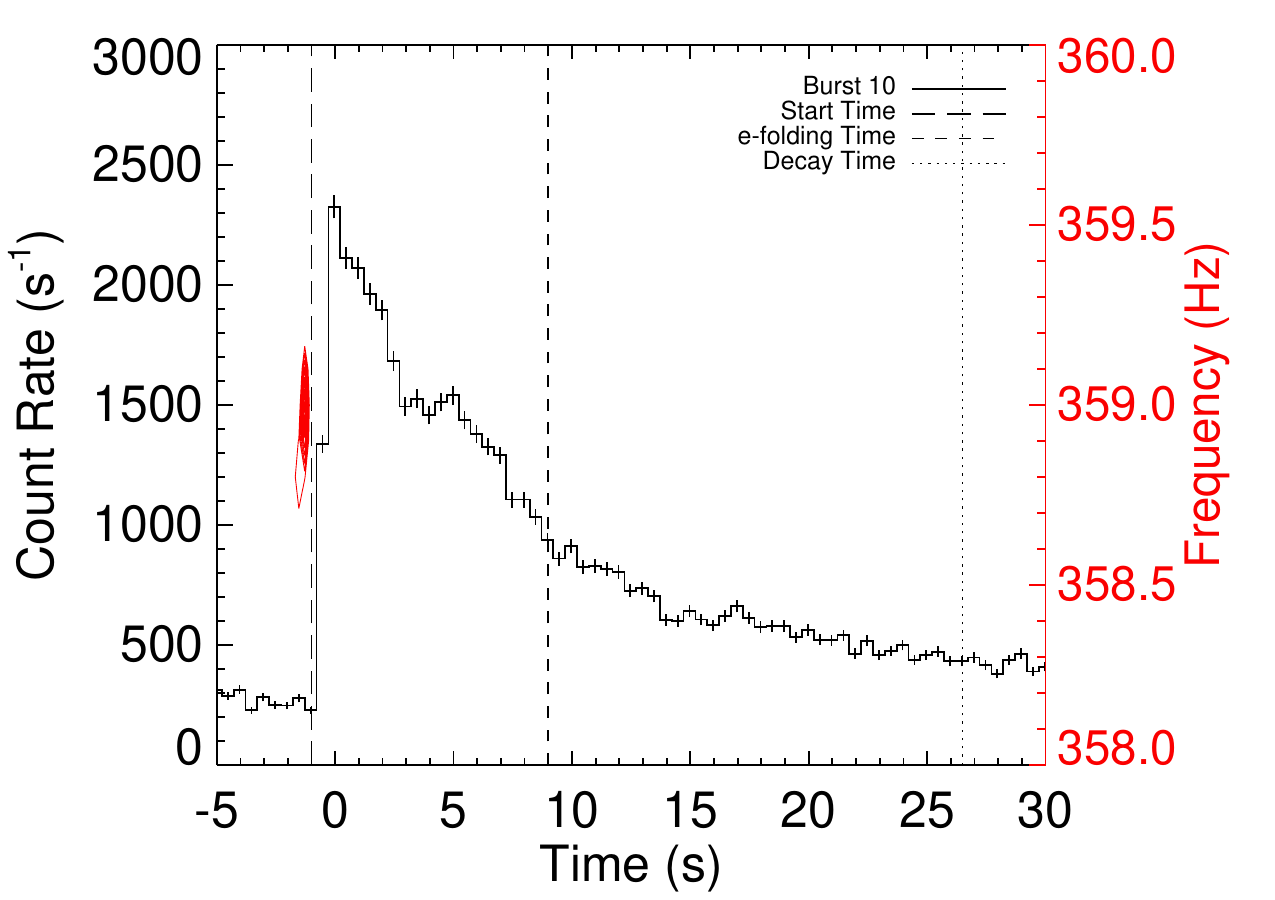}
    \includegraphics[scale=0.26]{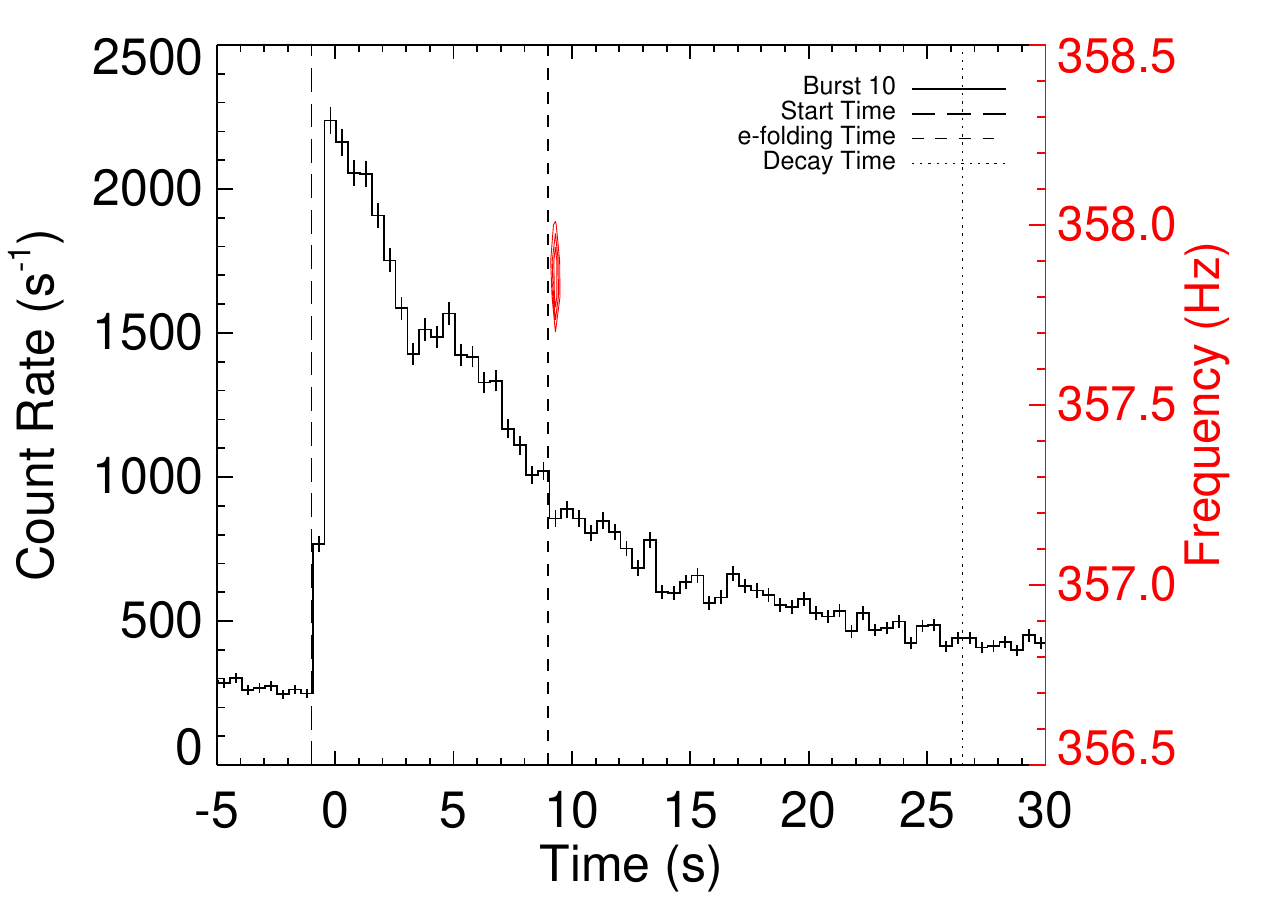}
    \includegraphics[scale=0.26]{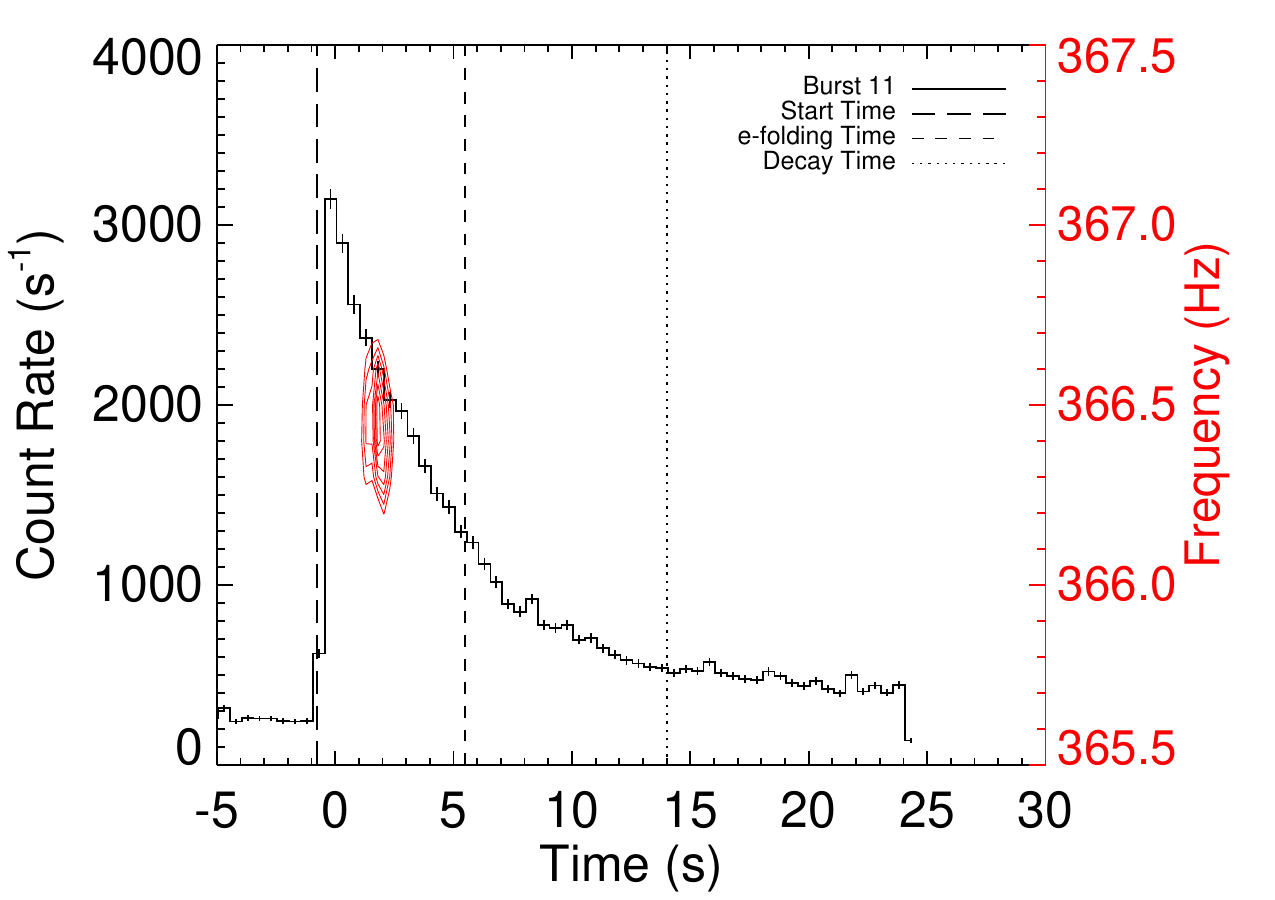}
    \caption{0.5$-$10~keV light curves of detected thermonuclear X-ray bursts with contours of the candidate oscillations listed in Table~\ref{tab:bursts_oscil}. For bursts 4 and 10 we detect oscillations at two different frequencies, we therefore show these bursts twice. The vertical lines show the start time, the e-folding time and the decay length, defined as the time the count rate declines to 10\% of the peak.}
    \label{fig:burst_lc2} 
\end{figure*}
%%%%%%%%%%%%%%%%%%%%%%%%%%%%%%%%%%%%

\end{document}